\documentclass[twocolumn,apj,numberedappendix,twocolappendix]{openjournal}

\usepackage{lipsum}

\usepackage{xcolor}
\usepackage{textgreek}
\usepackage[utf8]{inputenc}

\usepackage{graphicx}
\usepackage{amssymb}
\usepackage{appendix}
\usepackage{float}
\usepackage{hyperref}  
\usepackage{makecell}
\usepackage{amsmath}
\usepackage{txfonts}

\usepackage{hyperref}
\hypersetup{
    unicode, 
    colorlinks=true,
    linkcolor=linkcolor,
    citecolor=linkcolor,
    filecolor=linkcolor,
    urlcolor=linkcolor,
}
\usepackage{color,colortbl}
\definecolor{linkcolor}{rgb}{0.0,0.3,0.5}
\usepackage{tensind}
\tensordelimiter{?}
\DeclareGraphicsExtensions{.bmp,.png,.jpg,.pdf}
\usepackage{verbatim}
\usepackage[normalem]{ulem}
\usepackage{orcidlink}
\usepackage{soul}
\DeclareUnicodeCharacter{0306}{}

\newcommand{\orcidauthor}[3]{\author{\href{http://orcid.org/#1}{#2$^{#3}$}}}
  
\graphicspath{ {./figs/} }

\shorttitle{Quiescent and breathing galaxies with ASTRODEEP-JWST}
\shortauthors{Merlin et al.}

\begin{document}

\title{Witnessing downsizing in the making:\\ 
quiescent and breathing galaxies at the dawn of the Universe\vspace{-1.5cm}}

\orcidauthor{0000-0001-6870-8900}{Emiliano Merlin}{1}
\orcidauthor{0000-0001-6793-2572}{Flaminia Fortuni}{1}
\orcidauthor{0000-0003-2536-1614}{Antonello Calabr\'o}{1}
\orcidauthor{0000-0001-9875-8263}{Marco Castellano}{1}
\orcidauthor{0000-0002-9334-8705}{Paola Santini}{1}
\orcidauthor{0000-0003-3820-2823}{Adriano Fontana}{1}
\orcidauthor{0009-0006-8337-8712}{Lucas C. Kimmig}{2}
\orcidauthor{0000-0001-8973-5051}{Francesco Shankar}{3}
\orcidauthor{0000-0002-8951-4408}{Lorenzo Napolitano}{1}
\orcidauthor{0000-0002-6610-2048}{Anton M. Koekemoer}{4}
\orcidauthor{0000-0003-1581-7825}{Ray A. Lucas}{4}
\orcidauthor{0000-0001-9879-7780}{Fabio Pacucci}{5,6}
\orcidauthor{0000-0003-1371-6019}{Michael C. Cooper}{7}
\orcidauthor{0000-0002-3301-3321}{Michaela Hirschmann}{8}
\orcidauthor{0000-0003-4528-5639}{Pablo G. P\'erez-Gonz\'alez}{9}
\orcidauthor{0000-0000-0000-0000}{Guillermo Barro}{10}
\orcidauthor{0000-0001-5414-5131}{Mark Dickinson}{11}
\orcidauthor{0000-0003-3248-5666}{Giovanni Gandolfi}{1,12}
\orcidauthor{}{Diego Paris}{1}
\orcidauthor{0000-0001-9440-8872}{Norman A. Grogin}{4}
\orcidauthor{0000-0002-9373-3865}{Xin Wang}{13,14,15}

\affiliation{$^1$INAF Osservatorio Astronomico di Roma, Via Frascati 33, 00078 Monteporzio Catone, Rome, Italy}
\affiliation{$^2$Universitäts-Sternwarte, Fakultät für Physik, Ludwig-Maximilians-Universität München, Scheinerstr. 1, 81679 München, Germany}
\affiliation{$^3$School of Physics and Astronomy, University of Southampton, Highfield, Southampton, SO17 1BJ, UK}
\affiliation{$^4$Space Telescope Science Institute, 3700 San Martin Drive, Baltimore, MD 21218, USA}
\affiliation{$^5$Center for Astrophysics $\vert$ Harvard \& Smithsonian, 60 Garden St, Cambridge, MA 02138, USA}
\affiliation{$^6$Black Hole Initiative, Harvard University, 20 Garden St, Cambridge, MA 02138, USA}
\affiliation{$^7$Department of Physics \& Astronomy, University of California, Irvine, 4129 Reines Hall, Irvine, CA 92697, USA}
\affiliation{$^{8}$Institute of Physics, Laboratory for galaxy evolution, EPFL, Observatoire de Sauverny, Chemin Pegasi 51, 1290 Versoix, Switzerland}
\affiliation{$^{9}$Centro de Astrobiolog\'{\i}a (CAB), CSIC-INTA, Ctra. de Ajalvir km 4, Torrej\'on de Ardoz, E-28850, Madrid, Spain}
\affiliation{$^{10}$University of the Pacific, Stockton, CA 90340 USA}
\affiliation{$^{11}$NSF's National Optical-Infrared Astronomy Research Laboratory, 950 N. Cherry Ave., Tucson, AZ 85719, USA}
\affiliation{$^{12}$INAF-Osservatorio Astronomico di Padova, Vicolo dell’Osservatorio 5, 35122 Padova, Italy}
\affiliation{$^{13}$School of Astronomy and Space Science, University of Chinese Academy of Sciences (UCAS), Beijing 100049, China}
\affiliation{$^{14}$National Astronomical Observatories, Chinese Academy of Sciences, Beijing 100101, China}
\affiliation{$^{15}$Institute for Frontiers in Astronomy and Astrophysics, Beijing Normal University, Beijing 102206, China}

\thanks{\vspace{0.1cm}$^\dagger$Corresponding author: \href{mailto:emiliano.merlin@inaf.it}{emiliano.merlin@inaf.it}}

\begin{abstract}
The existence of a population of quiescent galaxies in the early Universe has been extensively confirmed observationally. 
We conduct a systematic search for $\log(M_*/M_\odot)\geq9.5$ quiescent galaxies at $z>3$ in six extragalactic deep fields observed with NIRCam, with the goal of extracting their physical and statistical features in a uniform and self-consistent manner, thus creating a robust baseline to compare with theoretical predictions.
We exploit the ASTRODEEP-JWST photometric catalogs to single out robust candidates, including sources quenched only a few tens of Myr before the observation. We apply a SED-fitting procedure which explores three functional forms of star formation history and the $\chi^2$ probabilities of the solutions, with additional checks to minimise the contamination from interlopers, tuning our selection criteria against available spectroscopic data from the DAWN archive and simulated catalogs.
We select 633 candidates, which we rank by a reliability parameter based on the probabilities of the quiescent and alternative star-forming solutions, with 291 candidates tagged as ``gold''. According to the best-fit models, 79\% of the massive ($\log(M_*/M_\odot)\geq10.5$) quiescent galaxies at $3<z<5$ stopped forming stars at least 150 Myr before the time of observation, while 89\% of low mass sources ($9.5\leq \log(M_*/M_\odot)<10.5$) have been quenched for less than 150 Myr. The abundance of low mass old quiescent systems does not increase significantly with time from $z=5$ to 3, which we interpret as an indication that the quenching must be temporary in this mass regime: only massive galaxies appear to be permanently quenched, while low mass objects seem to be experiencing a short episode of quenching followed by re-juvenation (``breathing''), consistent with a ``downsizing'' scenario of galaxy formation; rather than inferring it from the archaeological analysis of local galaxies, we can now witness it while it is happening. Interestingly, we also find an abrupt drop in the density of massive quiescent candidates at $z>5$. We derive  estimates for the number density of early passive galaxies up to $z=10$, including corrections for completeness and contamination, and compare it against the predictions of various models: while the global number density of quenched galaxies can be reproduced by recent numerical simulations, tensions with data remain in the modeling of the observed bimodality of time passed since quenching as a function of mass.
\end{abstract}


\maketitle


\section{Introduction}
\label{sec:intro}

It is now established that a population of massive, passively evolving galaxies exists at very early epochs ($z>3$, roughly the first two Gyr of cosmic time). In the past decade, a number of pioneering studies based on photometric data from Hubble observations coupled with Spitzer and ground based surveys have been published \citep[e.g.][M18 hereafter]{Labbe2005,Mobasher2005,Fontana2009,Muzzin2013,Straatman2014,Merlin2018}, finding robust candidates up to $z\approx\!5$, for a typical number density of $n\approx\!10^{-5}$ Mpc$^{-3}$ at $3<z<4$ \citep[][M19 hereafter]{Merlin2019}. Most of the candidates have been subsequently confirmed to be quiescent using sub-mm and radio data to exclude dust emission linked to obscured ongoing star-formation \citep[e.g.][]{Santini2019}, or spectroscopy \citep[e.g.][]{Glazebrook2017,Schreiber2018b,Forrest2020a,Forrest2020b}. These passively evolving objects are now known to dominate the massive end (M$_*>10^{11}$ M$_{\odot}$) of the galaxy stellar mass function up to $z\approx\!4$ \citep{Santini2021,Santini2022}.
These searches for high-redshift intrinsically red objects were limited by the observational characteristics of the surveys: the \textit{Hubble} data are deep and high-resolution but restricted to wavelengths below $\approx\!1.6$ $\mu$m, only supplemented by lower-resolution shallower data at longer wavelengths (IRAC fluxes from \textit{Spitzer}, and ground-based $K$ band). The situation dramatically changed for the better with the advent of JWST, with improved resolution and depth compared to previous studies, and extending to longer infrared wavelengths, including those covered by \textit{Spitzer}, allowing for exquisite measurements of the optical and near-infrared rest-frame at very early epochs. 
In the first three years of observations, many works have been published presenting selections of quiescent candidates based on  NIRCam photometric data \citep[e.g.][]{Carnall2023,Valentino2023,PerezGonzalez2023,Long2024,Russell2024,delaVega2025,Baker2025b}. Spectroscopic follow-ups with NIRSpec and NIRISS have confirmed the existence of quiescent sources up to $z\approx\!7$ \citep{Looser2024,Weibel2025,Jin2024,DeGraaff2025,Baker2025}. It is still being debated when and how these objects assembled their stellar mass, quenched their star-formation (SF) activity, and which feedback mechanisms can cause such early quenching events \citep[see e.g.][]{Carnall2018,Park2024,Xie2024,Kurinchi-Vendhan2024,DeLucia2024,Kimmig2025}. Moreover, it remains to be assessed how common these objects are. While each single galaxy with a high stellar mass or star formation efficiency (as inferred from the reconstruction of its stellar mass assembly history) at very high redshift has the potential to challenge the cosmological framework \citep{Glazebrook2024,Carnall2024,PerezGonzalez2025}, it is also crucial to investigate the statistical properties of the quiescent population as a whole.

Studying the stellar population of local systems, it has been inferred for many years that galaxies form in a ``downsizing'' trend, with massive objects forming earlier and in shorter timescales than low-mass systems, and quenching their SF activity at early times \citep{Matteucci1994,Cowie1996,Thomas2005,Bundy2006,Cimatti2006}. This appears to be at odds with a naive interpretation of the $\Lambda$CDM cosmological scenario, which seems to predict that large galaxies assemble their mass by forming stars across the whole Hubble time \citep[e.g.][]{Kauffmann1993,Bower2006}, and possibly quenching only after their merging history ceases, the galaxy gas reservoir is exhausted, and no cold gas from the cosmic web is inflowing and collapsing in the galactic potential well. However, hydrodynamical cosmological simulations have proven capable of broadly reproducing the observed ``downsizing'' trend by fine-tuning the modeling of feedback from stars and active galactic nuclei (AGN), although a relevant degree of tension with the observations remained \citep[e.g.][]{Merlin2019,Gould2023,Weller2025}. Recent developments in theoretical models seem to further mitigate these tensions (see Sect. \ref{models}). 

In this work we take advantage of the ASTRODEEP-JWST photometric catalogs \citep[][M24 hereafter]{Merlin2024} to search for quiescent galaxies with stellar mass M$_*>10^{9.5}$ M$_{\odot}$ (in the following we will define $lM_\ast \equiv \log(M_\ast/M_\odot)$, so the threshold can be concisely written as lM$_*>9.5$) at $z>3$, by means of a spectral energy distribution (SED) fitting technique which builds upon the one used in M18 and M19. The goal is to infer the physical properties of the candidates, in particular the duration of their SF activity and the epochs of quenching, and to compare the observational results to the theoretical predictions, exploiting the high quality of the photometric data and complementing it with available spectroscopic information from NIRSpec.

It is important to specify what we mean with the term ``quiescent'', as various definitions have been used in the literature. Generally speaking, a galaxy can be dubbed as quiescent if at the moment of observation it is forming a negligible amount of stars with respect to the total stellar mass it has already assembled. From the point of view of the modeling, here we require that the instantaneous specific star-formation rate (SSFR) of its best-fit model from the SED-fitting process is lower than $10^{-10}$ yr$^{-1}$ (lSSFR $<-10$ for short). Note that different criteria have been proposed in the literature. For example, a common choice in recent studies is SSFR $< 0.2/t_U(z)$ \citep[where $t_U(z)$ is the age of the Universe at the redshift of observation, see e.g.][]{Carnall2023,Russell2024,Baker2025b}; this is close to our criterion at $z\approx\!3$, but becomes less stringent at higher redshifts (see also Sect. \ref{othersel} for further discussion on this). Noticeably, we will not make any pre-selection based on colors, relying only on the properties of the best-fit model from the SED-fitting runs; so, our sample will be a census of quiescent sources regardless of the amount of time passed since their quenching, potentially including objects that have stopped forming stars for as little as 10 Myr. We will further distinguish between two sub-populations: (i) recently quenched galaxies, which we dub ``post-starburst'' (PS) and we heuristically define as objects that have been quiescent for less than 150 Myr at the moment of observation (concisely, $dt_q<150$~Myr; this threshold was chosen after analysing the trends of the full sample, see Sect. \ref{downsiz}); and (ii) ``red and dead'' (RD) galaxies, for which $dt_q\geq150$~Myr.\footnote{We did not make this distinction in M18 and M19, where we used the term ``red and dead'' to define any source with lSSFR $<-11$, regardless of its $dt_q$. Also, it is worth pointing out that historically the term ``post-starburst'' has been used with different meanings.} 

The paper is organised as follows. In Sect. \ref{sec:data} we summarise the main features of the adopted dataset. In Sect. \ref{sec:meth} we discuss the techniques adopted for the selection of quiescent candidates, with additional details given in the Appendix, and Sect. \ref{sec:results} presents the results of the selection. Finally, in Sect. \ref{sec:concl} we discuss the results and add some concluding remarks. We adopt a flat $\Lambda$CDM concordance model (H$_0$ = 70.0~km~s$^{-1}$ Mpc$^{-1}$, $\Omega_M=0.30$) and AB magnitudes \citep{Oke83}.

\section{Dataset}
\label{sec:data}


The ASTRODEEP-JWST catalogs collect photometric fluxes in 16 HST+JWST bands, for $\approx\!500,000$ sources over $\approx\!0.2$ sq.deg., in six public extragalactic deep fields: 
the A2744 field including JWST observations from GLASS-JWST \citep[ERS 1324, P.I. Treu,][]{Treu2022}, UNCOVER \citep[GO 2561, P.I. Labb\'e,][]{Bezanson2024}, DDT 2756 (P.I. Chen), and GO 3990 \citep[P.I. Morishita,][]{Morishita2025}; CEERS \citep[ERS 1345, P.I. Finkelstein, data release v0.51][]{Finkelstein2025} on the EGS field; the JADES-GS (data release v2.0) and JADES-GN (v1.0) fields on the GOODS-South and GOODS-North footprints, respectively \citep[GTO 1180 and GTO 1210, P.I. Eisenstein,][]{Eisenstein2023} including FRESCO data \citep[GO 1895, P.I. Oesch,][]{Oesch2023}; the first-epoch imaging of the NGDEEP field \citep[Co-PIs Finkelstein, Papovich, Pirzkal,][]{Bagley2024}; and the PRIMER (GO-1837, P.I. Dunlop) observations of the UDS and COSMOS fields in CANDELS \citep{Grogin2011,Koekemoer2011}. More than 200 thousand sources in the catalogs have spectroscopic or photometric redshift above 3. 

The 5$\sigma$ depths of the images in 0.2'' diameter apertures range from AB $\approx\!28$ in PRIMER-UDS to AB $\approx\! 30.5$ in NGDEEP and JADES-GS, depending on the considered NIRCam bands.
The detection was performed on weighted stacks of the 3.56 and 4.44 $\mu$m NIRCam images, and aperture photometry was measured using \textsc{a-phot} \citep{Merlin2019b} on PSF-matched images. See M24 for full information. 
Four photometric redshift estimates were provided, obtained with the two codes \textsc{zphot} \citep{Fontana2000} and \textsc{EAzY} \citep{Brammer2008} using various template sets. It is worth pointing out that in M24 for \textsc{EAzY} we included the ``\texttt{z\_best}'' rather than the ``\texttt{z\_chi2}'' estimate, while the \textsc{zphot} estimate is the one corresponding to the $\chi^2$ minimum.


\section{Methods}
\label{sec:meth}

We significantly updated the methods used in M18 and M19, given the profound differences between the CANDELS and the M24 datasets (redder detection band, higher resolution of the rest-frame infrared data, increased depths), as described in the following subsections.

\subsection{Selection technique} \label{select}

In short, we performed SED-fitting on the M24 photometry using the proprietary software \textsc{zphot} \citep{Fontana2000} with a custom library of galaxy SED models, described in Appendix \ref{lib}. We then explored the distribution of the solutions in the $\chi^2$ probability space, and ranked the sources on the basis of the reliability of the redshift estimate, and the existence and probability of a quiescent best-fit and alternative SF solutions. In the following we describe the procedure in full detail.

\subsubsection{Fiducial redshift and SED-fitting}

We started by assigning a fiducial redshift $z_{best}$ to each source. For the objects with a robust spectroscopic measurement, we set $z_{best}=z_{spec}$. In M24, we provided 16,666
spectroscopic redshifts with high quality flags in the original works, 3388 of which are at $z>3$. For the present work we updated the $z>3$ list adding the new robust (\texttt{GRADE}=3) redshifts available in the DAWN JWST Archive\footnote{\url{https://dawn-cph.github.io/dja/}} at the date of this work (July 2025), resulting in a total of 6592 \citep[$\approx\!85\%$ of which are from JWST; see e.g.][for an empirical assessment of the robustness of the DJA spec-$z$'s]{Napolitano2025}. The spectroscopic sample extends down to AB $\approx\! 30$ in F444W, but our mass threshold limit practically limits the useful spectra to AB $\approx\! 26$.
For the remaining sources, we started by considering the four photometric redshift estimates provided in M24. As done there, we first computed the median $z_{med}$ of three out of the four estimates, excluding the \textsc{EAzY} estimates with \citet{Larson2023} `Lya' templates to avoid over-weighting the results from runs with similar templates. Then, we evaluated the consistency of the three values: where \textsc{zphot} was consistent with at least one of the \textsc{EAzY} estimates, i.e. $|z_{\mbox{zphot}}-z_{\mbox{EAzY}}|\leq0.1 \times z_{med}$, we set $z_{best}=z_{zphot}$; where this was not the case, but the two \textsc{EAzY} estimates were consistent, we set $z_{best}=z_{EAzY\_v1.3}$. Finally, if no combination of estimates had $|\Delta z|\leq0.1 \times z_{med}$, we considered the photometric redshift as unreliable, and we excluded the source from further analysis.
We finally dubbed as ``$z$-robust'' the sources having all three M24 photo-$z$ estimates consistent with one another.

We then proceeded to SED-fit all the sources at $z_{best}>3$ with \textsc{zphot}, using a library of $\approx\!10$ million SED models, with three analytical star formation history (SFH) functions (top-hat, exponentially declining and delayed exponentially declining). The details of the library are given in Appendix \ref{lib}. 
To be conservative, for each object we discarded any solutions with lM$_*>12$, formation redshift higher than 30 (corresponding to an age of the Universe of $\approx\!100$ Myr), or a SFR peak 
higher than $10^4$ M$_{\odot}$/yr, unless no other solution was available; we point out that this threshold on SFR was chosen considering the highest values estimated for high-redshift starburst \citep[see e.g.][]{Ma2016}, but might be too strict considering some known low-redshift extreme sub-mm starburst with estimated SFRs reaching $\approx3\times10^4$ M$_{\odot}/yr$ \citep[see][]{Rowan-Robinson2018}. Finally, for the non ``$z$-robust'' sources we performed an alternative fit at the second most probable $z$ of the PDF($z$) functions, and kept as best-fit the solution with the minimum $\chi^2$ value (note that this might imply that the best solution with the current library is at a different redshift with respect to all the photo-$z$ solutions in M24).

\subsubsection{Selection of quiescent candidates} \label{selec}

Observationally, the presence of evolved stellar populations can be probed by the presence of significant Balmer (at rest-frame wavelength 3645 {\AA}) and/or 4000 {\AA} breaks in the SED of the galaxy. However, in recently quenched galaxies these breaks can be less pronounced (see also Sect. \ref{specomp}). Nevertheless, to consider a source as an eligible quiescent candidate we required that photometric data be available for the bands sampling the breaks at the redshifted wavelength, and for at least 4 NIRCam bands in general. We did not consider candidates at $z>12$, where the breaks are redshifted redward of the F444W filter. 
It is also worth mentioning that at $z>10.5$ the spectral continuum redward of the break is detectable only in F444W (and, at $z>9.5$, in F410M if the band is in the dataset); therefore, at these redshifts only one or two photometric points are available for faint sources having the rest-frame optical continuum below the detection limiting magnitude, making their fit inevitably less reliable.
We also excluded objects with signal-to-noise ratio (SNR) lower than 3 in F444W, or flagged as spurious, in M24. From the resulting sample, we manually removed some additional spurious detections after visual inspection; in particular, a relevant number of them were found in the PRIMER-COSMOS field at $z>8$ (typically hot pixels). We also required that the candidates have a physical half-light radius in the detection band smaller than 5 physical kpc; for this we used the \textsc{SExtractor} \citep{Bertin1996} \texttt{FLUX\_RADIUS} from the detection band as given in M24, and computed the corresponding physical length with the \texttt{astropy.cosmology.angular\_diameter\_distance} routine. Finally, we required that they are not in the brown dwarfs lists from \citet{Hainline2024} and \citet{Holwerda2024}, and we applied the criterion by \citet{Langeroodi2023} to exclude additional brown dwarfs interlopers, F200W-F150W $\leq$ 0.2; as a side note, we also considered the equivalent of the $BzK$ selection diagram to single out stars \citep{Daddi2004}, using F435W, F090W and F200W, but after visual inspection most of the selected objects looked extended, so we decided not to apply this criterion.

After all these checks, we considered the best-fit solutions at $z_{best}$ and pre-selected the possible candidates as those having (i) lSSFR $<-10$, (ii) lM$_*>9.5$, (iii) $dt_q>10$ Myr, and (iv) a reduced-$\chi^2$ probability\footnote{The probability of a model with a given reduced-$\chi^2$ and $\nu$ degrees of freedom is defined as $p(\chi^2,\nu)=c\int_{\chi^2}^{\infty}PDF(\chi^2,\nu)$, where $c$ is a normalisation factor to ensure $p=1$ for $\chi^2=0$, and $PDF(\chi^2,\nu)=(\nu\chi^2)^{\nu/2-1}e^{-\nu\chi^2/2}/[2^{\nu/2}\Gamma(\nu/2)]$.} of the best solution, $p_{best}$, larger than 30\%.
The resulting candidates were then ranked defining a reliability parameter $r=p_{best}-p_{SF}$, where $p_{SF}$ is the probability of the best alternative SF solution (or zero if no one exists), and we selected those with $r>0.1$, implying that for $p_{best}=0.3$ (the minimum allowed value) $p_{SF}$ must be lower than 0.2. 
This yielded a final list of 633 quiescent candidates, of which 164 fitted as RD (i.e. with $dt_q\geq150$ Myr), and the remaining 459 as PS; 34 candidates have $z_{best}>5$. 


Finally, we further selected a list of ``gold'' candidates requiring $r>0.4$  (which reduced the sample to 360 sources) plus the following additional safety qualities: the ``$z$-robust'' tag (i.e., spectroscopic or unambiguous photometric redshift), good data for at least 6 (rather than just 4) NIRCam bands, no evident flaw at visual inspection, and not being classified as a ``Little Red Dot'' \citep[LRDs,][]{Furtak2023,Labbe2023} in published lists (see the discussion on this point in Sect. \ref{limit}). These additional criteria reduced the list to 291 objects; of these, 95 were fitted as RDs, and 197 as PS. In the following we will also dub as ``silver'' selection the sample of 341 candidates excluded from the ``gold'' selection. 
We point out that the threshold values for the selections were chosen after extensive testing with simulated data and cross-checking against spectroscopic data, to obtain a good trade-off between completeness (i.e. the number of selected sources out of all the real quiescent galaxies) and contamination (i.e. the number of star-forming galaxies included in our selections out of the total number of selected galaxies): see Appendix \ref{simulations} for a discussion. 

\subsubsection{Discussion and known limitations} \label{limit}

As anticipated, the selection criteria based on the probability of  alternative SF solutions adopted in this work and described above are different from that adopted in M18 and M19, where we simply required $p_{SF}<5\%$. This can be understood considering that in those works the SED-fitting library only included top-hat SFH models: this implied that any SF but non-starburst object would be poorly fitted, so a hard cut on $p_{SF}$ was required to safely eliminate any possible SF interlopers. In contrast, to better characterise the physical properties of the galaxies, in this work we allowed for various SFHs, including many models with moderate SF activity: so a quiescent best-fit model is already robust enough to make a candidate reasonably reliable, because it has been preferred to any credible SF solutions. 

However, there are various possible sources of ambiguity. The SED models are certainly not fully adequate to describe these early galactic populations, for example because of the lack of population III stellar templates (although they might have already disappeared at $z<12$) and of AGN templates. We also adopt the standard \citet{Salpeter1959} initial mass function (IMF), which might be a sub-optimal choice for early, massive systems (see the discussion in Sect. \ref{sec:concl}. Furthermore, degeneracies of physical properties (age, extinction and metallicity in particular) are known to exist and are difficult to disentangle (e.g. see Sect. \ref{wow_objs_mass}). Also, while we tried to take into account possible errors on the redshift estimates, e.g. excluding from the ``gold'' selection sources with more than one peak in the PDF($z$), we might still have included lower redshift interlopers with wrong but consistent photometric redshift estimates from the three codes.

We decided not to exclude from our selection objects identified as LRDs in published lists, since the true nature of this class of sources is still being discussed. The origin of their uprising optical rest-frame emission is unclear and the presence of a substantial evolved stellar population has not yet been excluded \citep[see][]{PerezGonzalez2024,Baggen2024,Williams2024,Setton2025}, although it would lead to extremely high core stellar densities in most cases \citep{Guia2024}. However, we decided to exclude them from the ``gold'' sample. 
We used the lists presented in \citet{Kocevski2025}, \citet{Kokorev2024}, and \citet{PerezGonzalez2024}, supplemented with a sample based on the more exhaustive photometric selection of \citet[][priv. comm.]{Barro2024b}, which recovers additional sources. Eighteen candidates in our selection are matched, 9 of which at $z_{phot}>9$. With the exception of PRIMER-COSMOS ID104822, a low reliability candidate at $z_{spec}=8.29$ (but its photometric SED also presents features typical of LRDs), they are the only candidates we find at $z>7.29$, i.e. the redshift of Rubies-QG7 (PRIMER-UDS 117643 in M24), one of the two current farthest spectroscopically confirmed quiescent galaxy \citep[][W25]{Weibel2025}; the latter is also our ``gold'' farthest selected object (see Sect. \ref{wow_objs}). Interestingly, this galaxy would also be classified as a LRD with the color selection applied by \citet{Barro2024b}; however, the possibility of it actually being a LRD was already discussed and ruled out in W25, so we leave this galaxy out of the list of LRDs. Conversely, JADES-GS ID27908 at $z_{phot}=10.62$ is only classified as LRD in \citet{PerezGonzalez2024}, but in the absence of a more in-depth analysis we leave it in the list. 

We also did not exclude from the total selection the sources identified as AGN in the lists discussed in M24, Sect. 5, given that many known quiescent sources have spectral features indicating the presence of nuclear activity (see Sect. \ref{specomp}). A cross-match with existing catalogs \citep{Luo2017,Nandra2015,Kocevski2018,Kocevski2023,Taylor2025} revealed that 18 candidates are classified as X-ray emitters (9 in JADES-GS, 5 in PRIMER-UDS and 4 in CEERS), of which ten are in our ``gold'' selection, and six more (three of which as ``gold'') as broad line emitters.

Finally, we point out that the H$\alpha$ and [OIII] emission lines at $z\approx\!3.8$ fall in the gaps between the F277W and F356W filters, and F200W and F277W filters, respectively. This unfortunate coincidence can bias the SED-fitting procedure for objects at that specific redshift.

\subsection{Comparison with spectroscopic data} \label{specomp}

To assess the reliability of our selection, we compiled a list of 30 spectroscopically confirmed quiescent galaxies at $z>3$ that have a coordinate match with M24, including those from  \citet{Schreiber2018b,Barrufet2024,Carnall2024,DEugenio2024,Jin2024,Nanayakkara2024,Setton2024,Weibel2025,DeGraaff2025}, and \citet{Baker2025}. 
We then checked whether they were included in our sample; the results are shown in Tab.~\ref{spectrq}, where we also list the acronyms of the references used in the rest of the paper. Twenty-three galaxies are in our ``gold'' selection, including the aforementioned PS from W25 (PRIMER-UDS 117643 in M24 at $z_{spec}=7.29$), and two more are ``silver'' candidates. Of the five that we did not select, one (JADES-GN ID41598, ID25529 in B25), has only F444W data in our mosaics, so it was detected in M24 but it was not included in the current analysis. The other three were best-fitted with a SF model, although quiescent solutions exist with non-negligible probabilities. However, they were discussed in the original works as potentially not quiescent: in particular, IDs CEERS 56717 and PRIMER-COSMOS 86447 (EGS-26047 and ZF-COS-20133 in S18, respectively) were also consistent with star-forming models (the former from the strong emission lines, the latter was best fitted as recently rejuvenated), and CEERS ID79531 (EGS-34322 in S18) was actually best fitted by N24 as a mildly star-forming source (lSSFR$\approx\!$-9.22). Finally, the remaining missed source (PRIMER-UDS 26918) was best-fitted with a quiescent model in our runs, but was not selected because of a secondary SF solution with sufficiently high $p_{SF}$, albeit with lSSFR $\approx\!-9.87$, which is close to our quiescence threshold of $-10$. Given these considerations, we can conclude that on this spectroscopic sample we reach a completeness of $79^{+20}_{-16}$\% \citep[23 out of 29, uncertainties computed following][]{Gehrels1986} for the ``gold'' selection and of 86$^{+14}_{-17}$\% for the total selection, being conservative about the quiescent nature of the ambiguous sources discussed above; or, of $88^{+12}_{-10}$\% (23 out of 26) and 96$^{+4}_{-19}$\% for the ``gold'' and total selections respectively, if we exclude them. As a side note, we point out that we did not include in the list the PS at $z_{spec}=7.29$ singled out by \citet{Looser2024}, because its stellar mass is below our lower cut (see Sect. \ref{wow_objs}).

\begin{figure}
   \centering
   \includegraphics[width=\hsize]{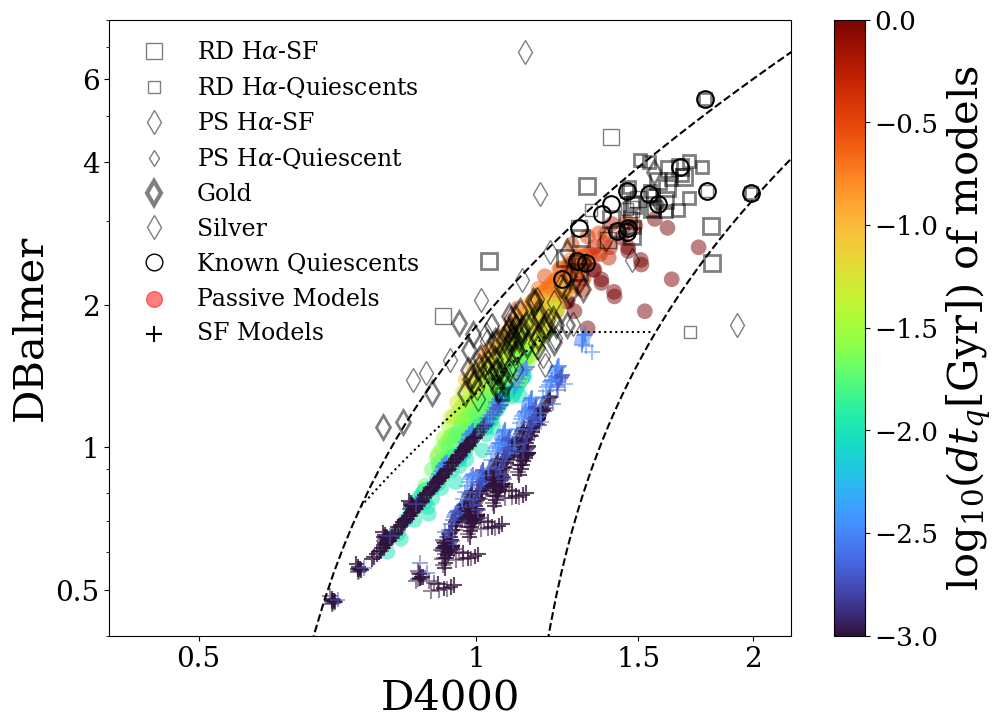}
    \caption{Values of the D4000 and DBalmer spectral indexes for theoretical models and the quiescent candidates from this work with available spectroscopic data. Passive models are color-coded by their quiescence time $dt_q$. The candidates are depicted with different symbols depending on their best-fit properties: squares for RDs, diamonds for PSs; larger symbols for the ``gold'' selection. Known quiescent sources from literature are marked with a black circle. The dashed and dotted lines are used to select the outliers with respect to the locus of the models, for visual inspection (see Sect. \ref{specomp} for details).}
    \label{breaks}
\end{figure}

We then performed a counter-check browsing the DJA for the spectra of our candidates (including those already in the list described above) to look at their main spectral features. Using a 0.3'' matching length and searching for prism data, at the time of this work we found 162 matches with our sample, 108 of which are in the ``gold'' selection. Note that some objects have already been included in published lists \citep{Schreiber2018b,Schmidt2021,Naidu2022b,ArrabalHaro2023b,Labbe2023,Bunker2024,Carnall2024,Kokorev2024,Maseda2024,Naidu2024,Simmons2024,Williams2024}.
A thorough spectro-photometric analysis of the candidates would be beyond the scope of this work; we opted for a simple test instead, which we describe in the following. A standard approach would be to check for the presence and strength of spectral emission lines associated with SF activity, e.g. H$\alpha$ and [OIII]. For example, the star formation rate of SF galaxies is typically estimated from the luminosity of the H$\alpha$ line, as SFR(H$\alpha$)=7.9$\times10^{-42} L(\mbox{H}\alpha)$ \citep{Kennicutt1998}. However, it is established that also nuclear activity can power these emission lines, with the accretion disk emitting UV radiation which heats up the surrounding gas \citep[e.g.][]{Ho2008}: therefore, this test alone is not sufficient to disentangle SF from quiescent galaxies. The same holds true for other lines classically associated with active SF activity like [OII], which have been shown to be produced also by AGN, fast shock waves, post-AGB stars, and cooling flows \citep[e.g.][]{Yan2006, Lemaux2010, Takahashi2025}.
As already mentioned, another useful indicator to check for the presence of a substantial evolved stellar population is the presence of breaks in the spectral continuum at 3645 {\AA} (Balmer break, DBalmer hereafter), defined as $<$$f_{\nu}[3400,3600]$$>$/$<$$f_{\nu}[4050,4250]$$>$, and at 4000 {\AA} rest-frame (D4000), defined as $<$$f_{\nu}[3750,3950]$$>$/$<$$f_{\nu}[4050,4250]$$>$ \citep[see e.g.][]{Binggeli2019}: they are created by massive O, B and A type stars moving away from the main sequence. 
However, this criterion is also ambiguous since the expected values of the breaks overlap significantly for recently quenched galaxies and reddened SF galaxies (see below). Furthermore, it is established that the Balmer break can also be created by extremely dense gas in the broad-line region of an AGN \citep[e.g.][]{DeGraaff2025b,Inayoshi2025}. 

All things considered, we decided to check how many sources in the spectroscopic subsample might be SF interlopers because they have both small values of the breaks and SSFR(H$\alpha$) $>10^{-10}$ yr$^{-1}$ (we used the photometric best fit stellar mass estimate to compute the SSFR). We checked the values of the breaks by means of a simple diagnostic diagram shown in Fig. \ref{breaks}. We analysed the values of the two breaks for $\approx\!500$ quiescent and $\approx\!1000$ SF models built using \citet{Gutkin2016} stellar models (which include emission lines), color-coded by $dt_q$ (we consider a model to be SF when it has $dt_q < 10$ Myr), and the values computed from the observed spectra of the candidates, shown with different symbols depending on their $dt_q$ and on the H$\alpha$ SF indicator. For reference, we also show the positions of the quiescent galaxies from the full known spectroscopic sample described above. 
It can be seen that no candidate lies in the region of the SF models, and most of them lie on the locus of the quiescent models. Note that PS models ($10<dt_q<100$ Myr, green to blue dots) can have D4000 and DBalmer values as low as 0.8 and 0.6, respectively: therefore, as one might expect, young PS candidates are the most ambiguous cases, because their breaks are small and emission lines can be present due to the many reasons discussed above (in fact, $\approx\!72\%$ of the full spectroscopic subsample should be discarded a SF considering only the criterion SSFR(H$\alpha$) $>10^{-10}$, and $\approx\!33\%$ at lM$_*>10.5$).
From the full sample, 156 sources have spectroscopic data in the wavelength range of the breaks. We empirically defined as ``outliers'' the sources having $\mbox{DBalmer} <3.67\times \mbox{D4000}-4.0$ or $\mbox{DBalmer}>4.2\times \mbox{D4000}-2.4$ (the dashed lines in the figure; note that one confirmed quiescent lies in the outlier region), and the ones below the dotted line, which are close to the locus of the SF models. There are 33 outliers, and visually checking them we found that 18 can be considered quiescent, while 15 have spectral features that cannot exclude SF activity. Based on this exercise, we can estimate a contamination of $10\pm3\%$ (15/156) at the bright end of the full sample. 
It is worth noticing that considering only the high-mass tail (lM$_*>10.5$) we have 54 matches, 41 of which are in the ``gold'' selection; 52 have data at the wavelength range of the breaks, nine of them are outliers (four of which also lack $L(\mbox{H}\alpha)$ information), and only two must be rejected as a quiescent candidate at visual inspection, implying a $4\pm3\%$ contamination level for massive sources. 

We also checked the consistency of our sample with previously published photometric selections \citep[M19,][]{Carnall2023,Valentino2023,Russell2024,delaVega2025,Baker2025b}. We discuss the results in Appendix \ref{othersel}.

\subsection{Completeness and contamination of simulated data} \label{simu}

Although the checks on spectroscopic data reinforce confidence in our selection procedure, they can only provide an approximate assessment of the accuracy of our selection technique. Reassuringly, they are in reasonable agreement with the estimates obtained with an extended set of simulated data which we created to quantitatively estimate the completeness and contamination of our sample. The detailed procedure and results are described in Appendix \ref{simulations}. In short, we created an extensive set of mock catalogs reproducing the observed statistical properties of the high-redshift galactic populations; we used the best-fits of real data as the input starting point, and added random perturbations both on their physical parameters and to the photometric values; then, we proceeded to fit the mock catalogs with the same procedure used on real data, and compared the fitted values to the input ones, estimating the completeness and contamination in bins of physical parameters. We found that the ``gold'' selection criteria provide completeness factors of $\approx\!70\%$ at $z < 6$ and SNR $> 100$, and the contamination factors typically stay below 5\%. On the other hand, the total selection reaches 90\% completeness at SNR $> 100$ and $z < 6$, and stays above 50\% for SNR $> 50$ at all redshifts; the contamination is typically very low but grows to $\approx\!20\%$ for SNR $< 50$, with some peaks at $\approx\!50\%$ at $z \approx\! 6$ and SNR $< 20$. While we are aware of the limits of this approach (we discuss them in the Appendix), we think it provides a reasonable proxy for the accuracy we can obtain on real data; therefore, we used these estimates corrected values to compute the number densities discussed in Sect. \ref{nd}.

\begin{figure*}
   \centering
   \includegraphics[width=0.9\hsize]{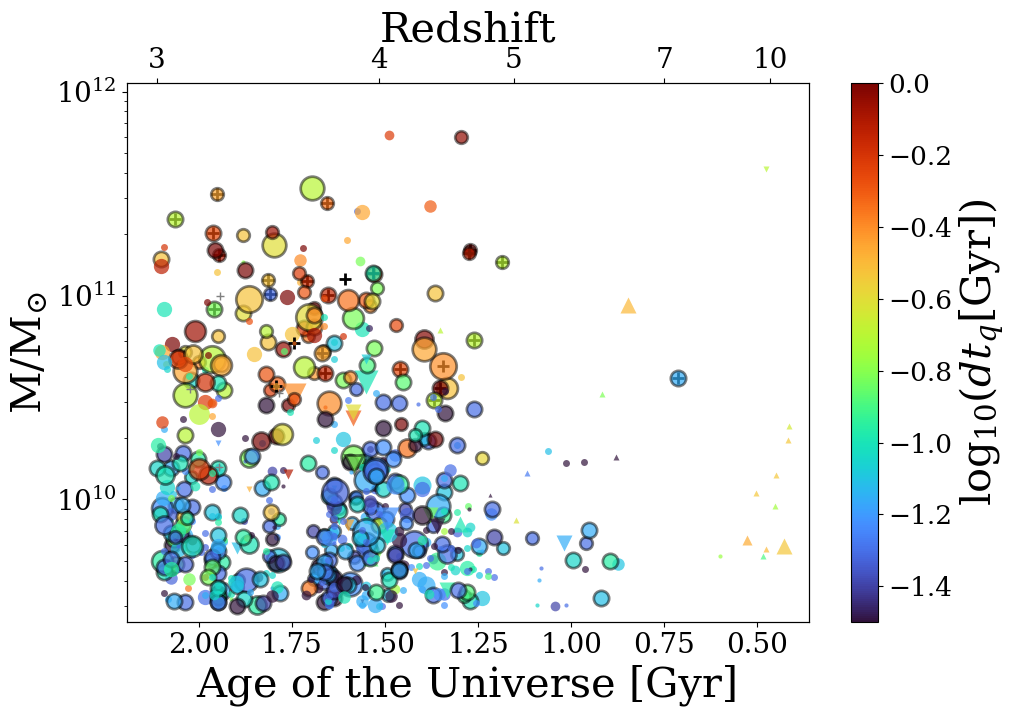}
    \caption{Best-fit stellar mass vs. redshift of the quiescent candidates. Each dot represents a galaxy, and it is color-coded by the logarithm of the time passed since quenching, $dt_q$; the size of the dot is a qualitative proxy for the reliability parameter $r$ (see Sect. \ref{selec}). ``Gold'' candidates are marked with a black border. Upward-pointing triangles indicate known LRDs from the lists cited in Sect. \ref{limit}, while downward-pointing triangles represent candidates with a secondary solution at lower redshift. Black crosses mark the sample of known spectroscopic quiescents discussed in Sect. \ref{specomp}.}
    \label{zm}
\end{figure*}

 \begin{figure}
    \centering
    \includegraphics[width=0.9\hsize]{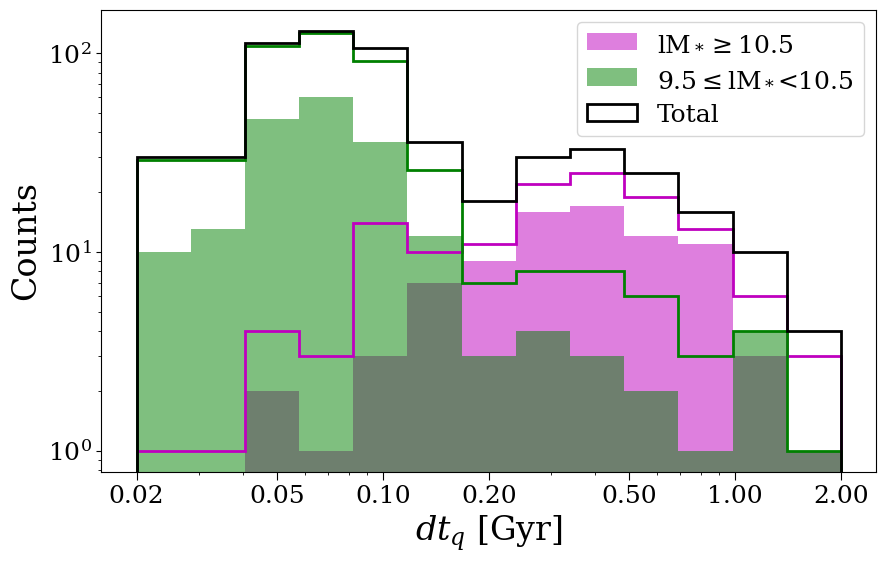}
     \caption{Distribution in time passed since quenching $dt_q$ of the high (lM$_*\geq10.5$, magenta) and low mass ($9.5\leq$ lM$_*<10.5$, green) quiescent candidates at $3<z<5$. The shaded histograms correspond to the ``gold'' selection (the grey area being the superposition of the red and blue histograms), the lines to the full sample.}
     \label{dtqmass}
 \end{figure}

\section{Results and analysis}
\label{sec:results}

\begin{figure}
   \centering
   \includegraphics[width=\hsize]{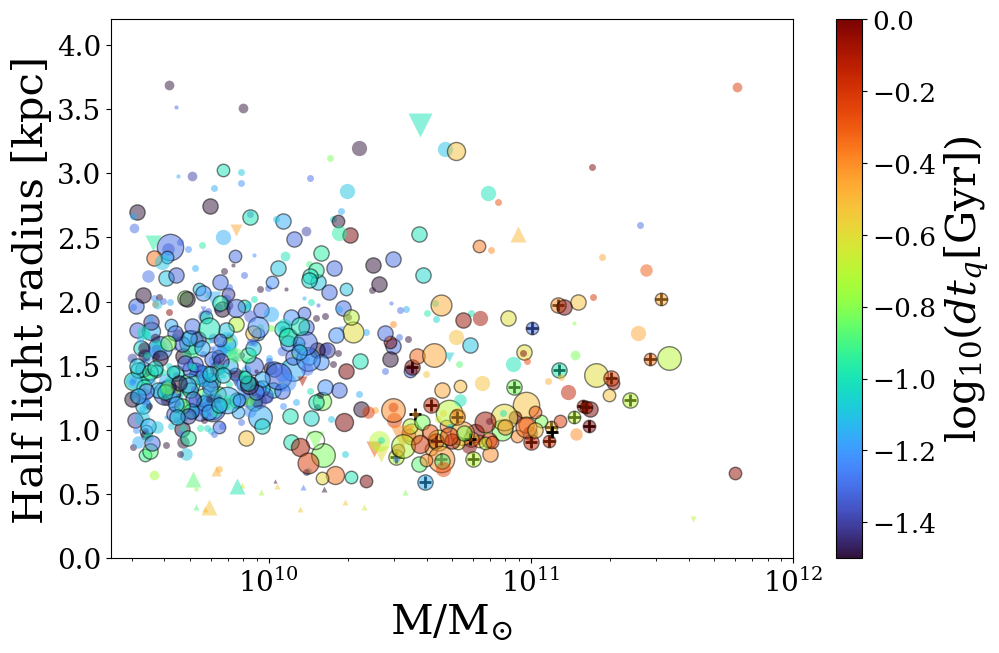}
    \caption{Half light radius vs stellar mass relation of the candidates. Symbols are as in Fig. \ref{zm}. Note that radii in M24 were estimated on the detection image, a F356W+F444W stack, so they sample different physical emissions at different redshifts. The two very massive and very compact source close to the bottom right corner PRIMER-UDS ID98234, most likely a peculiar AGN, and ``Capotauro'' \citep{Gandolfi2025}, barely visible because of its low reliability; see Sect. \ref{wow_objs_mass}.}
    \label{mr}
\end{figure}

The full list of quiescent candidates, along with the physical properties of their best-fit model and additional information, is available on the \textsc{ASTRODEEP} website\footnote{\url{http://www.astrodeep.eu/quiescents_merlin25}}. Of the 633 candidates, 535 (85\%) are best fitted with top-hat SFH models implying very rapid quenching times, 97 with exponentially declining SFH models, and one with a delayed exponentially declining SFH model. Concerning the distribution of the (gold) candidates in the fields, 46 (20) are in ABELL2744, 95 (53) in CEERS, 61 (28) in JADES-GN, 99 (53) in JADES-GS, 4 (1) in NGDEEP, 129 (44) in PRIMER-COSMOS, and 199 (92) in PRIMER-UDS. Considering the areas of the surveys (see M24), these numbers imply a reasonably homogeneous surface density of quiescent candidates of $\approx\!0.9\pm0.2$ ($0.4\pm0.1$) arcmin$^{-2}$, with the exception of NGDEEP for which it is $\approx\!0.4\pm0.3$ arcmin$^{-2}$ ($0.1\pm0.1$).

\vspace{0.5cm}

\subsection{Downsizing: a tale of two populations} \label{downsiz}

Figure \ref{zm} shows the best-fit stellar mass of the candidates as a function of their redshift (and the corresponding age of the Universe). The dots are colour-coded by $dt_q$, i.e. the time since when the SSFR of the model dropped below $10^{-10}$ yr$^{-1}$, and sized proportionally to the reliability parameter $r$; other symbols are explained in the caption of the Figure. 
A clear bimodality is immediately evident: at $3<z<5$, massive sources (lM$_*>10.5$) are predominantly RDs ($dt_q\geq150Myr$), while lower mass objects are mainly PSs. This is shown even more explicitly in Fig. \ref{dtqmass} were we plot the number counts in bins of $dt_q$ for the two mass regimes, obtaining two distinct peaks, particularly evident in the ``gold'' sample (we checked that the fraction of sources with a wrong fit, and in particular of RDs erroneously fitted as PSs, is not relevant; see Appendix \ref{pssim}). This implies that, at any moment in time since $z\approx\!5$, the most massive objects have formed their stellar mass in a remote and intense burst of SF activity, while lower mass systems have only recently stopped forming stars. In principle, one could expect the low mass PSs found at $z\approx\!5$ to be the progenitors of low mass RDs at lower redshifts, but this would require that they stay quenched, making the number of low mass RDs candidates increase going towards $z\approx\!3$. However, looking at Fig. \ref{zm}, it seems that this is not the case (see also the subsections \ref{assembly} and \ref{nddtq}). We might be missing low mass RDs because of our selection technique, but the results of the runs on simulated data that we used to assess the completeness of our sample seem to rule out this hypothesis (see Sect. \ref{simu}, and Appendix \ref{simulations}).

A possible explanation of the observed bimodality is that the low mass PSs are the progenitors of massive RDs, i.e. they will move upward and leftward in the plot after the observation. However, massive RDs are more compact than most low mass PSs, as shown in Fig. \ref{mr}, where we plot the the half light radius versus the best-fit mass of the candidates (in physical kpc, as inferred from the \textsc{SExtractor} pixel values given in M24 and the angular diameter distance of the source\footnote{We warn that considering the \textsc{SExtractor} flags as reported in M24, 38\% of the candidates are flagged as isolated (flag=0), 25\% have been deblended (flag=1), and 36\% are deblended and have close neighbors likely affecting aperture photometry (flag=3); however, in principle having a close neighbor should increase the estimate. Furthermore, since the values are measured on the detection F356W+F444W stack they might actually correspond to different physical radii depending on the redshift of the source, as different rest-frame wavelengths are sampled; however, this is true for both the RD and the PS candidates, so it does not impact our conclusions.}. To remain quiescent while increasing their mass by an order of magnitude, low mass PSs should undergo major dry merger events while becoming more compact; but collisionless systems such as gas-poor galaxies do not increase their concentration in dry mergers \citep[e.g.][]{Nipoti2009}, so we can exclude this scenario \citep[although PS galaxies may still retain gas reservoirs or further accrete gas thus undergoing wet mergers and some degree of compaction induced by gas dissipation, see e.g.][]{Hopkins2009,Lapi2018}. Incidentally, we do not find clear evidence for a correlation between the time since quenching $dt_q$ and the size of the candidates, as predicted by some models \citep[e.g.][]{Lapi2018}. We also point out that one source is fitted as very massive and compact: it is PRIMER-UDS ID98234, which we discuss in Sect. \ref{wow_objs_mass}.

A third hypothesis is that these PS candidates will become the low mass components of minor dry mergers with massive galaxies. To investigate this scenario, the clustering properties of these objects should be investigated, which is beyond the scope of this work; however, the candidates are scattered across the field of view (FoV) of the fields, making it unlikely that they belong to clustered systems.

The remaining, most credible hypothesis is that these low-mass PSs are only temporarily quenched, and will form new stars in their future. If this is the case, we can conclude that while massive RDs have formed in a burst of SF activity in a compact region where high density gas collapsed, PSs must have started assembling their mass in larger regions of lower density, and at the moment of observation they are experiencing a so-called ``breathing'' phase, in which bursts of SF alternate with short-lived quenched periods. This is expected to naturally happen in low mass galaxies \citep[e.g.][]{Stinson2007,Merlin2012}, and can be interpreted as a direct evidence of the so-called ``downsizing'' scenario, witnessed at its onset rather than inferred from archaeological studies of stellar populations in local galaxies. It is certainly worth pointing out that the mass threshold for this bimodality is at lM$_*\approx\!10.5$, which is the characteristic mass dividing the red and blue sequence as found and discussed long ago by e.g. \citet{Kauffmann2003}, \citet{Gallazzi2005}, \citet{Shankar2006}, \citet{Dekel2006}, and more recently by \citet{Haines2017} up to $z\approx\!1$; or the mass of maximum efficiency in the conversion of gas into stars \citep[e.g.][]{Silk2012}.

\subsubsection{No Red and Dead galaxies at $z>5$} \label{nordz5}

\begin{figure}
   \centering
   \includegraphics[width=\hsize]{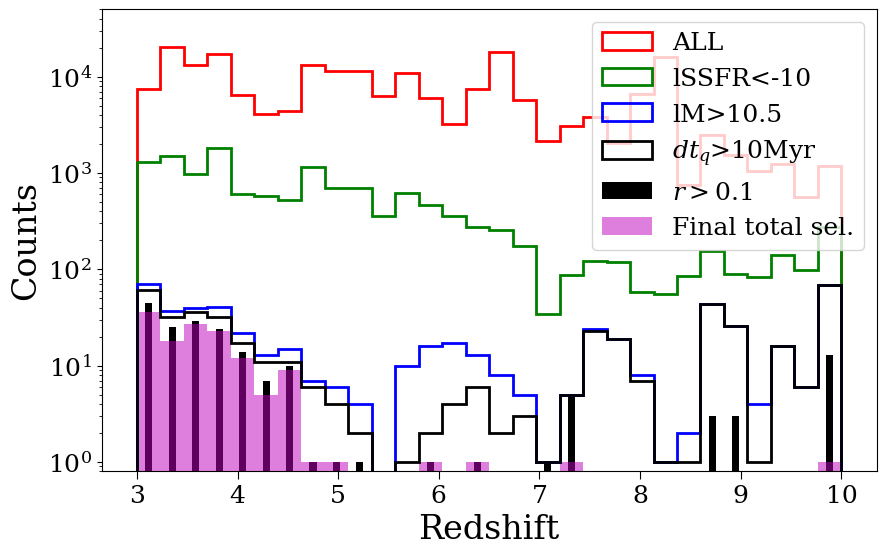}
    \caption{Number of candidates as a function of redshift adding increasingly stringent selection criteria, from the initial full catalog to the final selection (see Sect. \ref{nordz5}).}
    \label{selechist}
\end{figure}

\begin{figure*}
   \centering
   \includegraphics[width=0.95\hsize]{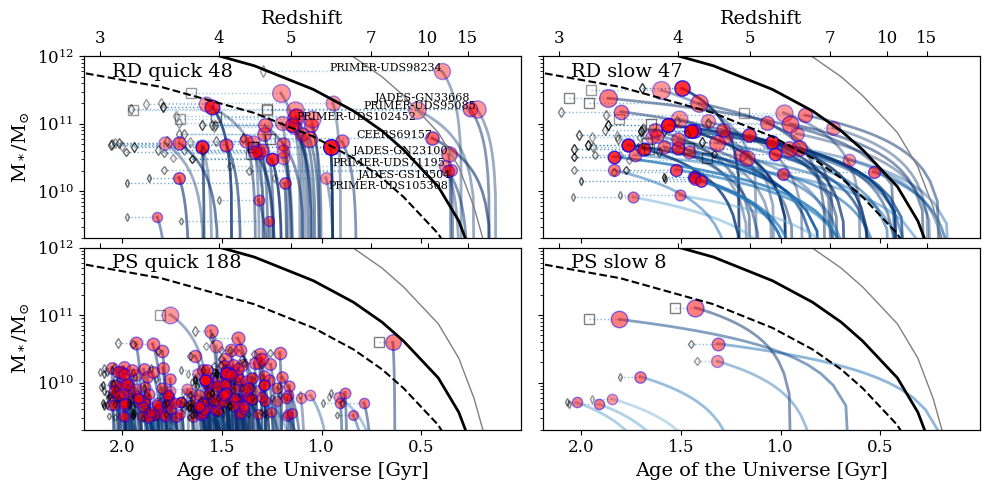}
    \caption{Stellar mass assembly history of the ``gold'' candidates according to their best fit model. Top panels: RDs (candidates quiescent since more than 150 Myr); bottom panels: PSs (candidates quiescent since less than 150 Myr); left panels: quick mass assembly (SFH shorter than 300 Myr); right panels: slow mass assembly (SFH shorter than 300 Myr). For each candidate, the red dots show the epoch of quenching, and the black empty symbols the epoch of observation (squares are candidates with spectroscopic redshift, diamonds are those with photometric redshifts); the two are connected by a black dotted line. The size of the symbols is a proxy for the SFR peak value (as computed from its best-fit SFH function, stellar mass, and age), and the transparency of the symbols is a qualitative proxy for the reliability parameter $r$ (more opaque being more reliable). Finally, the blue lines give an indicative depiction of the stellar mass assembly history as a function of time (bottom axis) and corresponding redshift (top axis), with the transparency of the line again being a proxy for the SFR peak value. The gray and black lines show the limits imposed by the $\Lambda$CDM cosmology: the thin grey line is the maximum halo mass expected in the survey volume at each redshift, as computed using the \textsc{Colossus} Python module \citep{Diemer2018}; the solid black line is the maximum galaxy mass assuming 100\% star-formation efficiency, i.e. $M_{gal}=f_{baryon}M_{halo}$ with $f_{baryon}=0.158$, and the dashed black line is the observed maximum galaxy mass with 20\% star formation efficiency. The IDs of the potential ``Universe breakers'' are reported in the Figure close to their quiescence marker.}
    \label{evol}
\end{figure*}

Another interesting outcome evident in Fig. \ref{zm} is the abrupt drop in the abundance of  robust candidates at $z>5$. Of the 34 candidates at $z>5$, 13 are fitted as RDs, but 12 of these are classified as LRDs in one of the lists cited in Sect. \ref{limit}, and ten of them are at $z>10$; the other two are PRIMER-COSMOS ID33783 at $z_{phot}=6.0$ and PRIMER-UDS ID51964 at $z_{phot}=6.38$, both fitted as lM$_*>10.5$ RDs. We then have only two more massive candidates at $z>5$: the spectroscopic PS from W25 at $z_{spec}=7.287$ (which we fit with lM$_*=10.6$), and ``Capotauro'' (CEERS ID17577), the source presented in \citet{Gandolfi2025}. See also Sect. \ref{wow_objs}; note that both ``Capotauro'' and PRIMER-COSMOS ID33783 are barely visible in Fig. \ref{zm} because of their  low reliability values.
In Fig. \ref{selechist} we show the number counts as a function of redshift of sources selected with increasingly stringent criteria, starting from the whole M24 updated list (red line), which does not show a similar drop. The number of galaxies with a quiescent best-fit decreases by an order of magnitude from $z=3$ to 7 (green line). The trend for massive sources is even stronger up to 5.5 (blue line), and then their counts oscillate between peaks and lows at higher redshifts, most likely indicating spurious selections (note that their counts are of the order of the tens in the most populated bins). The requirement for a minimum $dt_q$ of 10 Myr roughly halves those at $5.5<z<7$ (black line), but the additional criterion on the reliability factor $r$ has the strongest effect: most of the $z>4.5$ massive candidates have credible alternative SF solutions and are therefore excluded from our sample (black bars; the final selection, shown in magenta, further excludes some sources, especially at high redshift, because of additional checks and criteria).
Since it is impossible to discern when an alternative solution is actually the correct one, it remains unclear whether the drop is real or the result of a selection effect. However, we checked the completeness obtained with our selection procedure on a simulated dataset, and we did not find a significant decrease at $z>5$ (see Appendix \ref{simulations}). This seems to favour the conclusion that the drop is real, but the issue requires further verification; currently, there are few other studies that collect data for $z>5$ quiescent candidates (see Appendix \ref{othersel}), and the current lack of spectroscopically confirmed quiescent sources at $z>5$ with the only exception of the PS from W25 further strengthens the conclusion (see Tab. \ref{spectrq}). It might be worth pointing out that also the number density of the full galaxy population at lM$_*\geq10.5$ seems to drop by one order of magnitude from $z=4$ to 5 \citep[e.g.][their Fig. 6]{Weibel2024}. As a final remark, we recall that the stellar masses at $z>4$ might be overestimated up to $\approx\!0.4$dex without the inclusion of far-infrared  (MIRI) data \citep{Papovich2023}.

\subsection{Mass assembly history} \label{assembly}

Figure \ref{evol} illustrates the stellar mass assembly history of the ``gold'' candidates, dividing them into four sub-classes for clarity of visualisation. Each line represents a single galaxy, showing its mass growth in time (or redshift) according to the best fit model; of course this is a simplified and qualitative visualisation of the true histories, as it neglects the possibility of major mergers and the details of the real SFH, which we cannot know. However, given the short timescales, it is reasonable to assume that a monolithic-like mass assembly is a good approximation, perhaps via numerous early mergers of small substructures \citep[see e.g.][]{Merlin2012}. The meaning of the symbols is explained in the caption.
The upper left panel collects the 48 ``quick'' gold RD candidates: according to the best fit, these objects formed their stellar mass in a rapid burst of activity which lasted less than 300 Myr (note that this should not be confused with the $dt_q$ value, which is the time passed since quenching, and for these sources is greater than 150 Myr). Noticeably, many high mass sources fall into this category, implying very high SFR peaks. The upper right panel collects 47 ``slow'' RDs, with SFHs longer than 500 Myr. The bottom left panel collects the PSs with a ``quick'' mass assembly: they are almost all of the PS candidates. In fact, only 8 PSs have ``slow'' SFHs (bottom right panel). This plot emphasizes the conclusions of the previous subsection, since it can be seen even more clearly that there are very few low mass RDs at $3<z<5$, while from the abundance of PSs one could expect to find many approaching $z=3$, if the older ones stayed quenched for long times. However, the plot also highlights a further interesting point: in principle, one would expect to have a comparable number of ``quick'' and ``slow'' PSs, if the episodic bursts of SF were happening stochastically in the SFH of a galaxy. Instead, almost all of the PS candidates are very young, implying they have assembled their mass very quickly and recently. This might be caused by a bias of the SED-fitting technique, which might spuriously favour short SFH solutions; or, it might be a real feature, implying that after the first, short episode of quenching in which we are observing them, these sources will resume forming stars more steadily, without other relevant quenching episodes at least up to $z=3$. 

Based on their best-fit taken at face value, nine ``gold'' candidates (plus 14 in the ``silver'' sample) would have assembled their stellar mass too early with respect to the limits imposed by the $\Lambda$CDM framework. They are all fitted as ``quick'', massive RDs, with zero or very low extinction. They are highlighted with their ID in the Figure; they lay above the black solid line, which shows the maximum stellar mass allowed as a function of redshift, obtained using the \textsc{Colossus} package \citep{Diemer2018} to compute the maximum halo mass (thin grey line), and assuming a baryon fraction of 0.158 and 100\% star formation efficiency (the black dashed line corresponds to 20\% efficiency). These sources would be ``Universe breakers'' \citep[see][their Fig. 1]{Xiao2024}, so their fit has to be taken with particular caution. Again, they can actually be the result of early dry mergers, so that their stars actually belonged to smaller halos when the SF activity ceased; furthermore, their photometric masses might be overestimated \citep[e.g.][]{Papovich2023,Berger2025}, as well as their ages (see Appendix \ref{agesim}, also given the sharp drop in the number of candidates at $z\approx\!5$ discussed in the previous subsection). However, some of these objects have already been discussed in the literature. In particular, two of them (PRIMER-UDS IDs 95085 and 102452 in M24) are also presented as potential ``Universe breakers'' in \citet{Carnall2024}, as EXCELS 109760 and 117560, although we respectively infer formation redshifts of 26 and 17, and quenching redshifts of 17 and 9.2, while they were fitted with lower values in the original work (while stellar masses and metallicities are reasonably consistent). On the other hand, we fit the well known ZF-UDS-7329 galaxy from \citet[][PRIMER-UDS ID46683 in M24]{Glazebrook2024} as a high-metallicity ($Z/Z_{\odot}=2.5$), massive (lM$_*=11.3$) galaxy, which is consistent with their fit, but we infer a formation redshift of 7.3 rather than $\approx\!11$, and a quenching redshift of 4.25 rather than 6.3 \citep[see also][]{Turner2025}. Since the estimates in the cited works have been obtained from fits of spectro-photometric data with non-parametric SFHs, they should probably be considered more reliable.

\subsection{Colors, metallicity and extinction} \label{colmet}

\begin{figure}
   \centering
   \includegraphics[width=\hsize]{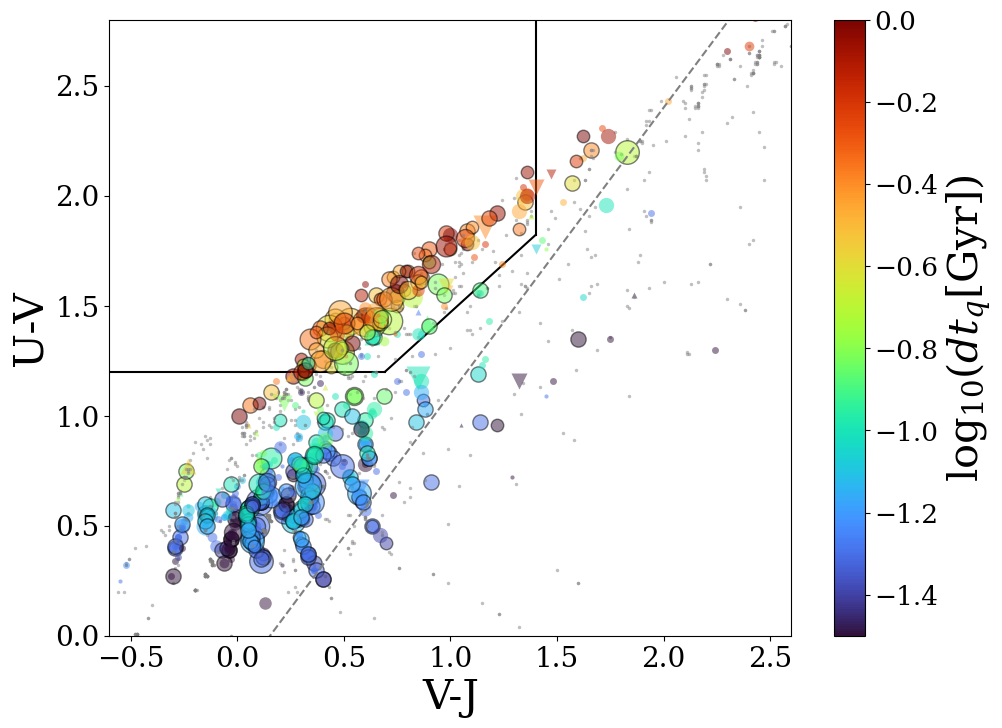}
    \caption{Position of the quiescent candidates in the $UVJ$ diagram according to their best fit model. Symbols are the same as in Fig. \ref{zm}, and small points are all the sources at $z_{best}>3$ with a quiescent best-fit.}
    \label{uvj}
\end{figure}

\begin{figure}
   \centering
   \includegraphics[width=\hsize]{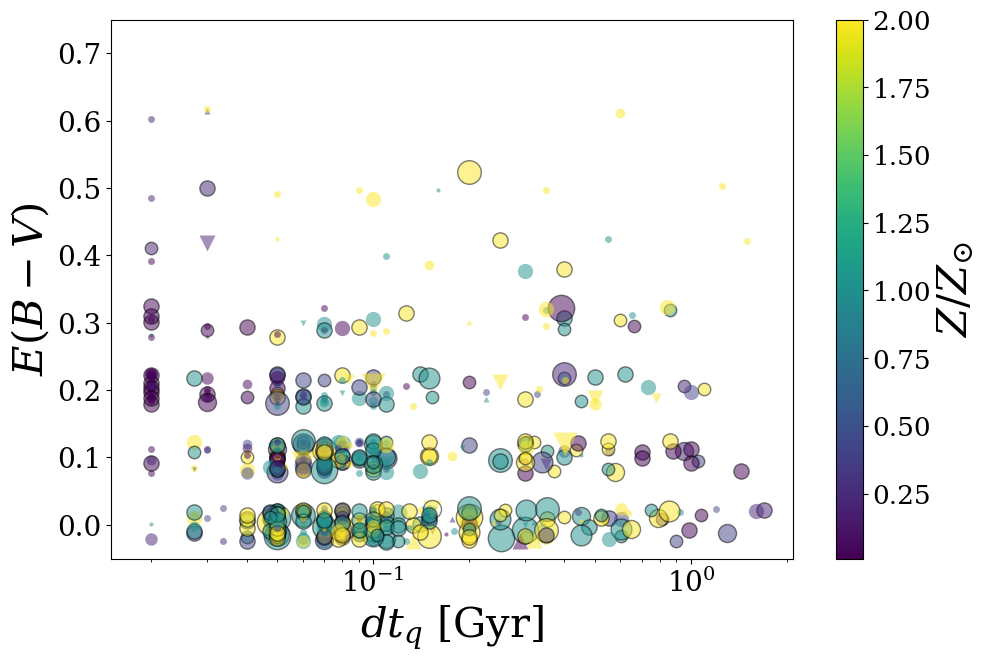}
    \caption{Extinction $E(B-V)$, time since quenching $dt_q$ and metalliticy $Z/Z_{\odot}$ of the quiescent candidates. Symbols are as in Fig. \ref{zm}, with the different color coding referring to metallicity; the crosses mark the sources that have best fit rest-frame colors $U-V<2$ and $V-J<1.5$. A random small offset on the $E(B-V)$ value has been applied for legibility.}
    \label{zebv}
\end{figure}

In Fig. \ref{uvj} we show the candidates on the $UVJ$ diagram \citep{Labbe2005,Wuyts2007}, with the same symbols as in Fig. \ref{zm}. Many (and almost all the PSs) fall outside of the standard quiescent area in the plot, as already pointed out by M18. Indeed, recent color selections have adopted less stringent criteria, e.g. the dotted line is the one adopted by \citet{Baker2025b}. A few of our young PS candidates actually fall even blue-ward of this threshold. The gap in the distribution at $U-V \approx\! 0.8$, $V-J \approx\! 0$ is due to the existence of SF solutions for the potential candidates with best-fit colors falling in the region. Note that not all the sources in M24 falling within the $UVJ$ quiescent region end up being selected as quiescent with our criteria, again because of the existence of reliable SF alternative solutions. It is worth reminding that the position on the $UVJ$ diagram is given by the rest-frame colors of the best-fit model, which are not necessarily an accurate proxy for the intrinsic colors of the real galaxy.

\begin{figure*}
   \centering
   \includegraphics[width=\hsize]{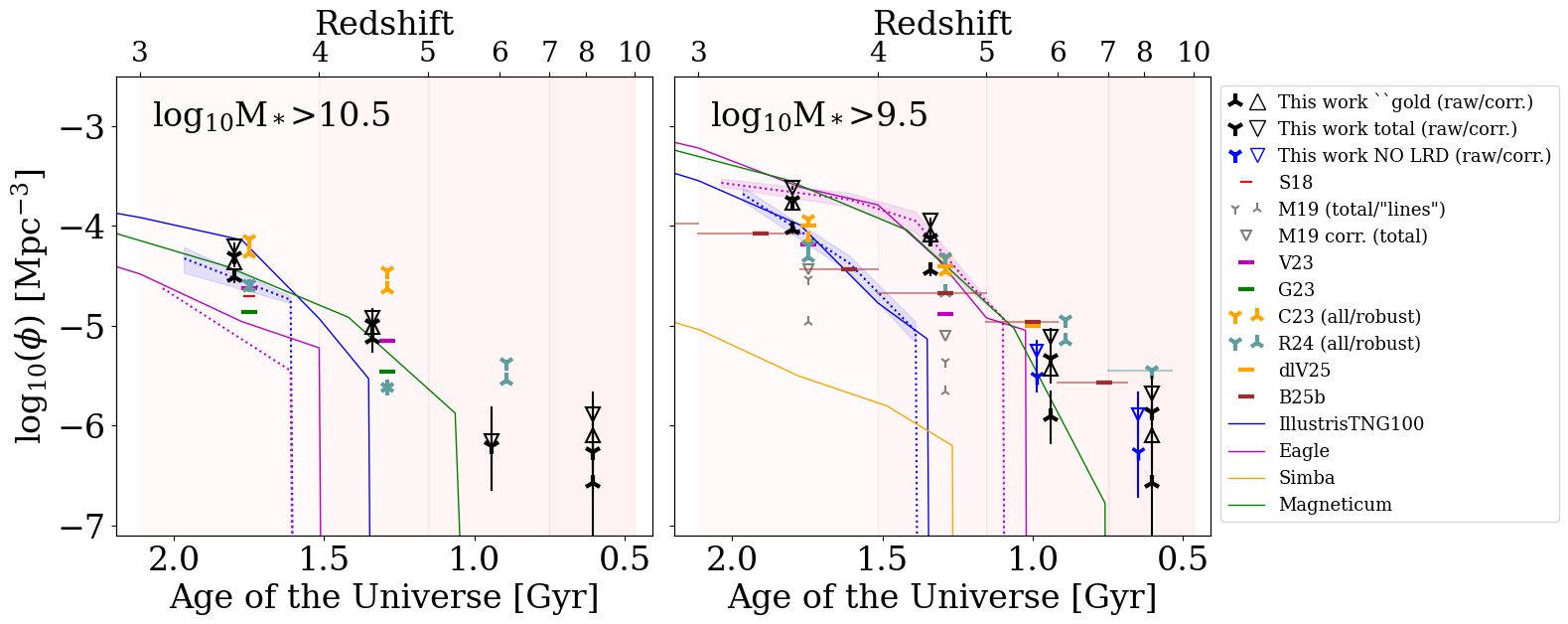}
   \caption{Number density $\phi$ of quiescent sources from this work, other works in literature, and theoretical models. The symbols are explained in the legends; the acronyms stand for \citet[][S18]{Schreiber2018b}, \citet[][M19]{Merlin2019}, \citet[][V23]{Valentino2023}, \citet[][G23]{Gould2023}, \citet[][C23]{Carnall2023}, \citet[][R24]{Russell2024}, \citet[][dlV25]{delaVega2025}, and \citet[][B25]{Baker2025b}. All data points refer to the same redshift intervals highlighted by the shaded vertical bands, and have been slightly offset to enhance readability, except for B25 and the $z=8$ point of R24, for which the intervals are shown by the horizontal lines. The number densities obtained without excluding LRD candidates are only shown where the difference with those including them, i.e. the $z>5$, lM$_*>9.5$ cases.
   The number densities predicted by theoretical models are shown with solid lines. For \textsc{TNG100} and \textsc{Eagle} we also show the prediction from four mock lightcones obtained with \textsc{Forecast}: the dotted lines are the median values and the shaded regions the 1$\sigma$ confidence levels.}
    \label{ndf}
\end{figure*}

Finally, Fig. \ref{zebv} shows the best-fit $E(B-V)$ values of the candidates as a function of their $dt_q$; here the dots are color-coded as a function of the stellar metallicity. Many candidates are fitted as chemically enriched galaxies. Local quiescent galaxies are known to be more metal rich than SF galaxies, reaching super-solar values at the high mass end \citep[e.g.][]{Peng2015}, and in principle the same physical mechanisms that cause this trend exist at high redshift as well; while this might be surprising, cases of high metallicity quiescent objects at high redshift have already been reported in the literature \citep[e.g.][]{Nanayakkara2024}. \citet{Wang2020} suggested that stellar mergers could be responsible for such early chemical enrichment.
Furthermore, a few candidates appear to have high extinctions, with $E(B-V)>0.4$, even after more than 100 Myr from their quenching. While this might also seem far-fetched, checking the SED-fitting output we found that most of these objects do not have reliable alternative solutions. Five of these sources are spectroscopically confirmed quiescents; the most striking case is ABELL2744 ID13197, characterised by \citet{Setton2024} as a two-component quiescent galaxy with an old ($\approx\!$1 Gyr) and dusty ($A_v$=1.5) bulge. For these reasons, we cannot rule out the possibility of old and dusty galaxies; they are few, and we keep them in our selection. 

\vspace{0.5cm}

\subsection{Number density of quiescent sources} \label{nd}

\begin{table*}
\caption{Number densities of quiescent candidates. Number densities are in log$_{10}$ Mpc$^{-3}$ (comoving). Values are given for the ``gold'' and total samples, both raw and corrected for completeness and contamination, and for two lower mass thresholds, lM$_*\geq9.5$ and lM$_*\geq10.5$. The numbers in parentheses are the actual number of RD ($dt_q\geq150$ Myr) and PS ($dt_q\leq 150$ Myr) candidates. Uncertainties are computed following \citet{Gehrels1986}.} \label{ndt}
\centering
\begin{tabular}{l  l  l }    

\hline\hline
Redshift & ``Gold'' raw / corrected (\#$_{RD}$, \#$_{PS}$) & Total raw / corrected (\#$_{RD}$, \#$_{PS}$) \\ \hline
\multicolumn{3}{c}{lM$_*\geq10.5$} \\
$3\leq z<4$ & $-4.52${\raisebox{0.5ex}{\tiny$^{+0.06}_{-0.06}$}} (56, 8) / $-4.37${\raisebox{0.5ex}{\tiny$^{+0.05}_{-0.05}$}} & $-4.30${\raisebox{0.5ex}{\tiny$^{+0.04}_{-0.04}$}} (82, 22) / $-4.21${\raisebox{0.5ex}{\tiny$^{+0.04}_{-0.04}$}} \\
$4\leq z<5$ & $-5.13${\raisebox{0.5ex}{\tiny$^{+0.13}_{-0.13}$}} (13, 1) / $-5.01${\raisebox{0.5ex}{\tiny$^{+0.11}_{-0.11}$}} & $-4.98${\raisebox{0.5ex}{\tiny$^{+0.11}_{-0.11}$}} (17, 3) / $-4.93${\raisebox{0.5ex}{\tiny$^{+0.10}_{-0.10}$}} \\
$5\leq z<7$ & $-$ (0, 0) / $-$ & $-6.21${\raisebox{0.5ex}{\tiny$^{+0.37}_{-0.45}$}} (2, 0) / $-6.15${\raisebox{0.5ex}{\tiny$^{+0.34}_{-0.41}$}} \\
$7\leq z<10$ & $-6.57${\raisebox{0.5ex}{\tiny$^{+0.52}_{-0.76}$}} (0, 1) / $-6.09${\raisebox{0.5ex}{\tiny$^{+0.30}_{-0.34}$}} & $-6.27${\raisebox{0.5ex}{\tiny$^{+0.37}_{-0.45}$}} (1, 1) / $-5.89${\raisebox{0.5ex}{\tiny$^{+0.23}_{-0.25}$}} \\
\multicolumn{3}{c}{lM$_*\geq9.5$} \\
$3\leq z<4$ & $-4.04${\raisebox{0.5ex}{\tiny$^{+0.03}_{-0.03}$}} (72, 120) / $-3.77${\raisebox{0.5ex}{\tiny$^{+0.02}_{-0.02}$}} & $-3.75${\raisebox{0.5ex}{\tiny$^{+0.02}_{-0.02}$}} (115, 260) / $-3.62${\raisebox{0.5ex}{\tiny$^{+0.02}_{-0.02}$}} \\
$4\leq z<5$ & $-4.45${\raisebox{0.5ex}{\tiny$^{+0.06}_{-0.06}$}} (19, 49) / $-4.09${\raisebox{0.5ex}{\tiny$^{+0.04}_{-0.04}$}} & $-4.12${\raisebox{0.5ex}{\tiny$^{+0.04}_{-0.04}$}} (28, 115) / $-3.95${\raisebox{0.5ex}{\tiny$^{+0.03}_{-0.03}$}} \\
$5\leq z<7$ & $-5.91${\raisebox{0.5ex}{\tiny$^{+0.25}_{-0.28}$}} (0, 4) / $-5.43${\raisebox{0.5ex}{\tiny$^{+0.14}_{-0.15}$}} & $-5.33${\raisebox{0.5ex}{\tiny$^{+0.12}_{-0.13}$}} (3, 12) / $-5.11${\raisebox{0.5ex}{\tiny$^{+0.09}_{-0.10}$}} \\
$7\leq z<10$ & $-6.57${\raisebox{0.5ex}{\tiny$^{+0.52}_{-0.76}$}} (0, 1) / $-6.09${\raisebox{0.5ex}{\tiny$^{+0.30}_{-0.34}$}} & $-5.87${\raisebox{0.5ex}{\tiny$^{+0.22}_{-0.25}$}} (3, 2) / $-5.68${\raisebox{0.5ex}{\tiny$^{+0.18}_{-0.19}$}} \\
\hline

\end{tabular}

\end{table*}

We estimated the comoving number density of the quiescent candidates considering two mass thresholds, lM$_*>9.5$ and lM$_*>10.5$, in the four redshift bins $3<z\leq4$, $4<z\leq5$, $5<z\leq7$, and $7<z\leq10$. We excluded the ABELL2744 field from this computation, since it is affected by lensing magnification by the foreground galaxy cluster. Figure \ref{ndf} shows the results, and we list the relevant values in Tab. \ref{ndt}. 
We plot the values obtained for the ``gold'' and the total (``gold'' plus ``silver'') selections.  The raw values (black triangular ticks) were obtained simply by dividing the number of candidates in each redshift bin by its comoving cosmological volume, normalised to the total observed area. The corresponding estimates corrected for contamination and completeness (black triangles) were obtained using coefficients estimated with the simulations described in Appendix \ref{simulations}. 
We point out that while in principle the corrected ``gold'' and total estimates should be close, in reality the additional criteria imposed on the ``gold'' selection (i.e. ``$z$-robust'', six valid NIRCam bands, no flaw at visual inspection) cause some differences in the corrected number densities, the ``gold'' selection resulting in lower values. It is also worth stressing that the points for the $7<z<10$ redshift bin are obtained with just seven candidates: one ``gold'' candidates (the confirmed PS from W25), one low reliability candidate at $<=8.26$, and five``silver'' sources, all identified as LRDs, at $9<z<10$ (ee Sect. \ref{wow_objs}). Finally, we point out that removing the LRDs even from the ``silver'' selection the number densities at $z>5$ would be significantly lower: in Fig. \ref{ndf} we  show the relevant points for $z>5$ and lM$_*>9.5$ (blue symbols), because for all the other cases the difference is negligible.

In Fig. \ref{ndf} we also show various estimates published in recent years (colored symbols), including: those from \citet{Schreiber2018b} and M19 from the pre-JWST era; the one by \citet{Gould2023} who used COSMOS2020 data \citep{Weaver2022}, with a lower mass cut at lM$_*$=10.6; and the recent samples described in Appendix \ref{othersel}, with the acronyms explained in the caption. C23 and R24 points are obtained by adapting the values reported in their work to our selection criteria (however, in C23 all candidates are fitted as RD, i.e. have $dt_q>100$ Myr). Publicly available data from V23, dlV25 and B25 do not provide all the necessary information, so we just plot the values given in their papers at face value (note that V23 also cuts at lM$_*$=10.6).

Our raw estimates are broadly in agreement with the other JWST-based ones at $3<z<5$, but tend to be higher because of the inclusion of many low mass, recently quenched candidates, which we are able to single out by exploring the SED-fitting solutions rather than selecting on colors. They are also significantly higher than the corresponding values from HST-based works due to the increased accuracy and sensitivity of the JWST observations: in particular, the total (``gold'') raw number density estimates for lM$_*>9.5$, in units of $10^{-5}$ cMpc$^{-3}$, respectively increase from 2.9 to 17.8 (9.1) at $3<z<4$, and from 0.4 to 7.6 (3.5) at $4<z<5$, compared to those from the ``lines'' selection in M19 (we remind that now emission lines are always included in the SED-fitting library). We point out that at lM$_*>10.5$ our estimates lay between those by C23, which were from the CEERS field, and those from other studies. At $z>5$, instead, our values are lower than those from other studies, most likely because of the stringent selection criteria we applied; indeed, when correction terms for completeness and contamination are included, the estimates are consistent with other studies at least in the full lM$_*>9.5$ sample. However, the drop of lM$_*>10.5$ candidates at $z\approx5$ discussed in Sect. \ref{nordz5} is clearly visible.

\subsubsection{Comparison with models} \label{models}

In Fig.~\ref{ndf} we also plot the number densities of quiescent galaxies from some hydrodynamical cosmological simulations (colored lines). For \textsc{Eagle} \citep{Schaye2015} and \textsc{TNG100} \citet{Pillepich2018} we computed the values from the publicly available output snapshots, adopting the same criteria used for the observed data (lSSFR $<-10$, $dt_q>10$ Myr, lM$_*\geq9.5$ and 10.5). The SFH of each galaxy was obtained from the formation times (age of the Universe at birth) and initial masses of the individual stellar particles in the snapshots, including only those within 2R$_{1/2}$, and we considered the SFR averaged on intervals of 10 Myr. We also show the values from four mock light-cones created with the code \textsc{Forecast} \citep{Fortuni2023}; the differences between the values from the snapshots and the light-cones can be explained by considering the details of the procedure, which we skip here. 
The data for the \textsc{Simba} \citep{Dave2019} and \textsc{Magneticum Pathfinder} \citep{Kimmig2025} models were kindly provided by the authors of the simulations. For \textsc{Simba} we used the same data as in M19, except that now we plot the selection with lSSFR $<-10$ instead of $-11$. 

As already known and discussed in M19, most models fail to reproduce the number densities of quiescent objects at $z>4$ (while typically overestimating them at $z<3$). \textsc{Eagle} is in broad agreement with the lower published estimates at $4<z<5$ for lM$_*>9.5$, but fails at producing massive quenched objects at $z>4$. In contrast, \textsc{Magneticum} seems in good agreement with the observations at $z<5$, and within some reasonable offset at $5<z<7$ for lM$_*>9.5$, but falls short of the observed values at $z>7$, and at $z>6$ for lM$_*>10.5$. However, as already pointed out the observed raw numbers at these redshifts are very small, so the comparison is less significant here (and again, the stellar masses estimated in the SED-fitting procedure might be overestimated). 
Taken at face value, this result alone would seem to imply that there is no intrinsic tension with the prediction of $\Lambda$CDM cosmology, provided tailored recipes are included in the simulations, in particular concerning the feedback mechanisms and the size of the simulated cosmological box. However, the details of the quenched galaxies reveal that the agreement is actually not satisfying yet, as shown in the next subsection.

\begin{figure*}
   \centering
   \includegraphics[width=0.95\hsize]{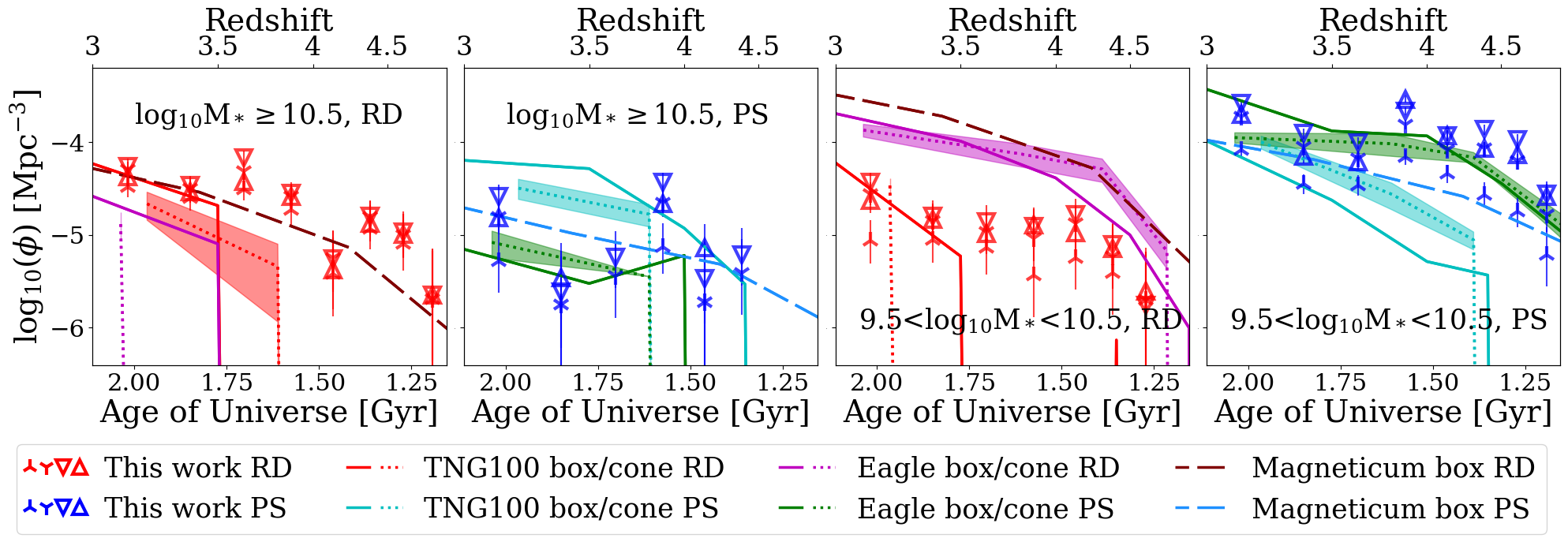}
    \caption{Number densities at $3 < z < 5$ in the two mass regimes M$_* \geq 10.5$ and $9.5 \leq $ M$_* < 10.5$ of the RD and PS candidates from this work (red and blue symbols, respectively; triangular ticks are the raw values, triangles are the values corrected for completeness and contamination; upward-pointing symbols are for the total selection, downward-pointing symbols are for the ``gold'' selection), and from the and \textsc{Magneticum}, \textsc{TNG100}, and \textsc{Eagle} simulations (with the colors of the lines resembling the same division between RDs and PSs), as estimated directly from the snapshots (solid and dashed lines), and from four \textsc{Forecast} light-cones (for \textsc{TNG} an \textsc{Eagle} only; the shaded areas give the standard deviation from the median values, shown as dotted lines).}
    \label{nd_dtq}
\end{figure*}

\subsubsection{Number densities of RD and PS, high and low mass candidates} \label{nddtq}

As discussed at the beginning of this Section, the observed raw number of low mass ($9.5<\mbox{lM}_*<10.5$) RD candidates does not increase significantly from $z=5$ to 3, implying that the PS candidates are most likely experiencing a phase of ``breathing'', i.e. a temporary quenching before re-juvenation. We now make this even cleared in Fig. \ref{nd_dtq}, where we plot the number densities (both raw and corrected for completeness and contamination) of our RD and PS candidates divided into the two mass regimes, showing how the values of low mass RDs always stays well below that of low mass PSs, while in principle one could expect to see them grow substantially. 

We also plot the trends for the \textsc{Eagle}, \textsc{TNG100}, and \textsc{Magneticum} simulations. For the first two, the data is publicly available, so we can apply the exact selection criteria used for the observations (we plot both the trends straightforwardly from the boxes, and those obtained with the \textsc{Forecast} light-cones); for the latter, the data was kindly provided by the research team. The difference between the three simulations is evident, especially in the low mass regime. At lM$_*>10.5$, only \textsc{Magneticum} reproduces the existence of RDs at $z>4.5$, also yielding the correct densities of RDs and PSs at lower redshifts; \textsc{Eagle} is also reasonably close, but falls short at $z>4$, while \textsc{TNG100} largely overestimates the abundance of PSs at $z<4.5$. However, at $9.5<\mbox{lM}_*<10.5$ the situation is even more complicated: \textsc{TNG} cannot reproduce the density of PSs at $z>3.5$ and of RDs at $z>4$, while \textsc{Eagle} is consistent with the density of PSs at all epochs but has too many RDs at $3<z<4$; finally, the trends in \textsc{Magneticum} are inverted with respect to the observations, with too many RDs and too few PSs, possibly hinting at an excessive efficiency of AGN feedback from low mass super-massive black holes. These comparisons suggest that even when the global trends can be reproduced, comparing in more details the properties of the simulated and observed galaxies still yields non negligible tensions.

\subsection{Interesting candidates} \label{wow_objs}

In this Section we discuss some interesting sources: robust candidates not included in previous selections, objects at $z>5$, and very massive galaxies (according to their best-fit). Some of them (snapshots, best-fit SEDs and parameters, PDF($z$), and observed spectrum when available) are shown as examples in Appendix \ref{objs}. 

\subsubsection{``Gold'' high reliability $z<5$ candidates}

Twenty sources not already included in the spectroscopic sample of Tab. \ref{spectrq} are ``gold'' with reliability parameter $r\geq0.8$; they all are at $3<z<5$. Eight examples are shown in Figs. \ref{objs1} (RDs) and \ref{objs1b} (PSs). Their best-fit SEDs clearly show the typical features of quiescent sources, with a pronounced Balmer/4000 break and a declining UV rest-frame emission. Noticeably, strong H$\alpha$ and [OIII] emission lines are evident in some cases where the observed spectrum is available; as already mentioned, they can be powered by nuclear activity or gas heated by shocks. Interestingly, 22 out of the 23 known spectroscopic quiescents in Tab. \ref{spectrq} included in our ``gold'' selection have $0.4<r<0.55$, in all cases because of the $p_{best}$ value (no one has SF alternative solutions). 

\subsubsection{Candidates at $z>5$}

As mentioned, we have a total of 34 candidates at $z>5$. Seven are ``gold'', although with reliability values $0.42<r<0.53$; five of them have spectroscopic redshifts. Of the remaining 27 ``silver'' candidates, 25 have photometric redshift, and only three do not have alternative SF solutions. 

The current record-holders as the farthest spectroscopically confirmed quiescent objects are the aforementioned PRIMERS-UDS ID117643 (W25) and JADES-GS ID47234 \citet{Looser2024}, both at $z_{spec}\approx\!7.29$. In our SED-fitting runs, the latter is best fitted as young ($dt_q$=20 Myr) PS of mass $M_*=4.6\times10^8M_{\odot}$ (in agreement with the value $4-6\times10^8$ from the original work), which is lower than our minimum threshold mass of $10^{9.5}M_{\odot}$, and it is therefore not included in our selection. W25 is in our ``gold'' sample, best fit as a massive (lM$_*$=10.6), mildly obscured ($E(B-V)$=0.3) galaxy with solar metallicity ($Z/Z_{\odot}=1$); it is formed at $z=7.97$, with a TH SF burst of 10 Myr followed by 70 Myr of quiescence, implying an extreme peak SFR of $\approx\!4000$ M$_{\odot}$/yr (the authors fit it at a lower lM$_*=10.2$, $A_v=0.25$, $t_{90}=70$ Myr). We notice that the spectrum available from the DJA archive appears contaminated at $\lambda_{obs}\leq2$ $\mu$m (it comes from the CAPERS survey while the original work is from the RUBIES survey).

JADES-GS ID57595, at $z_{spec}=6.1$, is the only other ``gold'' candidate at $z>6$; it is best-fitted as a lM$_*=9.70$ PS ($dtq=110$ Myr) formed at $z\approx\!7$. Unfortunately, despite having reliability $r=0.45$ it provides a good example of a wrong fit: the emission lines biased the photometry in the red bands, and in particular the OII line spuriously enhanced the estimated DBalmer, resulting in a quiescent fit (although a break is present, but smaller than the one estimated with the photometric SED-fitting).

We have two more RD candidates at $z>6$ with $r>0.4$: PRIMER-UDS ID51964, at $z_{phot}=6.38$ with no lower redshift solutions, is best fitted as a massive (lM$_*$ 10.95) galaxy with super-solar metallicity, formed at $z\approx\!9.5$ in a burst of 30 Myr followed by 320 Myr of quiescence; JADES-GS ID27908, with photometric solutions only at $z>10$, is best-fitted as a lM$_*=9.77$ galaxy at $z_{phot}=10.6$, formed at $z\approx\!25$ in a quick 10 Myr SF burst followed by 290 Myr of quiescence. Both fail the ``gold'' selection because they are identified as LRDs: the first in the new sample based on \citet{Barro2024b}, the second in the list by \citet{PerezGonzalez2024}. Indeed, these two objects appear as compact, almost point-like sources; and although their faintness make the HST photometry prone to large errors making it difficult to properly constrain the fit, an uprising in the rest-frame UV fluxes which is not reproduced by our models can be seen. As mentioned, we leave them both in our ``silver'' selection, because the real nature of these type of sources is still being debated (see cited references). We then have nine more LRDs in the ``silver'' selection at $z_{phot}>9$ \citep[five are from][where they have slightly lower redshift estimates, but three are above 9 anyway, one is at 8.32 at one at 6.01]{Kocevski2025}; they all have $r<0.2$ except for one having $r=0.234$ and the mentioned JADES-GS ID27908, which has $r=0.45$. 

Finally, we have only two candidates at $z>8$ not already classified as LRDs. PRIMER-COSMOS ID104822 is a ``silver'', but very low reliability ($r=0.11$) compact source at $z_{spec}=8.29$, fitted as a lM$_*=9.7$ formed at $z=11.2$ and quenched since 170 Myr; it looks faint and compact, and despite not being in any of the considered lists, its photometric SED (poorly reproduced by our best-fit model) shows the uprise in the optical rest-frame typical of LRDs. The other candidate is the very peculiar ``Capotauro'' \citep[][CEERS ID17577 in M24]{Gandolfi2025}, a faint, compact F356W drop-out. As discussed in their paper, it might be a distant cold brown dwarf or an orphan planet, with alternative solutions as a $z\approx\!30$ Lyman-break galaxy, or as a $z\approx\!9.8$ quiescent, very massive (lM$_*=11.6$) galaxy: the latter is the only acceptable solution with our library of models and the $z_{phot}$ estimate from M24 (where the maximum allowed redshift was 20). It is fitted with E(B-V)=1 and $Z=2.5Z_{\odot}$, formed at $z\approx\!20$ with a short (100 Myr) burst of SF (an alternative quiescent solution with lM$_*=10.6$ and E(B-V)=0.4 would require a formation redshift of $\approx\!$70). Admittedly, the quiescent galaxy solution is far-fetched, especially considering that the best-fit with our library fails the upper-limit in F356W. Nevertheless, it does enter our ``silver'' selection with a reliability value of 0.16, so we leave it in our sample, despite its extreme nature; a spectroscopic follow-up is needed to fully disentangle its mysteries.

The snapshots, best-fit SEDs and PDF($z$) of some of the objects discussed in this subsection are shown in Fig. \ref{objs2}. 


\subsubsection{Ultra-massive candidates} \label{wow_objs_mass}

Four candidates have best fit masses lM$_*>11.5$. One is the already mentioned ``Capotauro''. The other three have spectroscopic redshifts. Two are in our ``gold'' selection: ID98234 in PRIMER-UDS at $z_{spec}=4.56$ is point-like and the best-fit is certainly sub-optimal, despite having no reliable alternative solutions (it is indeed one of the outliers in Fig. \ref{breaks} discussed in Sect. \ref{specomp}, and its observed SED suggests it to be an AGN with strong H$\alpha$ and MgII broad emissions, and a peculiar extinction); while ID73230 in CEERS at $z_{spec}=3.64$ looks like an evolved spiral. The same holds for the last one, ID117344 in PRIMER-UDS, a ``silver'' candidate with $z_{spec}=4.06$. Both are best-fitted as RDs, but their morphology suggests that their real nature might be that of reddened SF sources with old bulges. However, ID177344 only has low probability SF alternative solutions, while ID73230 has none.  
The snapshots, best-fit SEDs and PDF($z$) of the four objects are shown in Fig. \ref{objs3}.
It is worth mentioning that many ultra-massive (lM$_*>11.7$) photometric candidates at intermediate redshift ($3<z<4$), including some in the quiescent wedge of the $UVJ$ diagram, have been shown to actually be lower mass, $z<3$ interlopers, suggesting difficulties for photometric redshift programs in fitting similarly red SEDs \citep{Forrest2024}; however, our ultra-massive candidates all have spectroscopic redshifts.

\section{Summary and discussion} \label{sec:concl}

We have searched for quiescent galaxies at $z>3$ exploiting the publicly available photometric catalog ASTRODEEP-JWST presented in \citet{Merlin2024}. Using a comprehensive library of galaxy SED models with the SED-fitting code \textsc{zphot} and exploring the $\chi^2$ probabilities of all fits to exclude objects with alternative reliable SF solutions, we select 633 candidates, among which 291 are tagged as ``gold'', having a high reliability value and a robust redshift determination (see Sect. \ref{selec}); of these, 164 (95 ``gold'') are best-fitted as red and dead (RD) sources quenched for more than 150 Myr ($dt_q\geq150$ Myr), while the rest are best-fitted as recently quenched post-starbursts (PS, $dt_q<150$ Myr). Since the detection is essentially complete down to AB$\approx\!28$, we can be confident that we did not lose faint potential candidates. We assess the completeness and the contamination of our sample by checking the spectra publicly available from the DAWN JWST Archive, both compiling a list of known quiescent sources and checking the spectra of our candidates having available spectroscopic data, and by means of simulated datasets, which show that with the total (``gold'') selection we reach $\approx70\%$ (90\%) completeness while staying below $\approx 5\%$ ($15\%$) contamination at F444W SNR $> 50$ and $z<6$. We used these estimates to compute corrective terms which we used to evaluate number densities. In the following we summarise our main findings.

\begin{itemize}
\item We found a strong bimodality in the mass distribution of the candidates at $3<z<5$, with $79\%$ of massive galaxies (lM$_*\geq10.5$) being RDs, and 89$\%$ of low mass systems ($9.5\leq$lM$_*<10.5$) being PSs. Massive RDs are also typically more compact than the low mass PSs. The number density of low mass RDs does not increase significantly from $z=5$ to 3, implying that most of the PSs do not evolve passively, and are most likely experiencing a temporary phase of quenching before starting again to form stars (``breathing''), consistently with the ``downsizing'' scenario of galaxy formation; we are now able to probe it while it is happening, rather then inferring it from the stellar populations of local galaxies.

\item Our raw estimates for the number densities of quiescent galaxies at lM$_*>9.5$ are in broad agreement with those from other recent studies, but tend to be higher, especially considering the corrected values; this is most likely because of the inclusion of many young PSs. We compared our values against theoretical predictions, also considering the distinction between RDs and PSs and two mass regimes; we confirm the relevant tensions for most of the models against the observations, although the \textsc{Magneticum} simulation yields values in agreement with our estimates for global number densities. However, considering the number densities of RDs and PSs and of high and low mass populations separately, we find that the agreement is still sub-optimal in most of the considered cases, suggesting that the recipes for feedback implemented in the simulations are not yet able to fully capture the complex physics at play in these systems, and to correctly reproduce their real quenching mechanisms.

\item We find an abrupt drop of RD candidates at $z>5$, which does not seem to be caused by our selection criteria, and might be consistent with the lack of spectroscopically confirmed quiescents at $z>5$ and an observed drop of the full galaxy population $z\approx4.5$.

\item A few of the candidates are best-fitted with extreme SFHs, implying peaks in the SFR of more than $10^3$ M$_{\odot}$/yr, a SF efficiency consistent with 100\%, and very early mass assembly, in tension with the timescales allowed in the $\Lambda$CDM cosmology (Fig. \ref{evol}). While the SED-fitting procedure cannot take into account their possible merging history, given their compactness it seems unlikely that they formed as the result of multiple mergers of smaller quiescent sub-units, which would result in more relaxed structures. Some of them are indeed already known as possible ``Universe breakers'' \citep[see][]{Glazebrook2024,Carnall2024}, while others are only in tension with $\Lambda$CDM predictions in our fit, and have more recent formation times in other works. We warn that our stellar masses and ages might be overestimated.

\item Even excluding these extreme objects, many of the candidates are best-fitted as systems forming most of their mass in a burst of SF activity in the first Gyr of the cosmic history, and most (85\%) are best fitted with a top-hat SFH, implying a quick and abrupt quenching. Interestingly, \citet{Barrufet2025} found that the number density of massive SF galaxies at $z>6$ is sufficient to yield the observed number densities of quiescent galaxies at $3<z<5$. 
\end{itemize}

While it is not impossible to form this many early, rapidly quenched massive systems within the $\Lambda$CDM scenario, it is also worth pointing out how these number densities seem to naturally agree with the qualitative predictions of structure formation in Modified Newtonian Dynamics (MOND), where the collapse of virialized baryonic over-densities of galactic mass (lM$_*\approx\!11$) is expected to happen at $z\approx\!10$ \citep[see][]{McGaugh2024}, in a monolithic-like structure formation \citep[see also][]{Yan2021,Eappen2022}. This would imply a rapid onset of strong SF activity, and consequently massive feedback from supernovae and young stars (winds and UV radiation). Among others, \citet{Merlin2012} have shown how this feedback from intense bursts of SF can be sufficient to heat and expel the gas reservoir from massive potential wells on a timescale of a few Myr. However, in $\approx\!33$\% of our massive RD candidates with available spectroscopic data we find strong emission lines (H$\alpha$, [OIII]) that can be associated with significant nuclear activity, implying that AGN feedback might be the main factor responsible for the inferred extremely rapid quenching \citep[see e.g.][]{Pacucci2024}. The halo quenching mechanism triggered by hot gas virial shocks in massive haloes \citep{Dekel2006} seems also consistent with our results, at least for galaxies with lM$_*>10.5$ (corresponding to a typical dark matter halo mass of $10^{12}$ M$_{\odot}$ assuming a 0.2 efficiency of star formation), whilst the quenched galaxies we observe below this mass may be temporarily quenched by stellar feedback or other baryonic processes \citep[see also][]{Cattaneo2008,Fu2024}. 


The compact, massive RDs might be the progenitors of so called ``relics'' galaxies found in the local Universe \citep{Trujillo2009,Trujillo2014,Ferre-Mateu2017,Spiniello2024,Hartmann2025}, which are a small fraction of the early-type population. As mentioned in Sect. \ref{limit}, the classic \citet{Salpeter1959} IMF that we adopted in this study might not be adequate to model early galaxies. While recent results suggest that a top-heavy IMF would be more appropriate for SF massive sources \citep[e.g.][]{Guo2024,Hutter2025}, \citet{Maksymowicz-Maciata2024} found a strong correlation between the ``degree of relicness'' (i.e., the age of the oldest stellar population) and the IMF, with quiescent older systems favouring a bottom-heavy IMF. Roughly speaking, a top-heavy IMF would increase early enrichment and reduce the observed mass-to-light ratio, while a bottom-heavy IMF would increase the estimated stellar masses, making many of our candidates even more ``extreme''; perhaps a time-varying IMF, i.e. an initial top-heavy phase followed by more bottom-heavy evolution, could produce partially compensating effects. The compact quiescent sources might also be the bulges of future M31-like disks, if they accrete gas from the cosmic web \citep[e.g.][]{Nipoti2025}.

The main issue that remains to be assessed is the real nature of the PS candidates, which are many and difficult to pinpoint given the fact that their photometric properties are similar to those of SF populations of similar age but different extinction or metallicity. However, in principle this would not affect our conclusions about the ``downsizing'' trend: the possibility that a fraction of the low mass PS candidates are actually obscured star-forming galaxies would just remove them from our sample, but the number of low mass RDs would still be too small to make the transition from PS to RD in the lM$_*<10.5$ regime a preferential evolutionary path. Analysis of available far-infrared MIRI data, and spectroscopic follow-ups especially for the dubious and the interesting candidates, are needed to fully assess their properties and to draw conclusions on the topics, and spectro-photometric analysis of the candidates is needed to disentangle the ambiguous nature of some of them and confirm the inferred physical properties. 

\newpage 

\section*{Acknowledgments}

Part of the research activities described in this paper were carried out with
contribution of the Next Generation EU funds within the National Recovery and
Resilience Plan (PNRR), Mission 4 - Education and Research, Component 2 -
From Research to Business (M4C2), Investment Line 3.1 - Strengthening and
creation of Research Infrastructures, Project IR0000034 – ``STILES -
Strengthening the Italian Leadership in ELT and SKA''.\\
EM acknowledges the the PRIN 2022 MUR project 2022CB3PJ3 - First Light And Galaxy aSsembly (FLAGS) funded by the European Union – Next Generation EU. \\
Some of the data products were retrieved from the Dawn JWST Archive (DJA). DJA is an initiative of the Cosmic Dawn Center (DAWN), which is funded by the Danish National Research Foundation under grant DNRF140.\\
LCK acknowledges support by the Deutsche Forschungsgemeinschaft (DFG, German Research Foundation) under project nr. 516355818 and under Germany's Excellence Strategy -- EXC-2096 -- 3900783311, as well as support for the COMPLEX project from the European Research Council (ERC) under the European Union’s Horizon 2020 research and innovation program grant agreement ERC-2019-AdG 882679.\\

\bibliographystyle{mn2e} 
\bibliography{biblio.bib} 

\begin{thebibliography}{154}
\expandafter\ifx\csname natexlab\endcsname\relax\def\natexlab#1{#1}\fi

\bibitem[{{Arrabal Haro} {et~al}\mbox{.}(2023){Arrabal Haro}, {Dickinson}, {Finkelstein}, {Kartaltepe}, {Donnan}, {Burgarella}, {Carnall}, {Cullen}, {Dunlop}, {Fern{\'a}ndez}, {Fujimoto}, {Jung}, {Krips}, {Larson}, {Papovich}, {P{\'e}rez-Gonz{\'a}lez}, {Amor{\'\i}n}, {Bagley}, {Buat}, {Casey}, {Chworowsky}, {Cohen}, {Ferguson}, {Giavalisco}, {Huertas-Company}, {Hutchison}, {Kocevski}, {Koekemoer}, {Lucas}, {McLeod}, {McLure}, {Pirzkal}, {Seill{\'e}}, {Trump}, {Weiner}, {Wilkins}, \& {Zavala}}]{ArrabalHaro2023b}
{Arrabal Haro} P. {et~al.}, 2023, \nat, 622, 707

\bibitem[{{Baggen} {et~al}\mbox{.}(2024){Baggen}, {van Dokkum}, {Brammer}, {de Graaff}, {Franx}, {Greene}, {Labb{\'e}}, {Leja}, {Maseda}, {Nelson}, {Rix}, {Wang}, \& {Weibel}}]{Baggen2024}
{Baggen} J. F.~W. {et~al.}, 2024, \apjl, 977, L13

\bibitem[{{Bagley} {et~al}\mbox{.}(2024){Bagley}, {Pirzkal}, {Finkelstein}, {Papovich}, {Berg}, {Lotz}, {Leung}, {Ferguson}, {Koekemoer}, {Dickinson}, {Kartaltepe}, {Kocevski}, {Somerville}, {Yung}, {Backhaus}, {Casey}, {Castellano}, {Ch{\'a}vez Ortiz}, {Chworowsky}, {Cox}, {Dav{\'e}}, {Davis}, {Estrada-Carpenter}, {Fontana}, {Fujimoto}, {Gardner}, {Giavalisco}, {Grazian}, {Grogin}, {Hathi}, {Hutchison}, {Jaskot}, {Jung}, {Kewley}, {Kirkpatrick}, {Larson}, {Matharu}, {Natarajan}, {Pentericci}, {P{\'e}rez-Gonz{\'a}lez}, {Ravindranath}, {Rothberg}, {Ryan}, {Shen}, {Simons}, {Snyder}, {Trump}, \& {Wilkins}}]{Bagley2024}
{Bagley} M.~B. {et~al.}, 2024, \apjl, 965, L6

\bibitem[{{Baker} {et~al}\mbox{.}(2025{\natexlab{a}}){Baker}, {Lim}, {D'Eugenio}, {Maiolino}, {Ji}, {Arribas}, {Bunker}, {Carniani}, {Charlot}, {de Graaff}, {Hainline}, {Looser}, {Lyu}, {Rinaldi}, {Robertson}, {Schaller}, {Schaye}, {Scholtz}, {{\"U}bler}, {Williams}, {Willmer}, {Willott}, \& {Zhu}}]{Baker2025}
{Baker} W.~M. {et~al.}, 2025{\natexlab{a}}, \mnras, 539, 557

\bibitem[{{Baker} {et~al}\mbox{.}(2025{\natexlab{b}}){Baker}, {Valentino}, {Lagos}, {Ito}, {Jespersen}, {Gottumukkala}, {Hjorth}, {Langeroodi}, \& {Sedgewick}}]{Baker2025b}
{Baker} W.~M. {et~al.}, 2025{\natexlab{b}}, arXiv e-prints, arXiv:2506.04119

\bibitem[{{Barro} {et~al}\mbox{.}(2024){Barro}, {Perez-Gonzalez}, {Kocevski}, {McGrath}, {Leung}, {Cullen}, {Dunlop}, {Ellis}, {Finkelstein}, {Grogin}, {Illingworth}, {Kartaltepe}, {Koekemoer}, {Lucas}, {McLure}, \& {Yang}}]{Barro2024b}
{Barro} G. {et~al.}, 2024, arXiv e-prints, arXiv:2412.01887

\bibitem[{{Barrufet} {et~al}\mbox{.}(2024){Barrufet}, {Arellano Cordova}, {Baggen}, {Begley}, {Carnall}, {Cullen}, {Donnan}, {Dunlop}, {Flury}, {Fudamoto}, {Gottumukkala}, {Illingworth}, {McLeod}, {McLure}, {Michalowski}, {Oesch}, {Stefanon}, {Valentino}, {Weibel}, {Williams}, \& {van Dokkum}}]{Barrufet2024}
{Barrufet} L. {et~al.}, 2024, {Dead or alive? Unveiling the nature of massive galaxies in the early Universe}. JWST Proposal. Cycle 3, ID. \#5545

\bibitem[{{Barrufet} {et~al}\mbox{.}(2025){Barrufet}, {Dunlop}, {Begley}, {Flury}, {McLeod}, {Arellano-Cordova}, {Carnall}, {Cullen}, {Donnan}, {Liu}, {McLure}, {Scholte}, {Stanton}, {Cochrane}, {Conselice}, {Ellis}, {P{\'e}rez-Gonz{\'a}lez}, {Gottumukkala}, {Grogin}, {Illingworth}, {Koekemoer}, {Magee}, \& {Michalowski}}]{Barrufet2025}
{Barrufet} L. {et~al.}, 2025, arXiv e-prints, arXiv:2508.05740

\bibitem[{{Berger} {et~al}\mbox{.}(2025){Berger}, {Marshall}, {Wyithe}, {di Matteo}, {Ni}, {Wilkins}, \& {Yue}}]{Berger2025}
{Berger} S., {Marshall} M.~A., {Wyithe} J. S.~B., {di Matteo} T., {Ni} Y., {Wilkins} S.~M., {Yue} M., 2025, arXiv e-prints, arXiv:2506.12130

\bibitem[{{Bertin} \& {Arnouts}(1996)}]{Bertin1996}
{Bertin} E., {Arnouts} S., 1996, \aaps, 117, 393

\bibitem[{{Bezanson} {et~al}\mbox{.}(2024){Bezanson}, {Labbe}, {Whitaker}, {Leja}, {Price}, {Franx}, {Brammer}, {Marchesini}, {Zitrin}, {Wang}, {Weaver}, {Furtak}, {Atek}, {Coe}, {Cutler}, {Dayal}, {van Dokkum}, {Feldmann}, {F{\"o}rster Schreiber}, {Fujimoto}, {Geha}, {Glazebrook}, {de Graaff}, {Greene}, {Juneau}, {Kassin}, {Kriek}, {Khullar}, {Maseda}, {Mowla}, {Muzzin}, {Nanayakkara}, {Nelson}, {Oesch}, {Pacifici}, {Pan}, {Papovich}, {Setton}, {Shapley}, {Smit}, {Stefanon}, {Taylor}, \& {Williams}}]{Bezanson2024}
{Bezanson} R. {et~al.}, 2024, \apj, 974, 92

\bibitem[{{Binggeli} {et~al}\mbox{.}(2019){Binggeli}, {Zackrisson}, {Ma}, {Inoue}, {Vikaeus}, {Hashimoto}, {Mawatari}, {Shimizu}, \& {Ceverino}}]{Binggeli2019}
{Binggeli} C. {et~al.}, 2019, \mnras, 489, 3827

\bibitem[{{Bower} {et~al}\mbox{.}(2006){Bower}, {Benson}, {Malbon}, {Helly}, {Frenk}, {Baugh}, {Cole}, \& {Lacey}}]{Bower2006}
{Bower} R.~G., {Benson} A.~J., {Malbon} R., {Helly} J.~C., {Frenk} C.~S., {Baugh} C.~M., {Cole} S., {Lacey} C.~G., 2006, 370, 645

\bibitem[{{Brammer}, {van Dokkum} \& {Coppi}(2008){Brammer}, {van Dokkum}, \& {Coppi}}]{Brammer2008}
{Brammer} G.~B., {van Dokkum} P.~G., {Coppi} P., 2008, \apj, 686, 1503

\bibitem[{{Bruzual} \& {Charlot}(2003)}]{Bruzual2003}
{Bruzual} G., {Charlot} S., 2003, \mnras, 344, 1000

\bibitem[{{Bundy} {et~al}\mbox{.}(2006){Bundy}, {Ellis}, {Conselice}, {Taylor}, {Cooper}, {Willmer}, {Weiner}, {Coil}, {Noeske}, \& {Eisenhardt}}]{Bundy2006}
{Bundy} K. {et~al.}, 2006, 651, 120

\bibitem[{{Bunker} {et~al}\mbox{.}(2024){Bunker}, {Cameron}, {Curtis-Lake}, {Jakobsen}, {Carniani}, {Curti}, {Witstok}, {Maiolino}, {D'Eugenio}, {Looser}, {Willott}, {Bonaventura}, {Hainline}, {{\"U}bler}, {Willmer}, {Saxena}, {Smit}, {Alberts}, {Arribas}, {Baker}, {Baum}, {Bhatawdekar}, {Bowler}, {Boyett}, {Charlot}, {Chen}, {Chevallard}, {Circosta}, {DeCoursey}, {de Graaff}, {Egami}, {Eisenstein}, {Endsley}, {Ferruit}, {Giardino}, {Hausen}, {Helton}, {Hviding}, {Ji}, {Johnson}, {Jones}, {Kumari}, {Laseter}, {L{\"u}tzgendorf}, {Maseda}, {Nelson}, {Parlanti}, {Perna}, {Rauscher}, {Rawle}, {Rix}, {Rieke}, {Robertson}, {Rodr{\'\i}guez Del Pino}, {Sandles}, {Scholtz}, {Sharpe}, {Skarbinski}, {Stark}, {Sun}, {Tacchella}, {Topping}, {Villanueva}, {Wallace}, {Williams}, \& {Woodrum}}]{Bunker2024}
{Bunker} A.~J. {et~al.}, 2024, \aap, 690, A288

\bibitem[{{Calzetti} {et~al}\mbox{.}(2000){Calzetti}, {Armus}, {Bohlin}, {Kinney}, {Koornneef}, \& {Storchi-Bergmann}}]{Calzetti2000}
{Calzetti} D., {Armus} L., {Bohlin} R.~C., {Kinney} A.~L., {Koornneef} J., {Storchi-Bergmann} T., 2000, \apj, 533, 682

\bibitem[{{Carnall} {et~al}\mbox{.}(2024){Carnall}, {Cullen}, {McLure}, {McLeod}, {Begley}, {Donnan}, {Dunlop}, {Shapley}, {Rowlands}, {Almaini}, {Arellano-C{\'o}rdova}, {Barrufet}, {Cimatti}, {Ellis}, {Grogin}, {Hamadouche}, {Illingworth}, {Koekemoer}, {Leung}, {Lovell}, {P{\'e}rez-Gonz{\'a}lez}, {Santini}, {Stanton}, \& {Wild}}]{Carnall2024}
{Carnall} A.~C. {et~al.}, 2024, \mnras, 534, 325

\bibitem[{{Carnall} {et~al}\mbox{.}(2023){Carnall}, {McLeod}, {McLure}, {Dunlop}, {Begley}, {Cullen}, {Donnan}, {Hamadouche}, {Jewell}, {Jones}, {Pollock}, \& {Wild}}]{Carnall2023}
{Carnall} A.~C. {et~al.}, 2023, \mnras, 520, 3974

\bibitem[{{Carnall} {et~al}\mbox{.}(2018){Carnall}, {McLure}, {Dunlop}, \& {Dav{\'e}}}]{Carnall2018}
{Carnall} A.~C., {McLure} R.~J., {Dunlop} J.~S., {Dav{\'e}} R., 2018, \mnras, 480, 4379

\bibitem[{{Castellano} {et~al}\mbox{.}(2014){Castellano}, {Sommariva}, {Fontana}, {Pentericci}, {Santini}, {Grazian}, {Amorin}, {Donley}, {Dunlop}, {Ferguson}, {Fiore}, {Galametz}, {Giallongo}, {Guo}, {Huang}, {Koekemoer}, {Maiolino}, {McLure}, {Paris}, {Schaerer}, {Troncoso}, \& {Vanzella}}]{Castellano2014}
{Castellano} M. {et~al.}, 2014, \aap, 566, A19

\bibitem[{{Cattaneo} {et~al}\mbox{.}(2008){Cattaneo}, {Dekel}, {Faber}, \& {Guiderdoni}}]{Cattaneo2008}
{Cattaneo} A., {Dekel} A., {Faber} S.~M., {Guiderdoni} B., 2008, \mnras, 389, 567

\bibitem[{{Cimatti}, {Daddi} \& {Renzini}(2006){Cimatti}, {Daddi}, \& {Renzini}}]{Cimatti2006}
{Cimatti} A., {Daddi} E., {Renzini} A., 2006, 453, L29

\bibitem[{{Cowie} {et~al}\mbox{.}(1996){Cowie}, {Songaila}, {Hu}, \& {Cohen}}]{Cowie1996}
{Cowie} L.~L., {Songaila} A., {Hu} E.~M., {Cohen} J.~G., 1996, 112, 839

\bibitem[{{Daddi} {et~al}\mbox{.}(2004){Daddi}, {Cimatti}, {Renzini}, {Fontana}, {Mignoli}, {Pozzetti}, {Tozzi}, \& {Zamorani}}]{Daddi2004}
{Daddi} E., {Cimatti} A., {Renzini} A., {Fontana} A., {Mignoli} M., {Pozzetti} L., {Tozzi} P., {Zamorani} G., 2004, 617, 746

\bibitem[{{Dav{\'e}} {et~al}\mbox{.}(2019){Dav{\'e}}, {Angl{\'e}s-Alc{\'a}zar}, {Narayanan}, {Li}, {Rafieferantsoa}, \& {Appleby}}]{Dave2019}
{Dav{\'e}} R., {Angl{\'e}s-Alc{\'a}zar} D., {Narayanan} D., {Li} Q., {Rafieferantsoa} M.~H., {Appleby} S., 2019, \mnras, 486, 2827

\bibitem[{{de Graaff} {et~al}\mbox{.}(2025{\natexlab{a}}){de Graaff}, {Rix}, {Naidu}, {Labbe}, {Wang}, {Leja}, {Matthee}, {Katz}, {Greene}, {Hviding}, {Baggen}, {Bezanson}, {Boogaard}, {Brammer}, {Dayal}, {van Dokkum}, {Goulding}, {Hirschmann}, {Maseda}, {McConachie}, {Miller}, {Nelson}, {Oesch}, {Setton}, {Shivaei}, {Weibel}, {Whitaker}, \& {Williams}}]{DeGraaff2025b}
{de Graaff} A. {et~al.}, 2025{\natexlab{a}}, arXiv e-prints, arXiv:2503.16600

\bibitem[{{de Graaff} {et~al}\mbox{.}(2025{\natexlab{b}}){de Graaff}, {Setton}, {Brammer}, {Cutler}, {Suess}, {Labb{\'e}}, {Leja}, {Weibel}, {Maseda}, {Whitaker}, {Bezanson}, {Boogaard}, {Cleri}, {De Lucia}, {Franx}, {Greene}, {Hirschmann}, {Matthee}, {McConachie}, {Naidu}, {Oesch}, {Price}, {Rix}, {Valentino}, {Wang}, \& {Williams}}]{DeGraaff2025}
{de Graaff} A. {et~al.}, 2025{\natexlab{b}}, Nature Astronomy, 9, 280

\bibitem[{{de la Vega} {et~al}\mbox{.}(2025){de la Vega}, {Babcock}, {Mobasher}, {Riemann}, {Chartab}, {Hemmati}, {Long}, \& {Sanjaripour}}]{delaVega2025}
{de la Vega} A., {Babcock} M.~D., {Mobasher} B., {Riemann} D.~A., {Chartab} N., {Hemmati} S., {Long} A.~S., {Sanjaripour} S., 2025, arXiv e-prints, arXiv:2501.09066

\bibitem[{{De Lucia} {et~al}\mbox{.}(2024){De Lucia}, {Fontanot}, {Xie}, \& {Hirschmann}}]{DeLucia2024}
{De Lucia} G., {Fontanot} F., {Xie} L., {Hirschmann} M., 2024, \aap, 687, A68

\bibitem[{{Dekel} \& {Birnboim}(2006)}]{Dekel2006}
{Dekel} A., {Birnboim} Y., 2006, \mnras, 368, 2

\bibitem[{{D'Eugenio} {et~al}\mbox{.}(2024){D'Eugenio}, {Cameron}, {Scholtz}, {Carniani}, {Willott}, {Curtis-Lake}, {Bunker}, {Parlanti}, {Maiolino}, {Willmer}, {Jakobsen}, {Robertson}, {Johnson}, {Tacchella}, {Cargile}, {Rawle}, {Arribas}, {Chevallard}, {Curti}, {Egami}, {Eisenstein}, {Kumari}, {Looser}, {Rieke}, {Rodr{\'\i}guez Del Pino}, {Saxena}, {{\"U}bler}, {Venturi}, {Witstok}, {Baker}, {Bhatawdekar}, {Bonaventura}, {Boyett}, {Charlot}, {Danhaive}, {Hainline}, {Hausen}, {Helton}, {Ji}, {Ji}, {Jones}, {Joud{\v{z}}balis}, {Maseda}, {P{\'e}rez-Gonz{\'a}lez}, {Perna}, {Pusk{\'a}s}, {Shivaei}, {Silcock}, {Simmonds}, {Smit}, {Sun}, {Villanueva}, {Williams}, \& {Zhu}}]{DEugenio2024}
{D'Eugenio} F. {et~al.}, 2024, submitted to ApJS, arXiv:2404.06531

\bibitem[{{Diemer}(2018)}]{Diemer2018}
{Diemer} B., 2018, \apjs, 239, 35

\bibitem[{{Eappen} {et~al}\mbox{.}(2022){Eappen}, {Kroupa}, {Wittenburg}, {Haslbauer}, \& {Famaey}}]{Eappen2022}
{Eappen} R., {Kroupa} P., {Wittenburg} N., {Haslbauer} M., {Famaey} B., 2022, \mnras, 516, 1081

\bibitem[{{Eisenstein} {et~al}\mbox{.}(2023){Eisenstein}, {Willott}, {Alberts}, {Arribas}, {Bonaventura}, {Bunker}, {Cameron}, {Carniani}, {Charlot}, {Curtis-Lake}, {D'Eugenio}, {Endsley}, {Ferruit}, {Giardino}, {Hainline}, {Hausen}, {Jakobsen}, {Johnson}, {Maiolino}, {Rieke}, {Rieke}, {Rix}, {Robertson}, {Stark}, {Tacchella}, {Williams}, {Willmer}, {Baker}, {Baum}, {Bhatawdekar}, {Boyett}, {Chen}, {Chevallard}, {Circosta}, {Curti}, {Danhaive}, {DeCoursey}, {de Graaff}, {Dressler}, {Egami}, {Helton}, {Hviding}, {Ji}, {Jones}, {Kumari}, {L{\"u}tzgendorf}, {Laseter}, {Looser}, {Lyu}, {Maseda}, {Nelson}, {Parlanti}, {Perna}, {Pusk{\'a}s}, {Rawle}, {Rodr{\'\i}guez Del Pino}, {Sandles}, {Saxena}, {Scholtz}, {Sharpe}, {Shivaei}, {Silcock}, {Simmonds}, {Skarbinski}, {Smit}, {Stone}, {Suess}, {Sun}, {Tang}, {Topping}, {{\"U}bler}, {Villanueva}, {Wallace}, {Whitler}, {Witstok}, \& {Woodrum}}]{Eisenstein2023}
{Eisenstein} D.~J. {et~al.}, 2023, submitted to ApJS, arXiv:2306.02465

\bibitem[{{Fan} {et~al}\mbox{.}(2006){Fan}, {Strauss}, {Becker}, {White}, {Gunn}, {Knapp}, {Richards}, {Schneider}, {Brinkmann}, \& {Fukugita}}]{Fan2006}
{Fan} X. {et~al.}, 2006, \aj, 132, 117

\bibitem[{{Ferr{\'e}-Mateu} {et~al}\mbox{.}(2017){Ferr{\'e}-Mateu}, {Trujillo}, {Mart{\'\i}n-Navarro}, {Vazdekis}, {Mezcua}, {Balcells}, \& {Dom{\'\i}nguez}}]{Ferre-Mateu2017}
{Ferr{\'e}-Mateu} A., {Trujillo} I., {Mart{\'\i}n-Navarro} I., {Vazdekis} A., {Mezcua} M., {Balcells} M., {Dom{\'\i}nguez} L., 2017, \mnras, 467, 1929

\bibitem[{{Finkelstein} {et~al}\mbox{.}(2025){Finkelstein}, {Bagley}, {Arrabal Haro}, {Dickinson}, {Ferguson}, {Kartaltepe}, {Kocevski}, {Koekemoer}, {Lotz}, {Papovich}, {P{\'e}rez-Gonz{\'a}lez}, {Pirzkal}, {Somerville}, {Trump}, {Yang}, {Yung}, {Fontana}, {Grazian}, {Grogin}, {Kewley}, {Kirkpatrick}, {Larson}, {Pentericci}, {Ravindranath}, {Wilkins}, {Almaini}, {Amor{\'\i}n}, {Barro}, {Bhatawdekar}, {Bisigello}, {Brooks}, {Buat}, {Buitrago}, {Burgarella}, {Calabr{\`o}}, {Castellano}, {Cheng}, {Cleri}, {Cole}, {Cooper}, {Cooper}, {Costantin}, {Cox}, {Croton}, {Daddi}, {Davis}, {Dekel}, {Elbaz}, {Fern{\'a}ndez}, {Fujimoto}, {Gandolfi}, {Gardner}, {Gawiser}, {Giavalisco}, {G{\'o}mez-Guijarro}, {Guo}, {Gupta}, {Hathi}, {Harish}, {Henry}, {Hirschmann}, {Hu}, {Hutchison}, {Iyer}, {Jaskot}, {Jha}, {Jung}, {Kassin}, {Kokorev}, {Kurczynski}, {Leung}, {Llerena}, {Long}, {Lucas}, {Lu}, {McGrath}, {McIntosh}, {Merlin}, {Mobasher}, {Morales}, {Napolitano}, {Pacucci}, {Pandya}, {Rafelski}, {Rodighiero}, {Rose}, {Santini},
  {Seill{\'e}}, {Simons}, {Shen}, {Straughn}, {Tacchella}, {Taylor}, {Vanderhoof}, {Vega-Ferrero}, {Weiner}, {Willmer}, {Zhu}, {Bell}, {Wuyts}, {Holwerda}, {Wang}, {Wang}, {Zavala}, \& {CEERS Collaboration}}]{Finkelstein2025}
{Finkelstein} S.~L. {et~al.}, 2025, \apjl, 983, L4

\bibitem[{{Fontana} {et~al}\mbox{.}(2000){Fontana}, {D'Odorico}, {Poli}, {Giallongo}, {Arnouts}, {Cristiani}, {Moorwood}, \& {Saracco}}]{Fontana2000}
{Fontana} A., {D'Odorico} S., {Poli} F., {Giallongo} E., {Arnouts} S., {Cristiani} S., {Moorwood} A., {Saracco} P., 2000, \aj, 120, 2206

\bibitem[{{Fontana} {et~al}\mbox{.}(2009){Fontana}, {Santini}, {Grazian}, {Pentericci}, {Fiore}, {Castellano}, {Giallongo}, {Menci}, {Salimbeni}, {Cristiani}, {Nonino}, \& {Vanzella}}]{Fontana2009}
{Fontana} A. {et~al.}, 2009, \aap, 501, 15

\bibitem[{{Forrest} {et~al}\mbox{.}(2020{\natexlab{a}}){Forrest}, {Annunziatella}, {Wilson}, {Marchesini}, {Muzzin}, {Cooper}, {Marsan}, {McConachie}, {Chan}, {Gomez}, {Kado-Fong}, {La Barbera}, {Labb{\'e}}, {Lange-Vagle}, {Nantais}, {Nonino}, {Pe{\~n}a}, {Saracco}, {Stefanon}, \& {van der Burg}}]{Forrest2020b}
{Forrest} B. {et~al.}, 2020{\natexlab{a}}, \apjl, 890, L1

\bibitem[{{Forrest} {et~al}\mbox{.}(2024){Forrest}, {Cooper}, {Muzzin}, {Wilson}, {Marchesini}, {McConachie}, {Gomez}, {Annunziatella}, {Marsan}, {Braspenning}, {Chang}, {de Lucia}, {Fontanot}, {Hirschmann}, {Nelson}, {Pillepich}, {Schaye}, {Urbano Stawinski}, {Stefanon}, \& {Xie}}]{Forrest2024}
{Forrest} B. {et~al.}, 2024, \apj, 977, 51

\bibitem[{{Forrest} {et~al}\mbox{.}(2020{\natexlab{b}}){Forrest}, {Marsan}, {Annunziatella}, {Wilson}, {Muzzin}, {Marchesini}, {Cooper}, {Chan}, {McConachie}, {Gomez}, {Kado-Fong}, {La Barbera}, {Lange-Vagle}, {Nantais}, {Nonino}, {Saracco}, {Stefanon}, \& {van der Burg}}]{Forrest2020a}
{Forrest} B. {et~al.}, 2020{\natexlab{b}}, \apj, 903, 47

\bibitem[{{Fortuni} {et~al}\mbox{.}(2023){Fortuni}, {Merlin}, {Fontana}, {Giocoli}, {Romelli}, {Graziani}, {Santini}, {Castellano}, {Charlot}, \& {Chevallard}}]{Fortuni2023}
{Fortuni} F. {et~al.}, 2023, \aap, 677, A102

\bibitem[{{Fu} {et~al}\mbox{.}(2024){Fu}, {Shankar}, {Ayromlou}, {Koutsouridou}, {Cattaneo}, {Bertemes}, {Bellstedt}, {Mart{\'\i}n-Navarro}, {Leja}, {Allevato}, {Bernardi}, {Boco}, {Dimauro}, {Gruppioni}, {Lapi}, {Menci}, {Rodr{\'\i}guez}, {Puglisi}, \& {Alonso-Tetilla}}]{Fu2024}
{Fu} H. {et~al.}, 2024, \mnras, 532, 177

\bibitem[{{Furtak} {et~al}\mbox{.}(2023){Furtak}, {Zitrin}, {Plat}, {Fujimoto}, {Wang}, {Nelson}, {Labb{\'e}}, {Bezanson}, {Brammer}, {van Dokkum}, {Endsley}, {Glazebrook}, {Greene}, {Leja}, {Price}, {Smit}, {Stark}, {Weaver}, {Whitaker}, {Atek}, {Chevallard}, {Curtis-Lake}, {Dayal}, {Feltre}, {Franx}, {Fudamoto}, {Marchesini}, {Mowla}, {Pan}, {Suess}, {Vidal-Garc{\'\i}a}, \& {Williams}}]{Furtak2023}
{Furtak} L.~J. {et~al.}, 2023, \apj, 952, 142

\bibitem[{{Gallazzi} {et~al}\mbox{.}(2005){Gallazzi}, {Charlot}, {Brinchmann}, {White}, \& {Tremonti}}]{Gallazzi2005}
{Gallazzi} A., {Charlot} S., {Brinchmann} J., {White} S.~D.~M., {Tremonti} C.~A., 2005, 362, 41

\bibitem[{{Gandolfi} {et~al}\mbox{.}(2025){Gandolfi}, {Rodighiero}, {Castellano}, {Fontana}, {Santini}, {Dickinson}, {Finkelstein}, {Catone}, {Calabr{\`o}}, {Merlin}, {Pentericci}, {Bisigello}, {Grazian}, {Napolitano}, {Vulcani}, {Taylor}, {Arrabal Haro}, {Kirkpatrick}, {Backhaus}, {Holwerda}, {Giulietti}, {Cleri}, {Daddi}, {Ferguson}, {Hirschmann}, {Koekemoer}, {Lapi}, {Pacucci}, {P{\'e}rez-Gonz{\'a}lez}, {de la Vega}, {Wilkins}, {Yung}, {Bagley}, {Bhatawdekar}, {Kartaltepe}, {Papovich}, \& {Pirzkal}}]{Gandolfi2025}
{Gandolfi} G. {et~al.}, 2025, arXiv e-prints, arXiv:2509.01664

\bibitem[{{Gehrels}(1986)}]{Gehrels1986}
{Gehrels} N., 1986, \apj, 303, 336

\bibitem[{{Glazebrook} {et~al}\mbox{.}(2024){Glazebrook}, {Nanayakkara}, {Schreiber}, {Lagos}, {Kawinwanichakij}, {Jacobs}, {Chittenden}, {Brammer}, {Kacprzak}, {Labbe}, {Marchesini}, {Marsan}, {Oesch}, {Papovich}, {Remus}, {Tran}, {Esdaile}, \& {Chandro-Gomez}}]{Glazebrook2024}
{Glazebrook} K. {et~al.}, 2024, \nat, 628, 277

\bibitem[{{Glazebrook} {et~al}\mbox{.}(2017){Glazebrook}, {Schreiber}, {Labb{\'e}}, {Nanayakkara}, {Kacprzak}, {Oesch}, {Papovich}, {Spitler}, {Straatman}, {Tran}, \& {Yuan}}]{Glazebrook2017}
{Glazebrook} K. {et~al.}, 2017, \nat, 544, 71

\bibitem[{{Gould} {et~al}\mbox{.}(2023){Gould}, {Brammer}, {Valentino}, {Whitaker}, {Weaver}, {Lagos}, {Rizzo}, {Franco}, {Hsieh}, {Ilbert}, {Jin}, {Magdis}, {McCracken}, {Mobasher}, {Shuntov}, {Steinhardt}, {Strait}, \& {Toft}}]{Gould2023}
{Gould} K. M.~L. {et~al.}, 2023, \aj, 165, 248

\bibitem[{{Grogin} {et~al}\mbox{.}(2011){Grogin}, {Kocevski}, {Faber}, {Ferguson}, {Koekemoer}, {Riess}, {Acquaviva}, {Alexander}, {Almaini}, {Ashby}, {Barden}, {Bell}, {Bournaud}, {Brown}, {Caputi}, {Casertano}, {Cassata}, {Castellano}, {Challis}, {Chary}, {Cheung}, {Cirasuolo}, {Conselice}, {Roshan Cooray}, {Croton}, {Daddi}, {Dahlen}, {Dav{\'e}}, {de Mello}, {Dekel}, {Dickinson}, {Dolch}, {Donley}, {Dunlop}, {Dutton}, {Elbaz}, {Fazio}, {Filippenko}, {Finkelstein}, {Fontana}, {Gardner}, {Garnavich}, {Gawiser}, {Giavalisco}, {Grazian}, {Guo}, {Hathi}, {H{\"a}ussler}, {Hopkins}, {Huang}, {Huang}, {Jha}, {Kartaltepe}, {Kirshner}, {Koo}, {Lai}, {Lee}, {Li}, {Lotz}, {Lucas}, {Madau}, {McCarthy}, {McGrath}, {McIntosh}, {McLure}, {Mobasher}, {Moustakas}, {Mozena}, {Nandra}, {Newman}, {Niemi}, {Noeske}, {Papovich}, {Pentericci}, {Pope}, {Primack}, {Rajan}, {Ravindranath}, {Reddy}, {Renzini}, {Rix}, {Robaina}, {Rodney}, {Rosario}, {Rosati}, {Salimbeni}, {Scarlata}, {Siana}, {Simard}, {Smidt}, {Somerville},
  {Spinrad}, {Straughn}, {Strolger}, {Telford}, {Teplitz}, {Trump}, {van der Wel}, {Villforth}, {Wechsler}, {Weiner}, {Wiklind}, {Wild}, {Wilson}, {Wuyts}, {Yan}, \& {Yun}}]{Grogin2011}
{Grogin} N.~A. {et~al.}, 2011, \apjs, 197, 35

\bibitem[{{Guia}, {Pacucci} \& {Kocevski}(2024){Guia}, {Pacucci}, \& {Kocevski}}]{Guia2024}
{Guia} C.~A., {Pacucci} F., {Kocevski} D.~D., 2024, Research Notes of the American Astronomical Society, 8, 207

\bibitem[{{Guo} {et~al}\mbox{.}(2024){Guo}, {Zhang}, {Yan}, {Gjergo}, {Man}, {Ivison}, {Fu}, \& {Shi}}]{Guo2024}
{Guo} Z., {Zhang} Z.-Y., {Yan} Z., {Gjergo} E., {Man} A. W.~S., {Ivison} R.~J., {Fu} X., {Shi} Y., 2024, \apj, 970, 136

\bibitem[{{Gutkin}, {Charlot} \& {Bruzual}(2016){Gutkin}, {Charlot}, \& {Bruzual}}]{Gutkin2016}
{Gutkin} J., {Charlot} S., {Bruzual} G., 2016, \mnras, 462, 1757

\bibitem[{{Haines} {et~al}\mbox{.}(2017){Haines}, {Iovino}, {Krywult}, {Guzzo}, {Davidzon}, {Bolzonella}, {Garilli}, {Scodeggio}, {Granett}, {de la Torre}, {De Lucia}, {Abbas}, {Adami}, {Arnouts}, {Bottini}, {Cappi}, {Cucciati}, {Franzetti}, {Fritz}, {Gargiulo}, {Le Brun}, {Le F{\`e}vre}, {Maccagni}, {Ma{\l}ek}, {Marulli}, {Moutard}, {Polletta}, {Pollo}, {Tasca}, {Tojeiro}, {Vergani}, {Zanichelli}, {Zamorani}, {Bel}, {Branchini}, {Coupon}, {Ilbert}, {Moscardini}, {Peacock}, \& {Siudek}}]{Haines2017}
{Haines} C.~P. {et~al.}, 2017, \aap, 605, A4

\bibitem[{{Hainline} {et~al}\mbox{.}(2024){Hainline}, {Helton}, {Johnson}, {Sun}, {Topping}, {Leisenring}, {Baker}, {Eisenstein}, {Hausen}, {Hviding}, {Lyu}, {Robertson}, {Tacchella}, {Williams}, {Willmer}, \& {Roellig}}]{Hainline2024}
{Hainline} K.~N. {et~al.}, 2024, \apj, 964, 66

\bibitem[{{Hartmann} {et~al}\mbox{.}(2025){Hartmann}, {Mart{\'\i}n-Navarro}, {Huertas-Company}, {Benedetti}, {Iglesias-Navarro}, {Vazdekis}, \& {Montes}}]{Hartmann2025}
{Hartmann} E.~A., {Mart{\'\i}n-Navarro} I., {Huertas-Company} M., {Benedetti} J. P.~V., {Iglesias-Navarro} P., {Vazdekis} A., {Montes} M., 2025, \aap, 694, L7

\bibitem[{{Ho}(2008)}]{Ho2008}
{Ho} L.~C., 2008, \araa, 46, 475

\bibitem[{{Holwerda} {et~al}\mbox{.}(2024){Holwerda}, {Hsu}, {Hathi}, {Bisigello}, {de la Vega}, {Haro}, {Bagley}, {Dickinson}, {Finkelstein}, {Kartaltepe}, {Koekemoer}, {Papovich}, {Pirzkal}, {Cook}, {Robertson}, {Casey}, {Aganze}, {P{\'e}rez-Gonz{\'a}lez}, {Lucas}, {Jogee}, {Wilkins}, {Burgarella}, \& {Kirkpatrick}}]{Holwerda2024}
{Holwerda} B.~W. {et~al.}, 2024, \mnras, 529, 1067

\bibitem[{{Hopkins} {et~al}\mbox{.}(2009){Hopkins}, {Somerville}, {Cox}, {Hernquist}, {Jogee}, {Kere{\v{s}}}, {Ma}, {Robertson}, \& {Stewart}}]{Hopkins2009}
{Hopkins} P.~F. {et~al.}, 2009, \mnras, 397, 802

\bibitem[{{Hutter} {et~al}\mbox{.}(2025){Hutter}, {Cueto}, {Dayal}, {Gottl{\"o}ber}, {Trebitsch}, \& {Yepes}}]{Hutter2025}
{Hutter} A., {Cueto} E.~R., {Dayal} P., {Gottl{\"o}ber} S., {Trebitsch} M., {Yepes} G., 2025, \aap, 694, A254

\bibitem[{{Inayoshi} \& {Maiolino}(2025)}]{Inayoshi2025}
{Inayoshi} K., {Maiolino} R., 2025, \apjl, 980, L27

\bibitem[{{Jin} {et~al}\mbox{.}(2024){Jin}, {Sillassen}, {Magdis}, {Brinch}, {Shuntov}, {Brammer}, {Gobat}, {Valentino}, {Carnall}, {Lee}, {Vijayan}, {Gillman}, {Kokorev}, {Le Bail}, {Greve}, {Gullberg}, {Gould}, \& {Toft}}]{Jin2024}
{Jin} S. {et~al.}, 2024, \aap, 683, L4

\bibitem[{{Kauffmann} {et~al}\mbox{.}(2003){Kauffmann}, {Heckman}, {White}, {Charlot}, {Tremonti}, {Peng}, {Seibert}, {Brinkmann}, {Nichol}, {SubbaRao}, \& {York}}]{Kauffmann2003}
{Kauffmann} G. {et~al.}, 2003, 341, 54

\bibitem[{{Kauffmann}, {White} \& {Guiderdoni}(1993){Kauffmann}, {White}, \& {Guiderdoni}}]{Kauffmann1993}
{Kauffmann} G., {White} S.~D.~M., {Guiderdoni} B., 1993, \mnras, 264, 201

\bibitem[{{Kennicutt}(1998)}]{Kennicutt1998}
{Kennicutt}, Jr. R.~C., 1998, \araa, 36, 189

\bibitem[{{Kimmig} {et~al}\mbox{.}(2025){Kimmig}, {Remus}, {Seidel}, {Valenzuela}, {Dolag}, \& {Burkert}}]{Kimmig2025}
{Kimmig} L.~C., {Remus} R.-S., {Seidel} B., {Valenzuela} L.~M., {Dolag} K., {Burkert} A., 2025, \apj, 979, 15

\bibitem[{{Kocevski} {et~al}\mbox{.}(2025){Kocevski}, {Finkelstein}, {Barro}, {Taylor}, {Calabr{\`o}}, {Laloux}, {Buchner}, {Trump}, {Leung}, {Yang}, {Dickinson}, {P{\'e}rez-Gonz{\'a}lez}, {Pacucci}, {Inayoshi}, {Somerville}, {McGrath}, {Akins}, {Bagley}, {Bowler}, {Bisigello}, {Carnall}, {Casey}, {Cheng}, {Cleri}, {Costantin}, {Cullen}, {Davis}, {Donnan}, {Dunlop}, {Ellis}, {Ferguson}, {Fujimoto}, {Fontana}, {Giavalisco}, {Grazian}, {Grogin}, {Hathi}, {Hirschmann}, {Huertas-Company}, {Holwerda}, {Illingworth}, {Juneau}, {Kartaltepe}, {Koekemoer}, {Li}, {Lucas}, {Magee}, {Mason}, {McLeod}, {McLure}, {Napolitano}, {Papovich}, {Pirzkal}, {Rodighiero}, {Santini}, {Wilkins}, \& {Yung}}]{Kocevski2025}
{Kocevski} D.~D. {et~al.}, 2025, \apj, 986, 126

\bibitem[{{Kocevski} {et~al}\mbox{.}(2018){Kocevski}, {Hasinger}, {Brightman}, {Nandra}, {Georgakakis}, {Cappelluti}, {Civano}, {Li}, {Li}, {Aird}, {Alexander}, {Almaini}, {Brusa}, {Buchner}, {Comastri}, {Conselice}, {Dickinson}, {Finoguenov}, {Gilli}, {Koekemoer}, {Miyaji}, {Mullaney}, {Papovich}, {Rosario}, {Salvato}, {Silverman}, {Somerville}, \& {Ueda}}]{Kocevski2018}
{Kocevski} D.~D. {et~al.}, 2018, \apjs, 236, 48

\bibitem[{{Kocevski} {et~al}\mbox{.}(2023){Kocevski}, {Onoue}, {Inayoshi}, {Trump}, {Arrabal Haro}, {Grazian}, {Dickinson}, {Finkelstein}, {Kartaltepe}, {Hirschmann}, {Aird}, {Holwerda}, {Fujimoto}, {Juneau}, {Amor{\'\i}n}, {Backhaus}, {Bagley}, {Barro}, {Bell}, {Bisigello}, {Calabr{\`o}}, {Cleri}, {Cooper}, {Ding}, {Grogin}, {Ho}, {Hutchison}, {Inoue}, {Jiang}, {Jones}, {Koekemoer}, {Li}, {Li}, {McGrath}, {Molina}, {Papovich}, {P{\'e}rez-Gonz{\'a}lez}, {Pirzkal}, {Wilkins}, {Yang}, \& {Yung}}]{Kocevski2023}
{Kocevski} D.~D. {et~al.}, 2023, \apjl, 954, L4

\bibitem[{{Koekemoer} {et~al}\mbox{.}(2011){Koekemoer}, {Faber}, {Ferguson}, {Grogin}, {Kocevski}, {Koo}, {Lai}, {Lotz}, {Lucas}, {McGrath}, {Ogaz}, {Rajan}, {Riess}, {Rodney}, {Strolger}, {Casertano}, {Castellano}, {Dahlen}, {Dickinson}, {Dolch}, {Fontana}, {Giavalisco}, {Grazian}, {Guo}, {Hathi}, {Huang}, {van der Wel}, {Yan}, {Acquaviva}, {Alexander}, {Almaini}, {Ashby}, {Barden}, {Bell}, {Bournaud}, {Brown}, {Caputi}, {Cassata}, {Challis}, {Chary}, {Cheung}, {Cirasuolo}, {Conselice}, {Roshan Cooray}, {Croton}, {Daddi}, {Dav{\'e}}, {de Mello}, {de Ravel}, {Dekel}, {Donley}, {Dunlop}, {Dutton}, {Elbaz}, {Fazio}, {Filippenko}, {Finkelstein}, {Frazer}, {Gardner}, {Garnavich}, {Gawiser}, {Gruetzbauch}, {Hartley}, {H{\"a}ussler}, {Herrington}, {Hopkins}, {Huang}, {Jha}, {Johnson}, {Kartaltepe}, {Khostovan}, {Kirshner}, {Lani}, {Lee}, {Li}, {Madau}, {McCarthy}, {McIntosh}, {McLure}, {McPartland}, {Mobasher}, {Moreira}, {Mortlock}, {Moustakas}, {Mozena}, {Nandra}, {Newman}, {Nielsen}, {Niemi}, {Noeske},
  {Papovich}, {Pentericci}, {Pope}, {Primack}, {Ravindranath}, {Reddy}, {Renzini}, {Rix}, {Robaina}, {Rosario}, {Rosati}, {Salimbeni}, {Scarlata}, {Siana}, {Simard}, {Smidt}, {Snyder}, {Somerville}, {Spinrad}, {Straughn}, {Telford}, {Teplitz}, {Trump}, {Vargas}, {Villforth}, {Wagner}, {Wandro}, {Wechsler}, {Weiner}, {Wiklind}, {Wild}, {Wilson}, {Wuyts}, \& {Yun}}]{Koekemoer2011}
{Koekemoer} A.~M. {et~al.}, 2011, \apjs, 197, 36

\bibitem[{{Kokorev} {et~al}\mbox{.}(2024){Kokorev}, {Caputi}, {Greene}, {Dayal}, {Trebitsch}, {Cutler}, {Fujimoto}, {Labb{\'e}}, {Miller}, {Iani}, {Navarro-Carrera}, \& {Rinaldi}}]{Kokorev2024}
{Kokorev} V. {et~al.}, 2024, \apj, 968, 38

\bibitem[{{Kurinchi-Vendhan} {et~al}\mbox{.}(2024){Kurinchi-Vendhan}, {Farcy}, {Hirschmann}, \& {Valentino}}]{Kurinchi-Vendhan2024}
{Kurinchi-Vendhan} S., {Farcy} M., {Hirschmann} M., {Valentino} F., 2024, \mnras, 534, 3974

\bibitem[{{Labb{\'e}} {et~al}\mbox{.}(2005){Labb{\'e}}, {Huang}, {Franx}, {Rudnick}, {Barmby}, {Daddi}, {van Dokkum}, {Fazio}, {Schreiber}, {Moorwood}, {Rix}, {R{\"o}ttgering}, {Trujillo}, \& {van der Werf}}]{Labbe2005}
{Labb{\'e}} I. {et~al.}, 2005, \apjl, 624, L81

\bibitem[{{Labb{\'e}} {et~al}\mbox{.}(2023){Labb{\'e}}, {van Dokkum}, {Nelson}, {Bezanson}, {Suess}, {Leja}, {Brammer}, {Whitaker}, {Mathews}, {Stefanon}, \& {Wang}}]{Labbe2023}
{Labb{\'e}} I. {et~al.}, 2023, \nat, 616, 266

\bibitem[{{Langeroodi} \& {Hjorth}(2023)}]{Langeroodi2023}
{Langeroodi} D., {Hjorth} J., 2023, \apjl, 957, L27

\bibitem[{{Lapi} {et~al}\mbox{.}(2018){Lapi}, {Pantoni}, {Zanisi}, {Shi}, {Mancuso}, {Massardi}, {Shankar}, {Bressan}, \& {Danese}}]{Lapi2018}
{Lapi} A. {et~al.}, 2018, \apj, 857, 22

\bibitem[{{Larson} {et~al}\mbox{.}(2023){Larson}, {Hutchison}, {Bagley}, {Finkelstein}, {Yung}, {Somerville}, {Hirschmann}, {Brammer}, {Holwerda}, {Papovich}, {Morales}, \& {Wilkins}}]{Larson2023}
{Larson} R.~L. {et~al.}, 2023, \apj, 958, 141

\bibitem[{{Lemaux} {et~al}\mbox{.}(2010){Lemaux}, {Lubin}, {Shapley}, {Kocevski}, {Gal}, \& {Squires}}]{Lemaux2010}
{Lemaux} B.~C., {Lubin} L.~M., {Shapley} A., {Kocevski} D., {Gal} R.~R., {Squires} G.~K., 2010, \apj, 716, 970

\bibitem[{{Long} {et~al}\mbox{.}(2024){Long}, {Antwi-Danso}, {Lambrides}, {Lovell}, {de la Vega}, {Valentino}, {Zavala}, {Casey}, {Wilkins}, {Yung}, {Arrabal Haro}, {Bagley}, {Bisigello}, {Chworowsky}, {Cooper}, {Cooper}, {Cooray}, {Croton}, {Dickinson}, {Finkelstein}, {Franco}, {Gould}, {Hirschmann}, {Hutchison}, {Kartaltepe}, {Kocevski}, {Koekemoer}, {Lucas}, {McKinney}, {Nere}, {Papovich}, {P{\'e}rez-Gonz{\'a}lez}, {Pirzkal}, \& {Santini}}]{Long2024}
{Long} A.~S. {et~al.}, 2024, \apj, 970, 68

\bibitem[{{Looser} {et~al}\mbox{.}(2024){Looser}, {D'Eugenio}, {Maiolino}, {Witstok}, {Sandles}, {Curtis-Lake}, {Chevallard}, {Tacchella}, {Johnson}, {Baker}, {Suess}, {Carniani}, {Ferruit}, {Arribas}, {Bonaventura}, {Bunker}, {Cameron}, {Charlot}, {Curti}, {de Graaff}, {Maseda}, {Rawle}, {Rix}, {Del Pino}, {Smit}, {{\"U}bler}, {Willott}, {Alberts}, {Egami}, {Eisenstein}, {Endsley}, {Hausen}, {Rieke}, {Robertson}, {Shivaei}, {Williams}, {Boyett}, {Chen}, {Ji}, {Jones}, {Kumari}, {Nelson}, {Perna}, {Saxena}, \& {Scholtz}}]{Looser2024}
{Looser} T.~J. {et~al.}, 2024, \nat, 629, 53

\bibitem[{{Luo} {et~al}\mbox{.}(2017){Luo}, {Brandt}, {Xue}, {Lehmer}, {Alexander}, {Bauer}, {Vito}, {Yang}, {Basu-Zych}, {Comastri}, {Gilli}, {Gu}, {Hornschemeier}, {Koekemoer}, {Liu}, {Mainieri}, {Paolillo}, {Ranalli}, {Rosati}, {Schneider}, {Shemmer}, {Smail}, {Sun}, {Tozzi}, {Vignali}, \& {Wang}}]{Luo2017}
{Luo} B. {et~al.}, 2017, \apjs, 228, 2

\bibitem[{{Ma} {et~al}\mbox{.}(2016){Ma}, {Gonzalez}, {Vieira}, {Aravena}, {Ashby}, {B{\'e}thermin}, {Bothwell}, {Brandt}, {de Breuck}, {Carlstrom}, {Chapman}, {Gullberg}, {Hezaveh}, {Litke}, {Malkan}, {Marrone}, {McDonald}, {Murphy}, {Spilker}, {Sreevani}, {Stark}, {Strandet}, \& {Wang}}]{Ma2016}
{Ma} J. {et~al.}, 2016, \apj, 832, 114

\bibitem[{{Maksymowicz-Maciata} {et~al}\mbox{.}(2024){Maksymowicz-Maciata}, {Spiniello}, {Mart{\'\i}n-Navarro}, {Ferr{\'e}-Mateu}, {Bevacqua}, {Cappellari}, {D'Ago}, {Tortora}, {Arnaboldi}, {Hartke}, {Napolitano}, {Saracco}, \& {Scognamiglio}}]{Maksymowicz-Maciata2024}
{Maksymowicz-Maciata} M. {et~al.}, 2024, \mnras, 531, 2864

\bibitem[{{Maseda} {et~al}\mbox{.}(2024){Maseda}, {de Graaff}, {Franx}, {Rix}, {Carniani}, {Laseter}, {Dudzevi{\v{c}}i{\={u}}t{\.{e}}}, {Rawle}, {Parlanti}, {Arribas}, {Bunker}, {Cameron}, {Charlot}, {Curti}, {D'Eugenio}, {Jones}, {Kumari}, {Maiolino}, {{\"U}bler}, {Saxena}, {Smit}, {Willott}, \& {Witstok}}]{Maseda2024}
{Maseda} M.~V. {et~al.}, 2024, \aap, 689, A73

\bibitem[{{Matteucci}(1994)}]{Matteucci1994}
{Matteucci} F., 1994, 288, 57

\bibitem[{{McGaugh} {et~al}\mbox{.}(2024){McGaugh}, {Schombert}, {Lelli}, \& {Franck}}]{McGaugh2024}
{McGaugh} S.~S., {Schombert} J.~M., {Lelli} F., {Franck} J., 2024, \apj, 976, 13

\bibitem[{{Merlin} {et~al}\mbox{.}(2012){Merlin}, {Chiosi}, {Piovan}, {Grassi}, {Buonomo}, \& {Barbera}}]{Merlin2012}
{Merlin} E., {Chiosi} C., {Piovan} L., {Grassi} T., {Buonomo} U., {Barbera} F.~L., 2012, \mnras, 427, 1530

\bibitem[{{Merlin} {et~al}\mbox{.}(2018){Merlin}, {Fontana}, {Castellano}, {Santini}, {Torelli}, {Boutsia}, {Wang}, {Grazian}, {Pentericci}, {Schreiber}, {Ciesla}, {McLure}, {Derriere}, {Dunlop}, \& {Elbaz}}]{Merlin2018}
{Merlin} E. {et~al.}, 2018, \mnras, 473, 2098

\bibitem[{{Merlin} {et~al}\mbox{.}(2019{\natexlab{a}}){Merlin}, {Fortuni}, {Torelli}, {Santini}, {Castellano}, {Fontana}, {Grazian}, {Pentericci}, {Pilo}, \& {Schmidt}}]{Merlin2019}
{Merlin} E. {et~al.}, 2019{\natexlab{a}}, \mnras, 490, 3309

\bibitem[{{Merlin} {et~al}\mbox{.}(2019{\natexlab{b}}){Merlin}, {Pilo}, {Fontana}, {Castellano}, {Paris}, {Roscani}, {Santini}, \& {Torelli}}]{Merlin2019b}
{Merlin} E., {Pilo} S., {Fontana} A., {Castellano} M., {Paris} D., {Roscani} V., {Santini} P., {Torelli} M., 2019{\natexlab{b}}, \aap, 622, A169

\bibitem[{{Merlin} {et~al}\mbox{.}(2024){Merlin}, {Santini}, {Paris}, {Castellano}, {Fontana}, {Treu}, {Finkelstein}, {Dunlop}, {Arrabal Haro}, {Bagley}, {Boyett}, {Calabr{\`o}}, {Correnti}, {Davis}, {Dickinson}, {Donnan}, {Ferguson}, {Fortuni}, {Giavalisco}, {Glazebrook}, {Grazian}, {Grogin}, {Hathi}, {Hirschmann}, {Kartaltepe}, {Kewley}, {Kirkpatrick}, {Kocevski}, {Koekemoer}, {Leung}, {Lotz}, {Lucas}, {Magee}, {Marchesini}, {Mascia}, {McLeod}, {McLure}, {Nanayakkara}, {Napolitano}, {Nonino}, {Papovich}, {Pentericci}, {P{\'e}rez-Gonz{\'a}lez}, {Pirzkal}, {Ravindranath}, {Roberts-Borsani}, {Somerville}, {Trenti}, {Trump}, {Vulcani}, {Wang}, {Watson}, {Wilkins}, {Yang}, \& {Yung}}]{Merlin2024}
{Merlin} E. {et~al.}, 2024, \aap, 691, A240

\bibitem[{{Mobasher} {et~al}\mbox{.}(2005){Mobasher}, {Dickinson}, {Ferguson}, {Giavalisco}, {Wiklind}, {Stark}, {Ellis}, {Fall}, {Grogin}, {Moustakas}, {Panagia}, {Sosey}, {Stiavelli}, {Bergeron}, {Casertano}, {Ingraham}, {Koekemoer}, {Labb{\'e}}, {Livio}, {Rodgers}, {Scarlata}, {Vernet}, {Renzini}, {Rosati}, {Kuntschner}, {K{\"u}mmel}, {Walsh}, {Chary}, {Eisenhardt}, {Pirzkal}, \& {Stern}}]{Mobasher2005}
{Mobasher} B. {et~al.}, 2005, 635, 832

\bibitem[{{Morishita} {et~al}\mbox{.}(2025){Morishita}, {Mason}, {Kreilgaard}, {Trenti}, {Treu}, {Vulcani}, {Zhang}, {Abdurro'uf}, {Alavi}, {Atek}, {Bah{\'e}}, {Brada{\v{c}}}, {Bradley}, {Bunker}, {Coe}, {Colbert}, {Gelli}, {Hayes}, {Jones}, {Kodama}, {Leethochawalit}, {Liu}, {Malkan}, {Mehta}, {Metha}, {Newman}, {Rafelski}, {Roberts-Borsani}, {Rutkowski}, {Scarlata}, {Stiavelli}, {Sutanto}, {Takahashi}, {Teplitz}, \& {Wang}}]{Morishita2025}
{Morishita} T. {et~al.}, 2025, \apj, 983, 152

\bibitem[{{Muzzin} {et~al}\mbox{.}(2013){Muzzin}, {Marchesini}, {Stefanon}, {Franx}, {McCracken}, {Milvang-Jensen}, {Dunlop}, {Fynbo}, {Brammer}, {Labb{\'e}}, \& {van Dokkum}}]{Muzzin2013}
{Muzzin} A. {et~al.}, 2013, \apj, 777, 18

\bibitem[{{Naidu} {et~al}\mbox{.}(2024){Naidu}, {Matthee}, {Kramarenko}, {Weibel}, {Brammer}, {Oesch}, {Lechner}, {Furtak}, {Di Cesare}, {Torralba}, {Kotiwale}, {Bezanson}, {Bouwens}, {Chandra}, {Claeyssens}, {Danhaive}, {Frebel}, {de Graaff}, {Greene}, {Heintz}, {Ji}, {Kashino}, {Katz}, {Labbe}, {Leja}, {Li}, {Maseda}, {Richard}, {Shivaei}, {Simcoe}, {Sobral}, {Suess}, {Tacchella}, \& {Williams}}]{Naidu2024}
{Naidu} R.~P. {et~al.}, 2024, arXiv e-prints, arXiv:2410.01874

\bibitem[{{Naidu} {et~al}\mbox{.}(2022){Naidu}, {Oesch}, {Setton}, {Matthee}, {Conroy}, {Johnson}, {Weaver}, {Bouwens}, {Brammer}, {Dayal}, {Illingworth}, {Barrufet}, {Belli}, {Bezanson}, {Bose}, {Heintz}, {Leja}, {Leonova}, {Marques-Chaves}, {Stefanon}, {Toft}, {van der Wel}, {van Dokkum}, {Weibel}, \& {Whitaker}}]{Naidu2022b}
{Naidu} R.~P. {et~al.}, 2022, arXiv e-prints, arXiv:2208.02794

\bibitem[{{Nanayakkara} {et~al}\mbox{.}(2024){Nanayakkara}, {Glazebrook}, {Jacobs}, {Kawinwanichakij}, {Schreiber}, {Brammer}, {Esdaile}, {Kacprzak}, {Labbe}, {Lagos}, {Marchesini}, {Marsan}, {Oesch}, {Papovich}, {Remus}, \& {Tran}}]{Nanayakkara2024}
{Nanayakkara} T. {et~al.}, 2024, Scientific Reports, 14, 3724

\bibitem[{{Nandra} {et~al}\mbox{.}(2015){Nandra}, {Laird}, {Aird}, {Salvato}, {Georgakakis}, {Barro}, {Perez-Gonzalez}, {Barmby}, {Chary}, {Coil}, {Cooper}, {Davis}, {Dickinson}, {Faber}, {Fazio}, {Guhathakurta}, {Gwyn}, {Hsu}, {Huang}, {Ivison}, {Koo}, {Newman}, {Rangel}, {Yamada}, \& {Willmer}}]{Nandra2015}
{Nandra} K. {et~al.}, 2015, \apjs, 220, 10

\bibitem[{{Napolitano} {et~al}\mbox{.}(2025){Napolitano}, {Pentericci}, {Dickinson}, {Arrabal Haro}, {Taylor}, {Calabr{\`o}}, {Bhagwat}, {Santini}, {Arevalo-Gonzalez}, {Begley}, {Castellano}, {Ciardi}, {Donnan}, {Dottorini}, {Dunlop}, {Finkelstein}, {Fontana}, {Giavalisco}, {Hirschmann}, {Jung}, {Koekemoer}, {Kokorev}, {Llerena}, {Lucas}, {Mascia}, {Merlin}, {P{\'e}rez-Gonz{\'a}lez}, {Stanton}, {Tripodi}, {Wang}, \& {Weiner}}]{Napolitano2025}
{Napolitano} L. {et~al.}, 2025, arXiv e-prints, arXiv:2508.14171

\bibitem[{{Nipoti}(2025)}]{Nipoti2025}
{Nipoti} C., 2025, \aap, 697, A74

\bibitem[{{Nipoti}, {Treu} \& {Bolton}(2009){Nipoti}, {Treu}, \& {Bolton}}]{Nipoti2009}
{Nipoti} C., {Treu} T., {Bolton} A.~S., 2009, \apj, 703, 1531

\bibitem[{{Oesch} {et~al}\mbox{.}(2023){Oesch}, {Brammer}, {Naidu}, {Bouwens}, {Chisholm}, {Illingworth}, {Matthee}, {Nelson}, {Qin}, {Reddy}, {Shapley}, {Shivaei}, {van Dokkum}, {Weibel}, {Whitaker}, {Wuyts}, {Covelo-Paz}, {Endsley}, {Fudamoto}, {Giovinazzo}, {Herard-Demanche}, {Kerutt}, {Kramarenko}, {Labbe}, {Leonova}, {Lin}, {Magee}, {Marchesini}, {Maseda}, {Mason}, {Matharu}, {Meyer}, {Neufeld}, {Prieto Lyon}, {Schaerer}, {Sharma}, {Shuntov}, {Smit}, {Stefanon}, {Wyithe}, \& {Xiao}}]{Oesch2023}
{Oesch} P.~A. {et~al.}, 2023, \mnras, 525, 2864

\bibitem[{{Oke} \& {Gunn}(1983)}]{Oke83}
{Oke} J.~B., {Gunn} J.~E., 1983, \apj, 266, 713

\bibitem[{{Pacucci} \& {Loeb}(2024)}]{Pacucci2024}
{Pacucci} F., {Loeb} A., 2024, \apj, 964, 154

\bibitem[{{Papovich} {et~al}\mbox{.}(2023){Papovich}, {Cole}, {Yang}, {Finkelstein}, {Barro}, {Buat}, {Burgarella}, {P{\'e}rez-Gonz{\'a}lez}, {Santini}, {Seill{\'e}}, {Shen}, {Arrabal Haro}, {Bagley}, {Bell}, {Bisigello}, {Calabr{\`o}}, {Casey}, {Castellano}, {Chworowsky}, {Cleri}, {Costantin}, {Cooper}, {Dickinson}, {Ferguson}, {Fontana}, {Giavalisco}, {Grazian}, {Grogin}, {Hathi}, {Holwerda}, {Hutchison}, {Kartaltepe}, {Kewley}, {Kirkpatrick}, {Kocevski}, {Koekemoer}, {Larson}, {Long}, {Lucas}, {Pentericci}, {Pirzkal}, {Ravindranath}, {Somerville}, {Trump}, {Urbano Stawinski}, {Weiner}, {Wilkins}, {Yung}, \& {Zavala}}]{Papovich2023}
{Papovich} C. {et~al.}, 2023, \apjl, 949, L18

\bibitem[{{Park} {et~al}\mbox{.}(2024){Park}, {Belli}, {Conroy}, {Johnson}, {Davies}, {Leja}, {Tacchella}, {Mendel}, {Benton}, {Bugiani}, {Emami}, {Khoram}, {Li}, {Maheson}, {Mathews}, {Naidu}, {Nelson}, {Terrazas}, \& {Weinberger}}]{Park2024}
{Park} M. {et~al.}, 2024, \apj, 976, 72

\bibitem[{{Peng}, {Maiolino} \& {Cochrane}(2015){Peng}, {Maiolino}, \& {Cochrane}}]{Peng2015}
{Peng} Y., {Maiolino} R., {Cochrane} R., 2015, \nat, 521, 192

\bibitem[{{P{\'e}rez-Gonz{\'a}lez} {et~al}\mbox{.}(2023){P{\'e}rez-Gonz{\'a}lez}, {Barro}, {Annunziatella}, {Costantin}, {Garc{\'\i}a-Argum{\'a}nez}, {McGrath}, {M{\'e}rida}, {Zavala}, {Arrabal Haro}, {Bagley}, {Backhaus}, {Behroozi}, {Bell}, {Bisigello}, {Buat}, {Calabr{\`o}}, {Casey}, {Cleri}, {Coogan}, {Cooper}, {Cooray}, {Dekel}, {Dickinson}, {Elbaz}, {Ferguson}, {Finkelstein}, {Fontana}, {Franco}, {Gardner}, {Giavalisco}, {G{\'o}mez-Guijarro}, {Grazian}, {Grogin}, {Guo}, {Huertas-Company}, {Jogee}, {Kartaltepe}, {Kewley}, {Kirkpatrick}, {Kocevski}, {Koekemoer}, {Long}, {Lotz}, {Lucas}, {Papovich}, {Pirzkal}, {Ravindranath}, {Somerville}, {Tacchella}, {Trump}, {Wang}, {Wilkins}, {Wuyts}, {Yang}, \& {Yung}}]{PerezGonzalez2023}
{P{\'e}rez-Gonz{\'a}lez} P.~G. {et~al.}, 2023, \apjl, 946, L16

\bibitem[{{P{\'e}rez-Gonz{\'a}lez} {et~al}\mbox{.}(2024){P{\'e}rez-Gonz{\'a}lez}, {Barro}, {Rieke}, {Lyu}, {Rieke}, {Alberts}, {Williams}, {Hainline}, {Sun}, {Pusk{\'a}s}, {Annunziatella}, {Baker}, {Bunker}, {Egami}, {Ji}, {Johnson}, {Robertson}, {Rodr{\'\i}guez Del Pino}, {Rujopakarn}, {Shivaei}, {Tacchella}, {Willmer}, \& {Willott}}]{PerezGonzalez2024}
{P{\'e}rez-Gonz{\'a}lez} P.~G. {et~al.}, 2024, \apj, 968, 4

\bibitem[{{P{\'e}rez-Gonz{\'a}lez} {et~al}\mbox{.}(2025){P{\'e}rez-Gonz{\'a}lez}, {D'Eugenio}, {Rodr{\'\i}guez del Pino}, {Perna}, {{\"U}bler}, {Maiolino}, {Arribas}, {Cresci}, {Lamperti}, {Bunker}, {Carniani}, {Charlot}, {Willott}, {B{\"o}ker}, {Parlanti}, {Scholtz}, {Venturi}, {Barro}, {Costantin}, {Mart{\'\i}n-Navarro}, {Dunlop}, \& {Magee}}]{PerezGonzalez2025}
{P{\'e}rez-Gonz{\'a}lez} P.~G. {et~al.}, 2025, Nature Astronomy

\bibitem[{{Pillepich} {et~al}\mbox{.}(2018){Pillepich}, {Springel}, {Nelson}, {Genel}, {Naiman}, {Pakmor}, {Hernquist}, {Torrey}, {Vogelsberger}, {Weinberger}, \& {Marinacci}}]{Pillepich2018}
{Pillepich} A. {et~al.}, 2018, \mnras, 473, 4077

\bibitem[{{Rowan-Robinson} {et~al}\mbox{.}(2018){Rowan-Robinson}, {Wang}, {Farrah}, {Rigopoulou}, {Gruppioni}, {Vaccari}, {Marchetti}, {Clements}, \& {Pearson}}]{Rowan-Robinson2018}
{Rowan-Robinson} M. {et~al.}, 2018, \aap, 619, A169

\bibitem[{{Russell} {et~al}\mbox{.}(2024){Russell}, {Dobric}, {Adams}, {Conselice}, {Austin}, {Harvey}, {Trussler}, {Ferreira}, {Westcott}, {Harris}, {Windhorst}, {Coe}, {Cohen}, {Driver}, {Frye}, {Grogin}, {Hathi}, {Jansen}, {Koekemoer}, {Marshall}, {Ortiz}, {Pirzkal}, {Robotham}, {Ryan}, {Summers}, {D'Silva}, {Willmer}, \& {Yan}}]{Russell2024}
{Russell} T.~A. {et~al.}, 2024, arXiv e-prints, arXiv:2412.11861

\bibitem[{{Salpeter}(1955)}]{Salpeter1955}
{Salpeter} E.~E., 1955, 121, 161

\bibitem[{{Salpeter}(1959)}]{Salpeter1959}
{Salpeter} E.~E., 1959, 129, 608

\bibitem[{{Santini} {et~al}\mbox{.}(2022){Santini}, {Castellano}, {Fontana}, {Fortuni}, {Menci}, {Merlin}, {Pagul}, {Testa}, {Calabr{\`o}}, {Paris}, \& {Pentericci}}]{Santini2022}
{Santini} P. {et~al.}, 2022, \apj, 940, 135

\bibitem[{{Santini} {et~al}\mbox{.}(2021){Santini}, {Castellano}, {Merlin}, {Fontana}, {Fortuni}, {Kodra}, {Magnelli}, {Menci}, {Calabr{\`o}}, {Lovell}, {Pentericci}, {Testa}, \& {Wilkins}}]{Santini2021}
{Santini} P. {et~al.}, 2021, \aap, 652, A30

\bibitem[{{Santini} {et~al}\mbox{.}(2019){Santini}, {Merlin}, {Fontana}, {Magnelli}, {Paris}, {Castellano}, {Grazian}, {Pentericci}, {Pilo}, \& {Torelli}}]{Santini2019}
{Santini} P. {et~al.}, 2019, arXiv e-prints

\bibitem[{{Schaerer} \& {de Barros}(2009)}]{Schaerer2009}
{Schaerer} D., {de Barros} S., 2009, \aap, 502, 423

\bibitem[{{Schaye} {et~al}\mbox{.}(2015){Schaye}, {Crain}, {Bower}, {Furlong}, {Schaller}, {Theuns}, {Dalla Vecchia}, {Frenk}, {McCarthy}, {Helly}, {Jenkins}, {Rosas-Guevara}, {White}, {Baes}, {Booth}, {Camps}, {Navarro}, {Qu}, {Rahmati}, {Sawala}, {Thomas}, \& {Trayford}}]{Schaye2015}
{Schaye} J. {et~al.}, 2015, \mnras, 446, 521

\bibitem[{{Schmidt} {et~al}\mbox{.}(2021){Schmidt}, {Kerutt}, {Wisotzki}, {Urrutia}, {Feltre}, {Maseda}, {Nanayakkara}, {Bacon}, {Boogaard}, {Conseil}, {Contini}, {Herenz}, {Kollatschny}, {Krumpe}, {Leclercq}, {Mahler}, {Matthee}, {Mauerhofer}, {Richard}, \& {Schaye}}]{Schmidt2021}
{Schmidt} K.~B. {et~al.}, 2021, \aap, 654, A80

\bibitem[{{Schreiber} {et~al}\mbox{.}(2018{\natexlab{a}}){Schreiber}, {Glazebrook}, {Nanayakkara}, {Kacprzak}, {Labb{\'e}}, {Oesch}, {Yuan}, {Tran}, {Papovich}, {Spitler}, \& {Straatman}}]{Schreiber2018b}
{Schreiber} C. {et~al.}, 2018{\natexlab{a}}, \aap, 618, A85

\bibitem[{{Schreiber} {et~al}\mbox{.}(2018{\natexlab{b}}){Schreiber}, {Labb{\'e}}, {Glazebrook}, {Bekiaris}, {Papovich}, {Costa}, {Elbaz}, {Kacprzak}, {Nanayakkara}, {Oesch}, {Pannella}, {Spitler}, {Straatman}, {Tran}, \& {Wang}}]{Schreiber2018a}
{Schreiber} C. {et~al.}, 2018{\natexlab{b}}, \aap, 611, A22

\bibitem[{{Setton} {et~al}\mbox{.}(2025){Setton}, {Greene}, {Spilker}, {Williams}, {Labbe}, {Ma}, {Wang}, {Whitaker}, {Leja}, {de Graaff}, {Alberts}, {Bezanson}, {Boogaard}, {Brammer}, {Cutler}, {Cleri}, {Cooper}, {Dayal}, {Fujimoto}, {Furtak}, {Goulding}, {Hirschmann}, {Kokorev}, {Maseda}, {McConachie}, {Matthee}, {Miller}, {Naidu}, {Oesch}, {Pan}, {Price}, {Suess}, {Weaver}, {Xiao}, {Zhang}, \& {Zitrin}}]{Setton2025}
{Setton} D.~J. {et~al.}, 2025, arXiv e-prints, arXiv:2503.02059

\bibitem[{{Setton} {et~al}\mbox{.}(2024){Setton}, {Khullar}, {Miller}, {Bezanson}, {Greene}, {Suess}, {Whitaker}, {Antwi-Danso}, {Atek}, {Brammer}, {Cutler}, {Dayal}, {Feldmann}, {Fujimoto}, {Furtak}, {Glazebrook}, {Goulding}, {Kokorev}, {Labbe}, {Leja}, {Ma}, {Marchesini}, {Nanayakkara}, {Pan}, {Price}, {Siegel}, {Shipley}, {Weaver}, {van Dokkum}, {Wang}, \& {Williams}}]{Setton2024}
{Setton} D.~J. {et~al.}, 2024, \apj, 974, 145

\bibitem[{{Shankar} {et~al}\mbox{.}(2006){Shankar}, {Lapi}, {Salucci}, {De Zotti}, \& {Danese}}]{Shankar2006}
{Shankar} F., {Lapi} A., {Salucci} P., {De Zotti} G., {Danese} L., 2006, \apj, 643, 14

\bibitem[{{Silk} \& {Mamon}(2012)}]{Silk2012}
{Silk} J., {Mamon} G.~A., 2012, Research in Astronomy and Astrophysics, 12, 917

\bibitem[{{Simmonds} {et~al}\mbox{.}(2024){Simmonds}, {Tacchella}, {Hainline}, {Johnson}, {McClymont}, {Robertson}, {Saxena}, {Sun}, {Witten}, {Baker}, {Bhatawdekar}, {Boyett}, {Bunker}, {Charlot}, {Curtis-Lake}, {Egami}, {Eisenstein}, {Hausen}, {Maiolino}, {Maseda}, {Scholtz}, {Williams}, {Willott}, \& {Witstok}}]{Simmons2024}
{Simmonds} C. {et~al.}, 2024, \mnras, 527, 6139

\bibitem[{{Spiniello} {et~al}\mbox{.}(2024){Spiniello}, {D'Ago}, {Coccato}, {Hartke}, {Tortora}, {Ferr{\'e}-Mateu}, {Pulsoni}, {Cappellari}, {Maksymowicz-Maciata}, {Arnaboldi}, {Bevacqua}, {Gallazzi}, {Hunt}, {La Barbera}, {Mart{\'\i}n-Navarro}, {Napolitano}, {Radovich}, {Saracco}, {Scognamiglio}, {Spavone}, \& {Zibetti}}]{Spiniello2024}
{Spiniello} C. {et~al.}, 2024, \mnras, 527, 8793

\bibitem[{{Stinson} {et~al}\mbox{.}(2007){Stinson}, {Dalcanton}, {Quinn}, {Kaufmann}, \& {Wadsley}}]{Stinson2007}
{Stinson} G.~S., {Dalcanton} J.~J., {Quinn} T., {Kaufmann} T., {Wadsley} J., 2007, \apj, 667, 170

\bibitem[{{Straatman} {et~al}\mbox{.}(2014){Straatman}, {Labb{\'e}}, {Spitler}, {Allen}, {Altieri}, {Brammer}, {Dickinson}, {van Dokkum}, {Inami}, {Glazebrook}, {Kacprzak}, {Kawinwanichakij}, {Kelson}, {McCarthy}, {Mehrtens}, {Monson}, {Murphy}, {Papovich}, {Persson}, {Quadri}, {Rees}, {Tomczak}, {Tran}, \& {Tilvi}}]{Straatman2014}
{Straatman} C.~M.~S. {et~al.}, 2014, \apjl, 783, L14

\bibitem[{{Takahashi} {et~al}\mbox{.}(2025){Takahashi}, {Morishita}, {Kodama}, {Liu}, {Daikuhara}, \& {Chen}}]{Takahashi2025}
{Takahashi} K., {Morishita} T., {Kodama} T., {Liu} Z., {Daikuhara} K., {Chen} N., 2025, arXiv e-prints, arXiv:2505.05942

\bibitem[{{Taylor} {et~al}\mbox{.}(2025){Taylor}, {Finkelstein}, {Kocevski}, {Jeon}, {Bromm}, {Amor{\'\i}n}, {Arrabal Haro}, {Backhaus}, {Bagley}, {Banados}, {Bhatawdekar}, {Brooks}, {Calabr{\`o}}, {Ch{\'a}vez Ortiz}, {Cheng}, {Cleri}, {Cole}, {Davis}, {Dickinson}, {Donnan}, {Dunlop}, {Ellis}, {Fern{\'a}ndez}, {Fontana}, {Fujimoto}, {Giavalisco}, {Grazian}, {Guo}, {Hathi}, {Holwerda}, {Hirschmann}, {Inayoshi}, {Kartaltepe}, {Khusanova}, {Koekemoer}, {Kokorev}, {Larson}, {Leung}, {Lucas}, {McLeod}, {Napolitano}, {Onoue}, {Pacucci}, {Papovich}, {P{\'e}rez-Gonz{\'a}lez}, {Pirzkal}, {Somerville}, {Trump}, {Wilkins}, {Yung}, \& {Zhang}}]{Taylor2025}
{Taylor} A.~J. {et~al.}, 2025, \apj, 986, 165

\bibitem[{{Thomas} {et~al}\mbox{.}(2005){Thomas}, {Maraston}, {Bender}, \& {Mendes de Oliveira}}]{Thomas2005}
{Thomas} D., {Maraston} C., {Bender} R., {Mendes de Oliveira} C., 2005, 621, 673

\bibitem[{{Treu} {et~al}\mbox{.}(2022){Treu}, {Roberts-Borsani}, {Bradac}, {Brammer}, {Fontana}, {Henry}, {Mason}, {Morishita}, {Pentericci}, {Wang}, {Acebron}, {Bagley}, {Bergamini}, {Belfiori}, {Bonchi}, {Boyett}, {Boutsia}, {Calabr{\'o}}, {Caminha}, {Castellano}, {Dressler}, {Glazebrook}, {Grillo}, {Jacobs}, {Jones}, {Kelly}, {Leethochawalit}, {Malkan}, {Marchesini}, {Mascia}, {Mercurio}, {Merlin}, {Nanayakkara}, {Nonino}, {Paris}, {Poggianti}, {Rosati}, {Santini}, {Scarlata}, {Shipley}, {Strait}, {Trenti}, {Tubthong}, {Vanzella}, {Vulcani}, \& {Yang}}]{Treu2022}
{Treu} T. {et~al.}, 2022, \apj, 935, 110

\bibitem[{{Trujillo} {et~al}\mbox{.}(2009){Trujillo}, {Cenarro}, {de Lorenzo-C{\'a}ceres}, {Vazdekis}, {de la Rosa}, \& {Cava}}]{Trujillo2009}
{Trujillo} I., {Cenarro} A.~J., {de Lorenzo-C{\'a}ceres} A., {Vazdekis} A., {de la Rosa} I.~G., {Cava} A., 2009, \apjl, 692, L118

\bibitem[{{Trujillo} {et~al}\mbox{.}(2014){Trujillo}, {Ferr{\'e}-Mateu}, {Balcells}, {Vazdekis}, \& {S{\'a}nchez-Bl{\'a}zquez}}]{Trujillo2014}
{Trujillo} I., {Ferr{\'e}-Mateu} A., {Balcells} M., {Vazdekis} A., {S{\'a}nchez-Bl{\'a}zquez} P., 2014, \apjl, 780, L20

\bibitem[{{Turner} {et~al}\mbox{.}(2025){Turner}, {Tacchella}, {D'Eugenio}, {Carniani}, {Curti}, {Glazebrook}, {Johnson}, {Lim}, {Looser}, {Maiolino}, {Nanayakkara}, \& {Wan}}]{Turner2025}
{Turner} C. {et~al.}, 2025, \mnras, 537, 1826

\bibitem[{{Valentino} {et~al}\mbox{.}(2023){Valentino}, {Brammer}, {Gould}, {Kokorev}, {Fujimoto}, {Jespersen}, {Vijayan}, {Weaver}, {Ito}, {Tanaka}, {Ilbert}, {Magdis}, {Whitaker}, {Faisst}, {Gallazzi}, {Gillman}, {Gim{\'e}nez-Arteaga}, {G{\'o}mez-Guijarro}, {Kubo}, {Heintz}, {Hirschmann}, {Oesch}, {Onodera}, {Rizzo}, {Lee}, {Strait}, \& {Toft}}]{Valentino2023}
{Valentino} F. {et~al.}, 2023, \apj, 947, 20

\bibitem[{{Wang} {et~al}\mbox{.}(2020){Wang}, {Kroupa}, {Takahashi}, \& {Jerabkova}}]{Wang2020}
{Wang} L., {Kroupa} P., {Takahashi} K., {Jerabkova} T., 2020, \mnras, 491, 440

\bibitem[{{Weaver} {et~al}\mbox{.}(2022){Weaver}, {Kauffmann}, {Ilbert}, {McCracken}, {Moneti}, {Toft}, {Brammer}, {Shuntov}, {Davidzon}, {Hsieh}, {Laigle}, {Anastasiou}, {Jespersen}, {Vinther}, {Capak}, {Casey}, {McPartland}, {Milvang-Jensen}, {Mobasher}, {Sanders}, {Zalesky}, {Arnouts}, {Aussel}, {Dunlop}, {Faisst}, {Franx}, {Furtak}, {Fynbo}, {Gould}, {Greve}, {Gwyn}, {Kartaltepe}, {Kashino}, {Koekemoer}, {Kokorev}, {Le F{\`e}vre}, {Lilly}, {Masters}, {Magdis}, {Mehta}, {Peng}, {Riechers}, {Salvato}, {Sawicki}, {Scarlata}, {Scoville}, {Shirley}, {Silverman}, {Sneppen}, {Smolc̆i{\'c}}, {Steinhardt}, {Stern}, {Tanaka}, {Taniguchi}, {Teplitz}, {Vaccari}, {Wang}, \& {Zamorani}}]{Weaver2022}
{Weaver} J.~R. {et~al.}, 2022, \apjs, 258, 11

\bibitem[{{Weibel} {et~al}\mbox{.}(2025){Weibel}, {de Graaff}, {Setton}, {Miller}, {Oesch}, {Brammer}, {Lagos}, {Whitaker}, {Williams}, {Baggen}, {Bezanson}, {Boogaard}, {Cleri}, {Greene}, {Hirschmann}, {Hviding}, {Kuruvanthodi}, {Labb{\'e}}, {Leja}, {Maseda}, {Matthee}, {McConachie}, {Naidu}, {Roberts-Borsani}, {Schaerer}, {Suess}, {Valentino}, {van Dokkum}, \& {Wang}}]{Weibel2025}
{Weibel} A. {et~al.}, 2025, \apj, 983, 11

\bibitem[{{Weibel} {et~al}\mbox{.}(2024){Weibel}, {Oesch}, {Barrufet}, {Gottumukkala}, {Ellis}, {Santini}, {Weaver}, {Allen}, {Bouwens}, {Bowler}, {Brammer}, {Carnall}, {Cullen}, {Dayal}, {Dickinson}, {Donnan}, {Dunlop}, {Giavalisco}, {Grogin}, {Illingworth}, {Koekemoer}, {Labbe}, {Marchesini}, {McLeod}, {McLure}, {Naidu}, {P{\'e}rez-Gonz{\'a}lez}, {Shuntov}, {Stefanon}, {Toft}, \& {Xiao}}]{Weibel2024}
{Weibel} A. {et~al.}, 2024, \mnras, 533, 1808

\bibitem[{{Weller} {et~al}\mbox{.}(2025){Weller}, {Pacucci}, {Ni}, {Hernquist}, \& {Park}}]{Weller2025}
{Weller} E.~J., {Pacucci} F., {Ni} Y., {Hernquist} L., {Park} M., 2025, \apj, 979, 181

\bibitem[{{Williams} {et~al}\mbox{.}(2024){Williams}, {Alberts}, {Ji}, {Hainline}, {Lyu}, {Rieke}, {Endsley}, {Suess}, {Sun}, {Johnson}, {Florian}, {Shivaei}, {Rujopakarn}, {Baker}, {Bhatawdekar}, {Boyett}, {Bunker}, {Cameron}, {Carniani}, {Charlot}, {Curtis-Lake}, {DeCoursey}, {de Graaff}, {Egami}, {Eisenstein}, {Gibson}, {Hausen}, {Helton}, {Maiolino}, {Maseda}, {Nelson}, {P{\'e}rez-Gonz{\'a}lez}, {Rieke}, {Robertson}, {Saxena}, {Tacchella}, {Willmer}, \& {Willott}}]{Williams2024}
{Williams} C.~C. {et~al.}, 2024, \apj, 968, 34

\bibitem[{{Wuyts} {et~al}\mbox{.}(2007){Wuyts}, {Labb{\'e}}, {Franx}, {Rudnick}, {van Dokkum}, {Fazio}, {F{\"o}rster Schreiber}, {Huang}, {Moorwood}, {Rix}, {R{\"o}ttgering}, \& {van der Werf}}]{Wuyts2007}
{Wuyts} S. {et~al.}, 2007, \apj, 655, 51

\bibitem[{{Xiao} {et~al}\mbox{.}(2024){Xiao}, {Oesch}, {Elbaz}, {Bing}, {Nelson}, {Weibel}, {Illingworth}, {van Dokkum}, {Naidu}, {Daddi}, {Bouwens}, {Matthee}, {Wuyts}, {Chisholm}, {Brammer}, {Dickinson}, {Magnelli}, {Leroy}, {Schaerer}, {Herard-Demanche}, {Lim}, {Barrufet}, {Endsley}, {Fudamoto}, {G{\'o}mez-Guijarro}, {Gottumukkala}, {Labb{\'e}}, {Magee}, {Marchesini}, {Maseda}, {Qin}, {Reddy}, {Shapley}, {Shivaei}, {Shuntov}, {Stefanon}, {Whitaker}, \& {Wyithe}}]{Xiao2024}
{Xiao} M. {et~al.}, 2024, \nat, 635, 311

\bibitem[{{Xie} {et~al}\mbox{.}(2024){Xie}, {De Lucia}, {Fontanot}, {Hirschmann}, {Bah{\'e}}, {Balogh}, {Muzzin}, {Vulcani}, {Baxter}, {Forrest}, {Wilson}, {Rudnick}, {Cooper}, \& {Rescigno}}]{Xie2024}
{Xie} L. {et~al.}, 2024, \apjl, 966, L2

\bibitem[{{Yan} {et~al}\mbox{.}(2006){Yan}, {Newman}, {Faber}, {Konidaris}, {Koo}, \& {Davis}}]{Yan2006}
{Yan} R., {Newman} J.~A., {Faber} S.~M., {Konidaris} N., {Koo} D., {Davis} M., 2006, \apj, 648, 281

\bibitem[{{Yan}, {Je{\v{r}}{\'a}bkov{\'a}} \& {Kroupa}(2021){Yan}, {Je{\v{r}}{\'a}bkov{\'a}}, \& {Kroupa}}]{Yan2021}
{Yan} Z., {Je{\v{r}}{\'a}bkov{\'a}} T., {Kroupa} P., 2021, \aap, 655, A19

\end{thebibliography}

\begin{appendix}

\section{SED-fitting library} \label{lib}

\begin{figure*}
   \centering
   \includegraphics[width=\hsize]{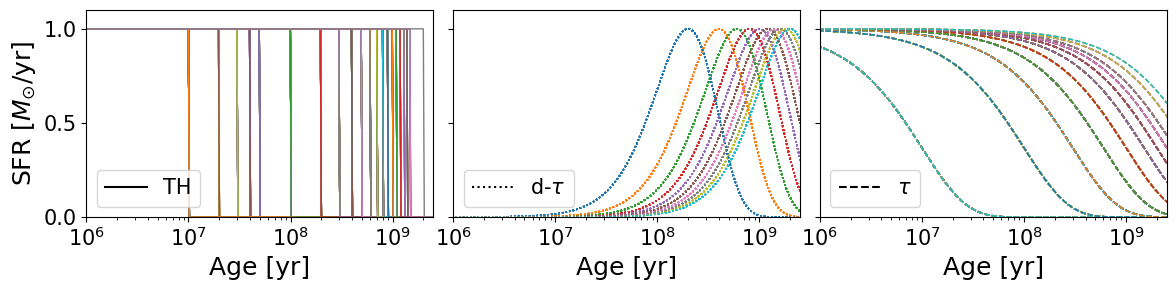}
    \caption{SFHs of the models in the library. Left to right: top-hat (TH), delayed declining exponential (d-$\tau$) and declining exponential ($\tau$) SFHs. Each line of a given color represents a set of models with the same $\tau$ parameter.}
    \label{lib_sfh}
\end{figure*}


The SED-fitting runs with \textsc{zphot} were performed using a library of galaxy models created with the parameters listed in Table \ref{zphotlib}. We used \citet{Bruzual2003} stellar libraries and included emission lines as described in \citet{Castellano2014} and \citet{Schaerer2009}, with \citet{Calzetti2000} extinction law, \citet{Salpeter1955} IMF and \citet{Fan2006} IGM absorption. We included three SFH functions: namely, top-hat (TH in short, with constant SFR before the epoch of quenching and SFR=0 after it), exponentially declining ($\tau$, with SFR$\propto$$e^{-t/\tau}$), and delayed exponentially declining (d-$\tau$, with SFR$\propto$$t^2 e^{-t/\tau}$). 
We only included models with meaningful configurations (e.g., the age must be lower than the age of the Universe at the considered redshift, etc.). We also removed redundant models (for example, at early ages exponentially declining SFHs with large values of $\tau$ are essentially equivalent to top-hat SFHs). Figure \ref{lib_sfh} shows the SFHs of the models in the library, with each line representing a model with a given value of the parameter $\tau$ in the functional form of the SFH.
We included five metallicities, with values ranging from 2\% to 250\% $Z_{\odot}$ (see Sect. \ref{colmet}), and we allowed for $E(B-V)$ values from 0 to 1, with a step of 0.1.


\begin{table}
\caption{Parameters of the SED-fitting library.} \label{zphotlib}
\centering
\begin{tabular}{l p{5cm}}    
\hline\hline
Parameter & Values \\ \hline
Redshift & From 3 to 12, $dz$=0.025 \\
TH [Gyr] & 0.01, 0.02, 0.03, 0.04, 0.05,\\
& 0.1, 0.2, 0.3, 0.4, 0.5,\\
& 0.6, 0.7, 0.8, 0.9, 1.0,\\
& 1.1, 1.2, 1.3, 1.4, 1.5, \\
& 2.0, 3.0\\
$\tau$ $\tau$ [Gyr] & 0.01, 0.1, 0.3, 0.6, 1.0,\\
& 1.5, 2.0, 2.5, 3.0, 4.0,\\
& 5.0, 7.0, 9.0, 11.0, 13.0,\\
& 15.0, 20.0, 30.0\\
d-$\tau$ [Gyr] & 0.1, 0.2, 0.3, 0.4, 0.5,\\
& 0.6, 0.7, 0.8, 0.9, 1.0 \\
Age [Gyr] & 0.03, 0.04, 0.05, 0.06, 0.08,\\
& 0.1, 0.125, 0.15, 0.2, 0.25,\\
& 0.3, 0.35, 0.4, 0.45, 0.5,\\
& 0.55, 0.6, 0.65, 0.7, 0.75,\\
& 0.8, 0.85, 0.9, 0.95, 1.0,\\
& 1.05, 1.1, 1.15, 1.2, 1.25,\\
& 1.3, 1.35, 1.4, 1.45, 1.5,\\
& 1.6, 1.7, 1.8, 1.9, 2.0\\
Metallicity [$Z_{\odot}$] & 0.02, 0.2, 0.4, 1, 2.5 \\
$E(B-V)$ & 0.0, 0.1, 0.2, 0.3, 0.4, 0.5,\\
& 0.6, 0.7, 0.8, 0.9, 1.0 \\ \hline
\end{tabular}
\end{table}

\begin{figure}
   \centering
   \includegraphics[width=\hsize]{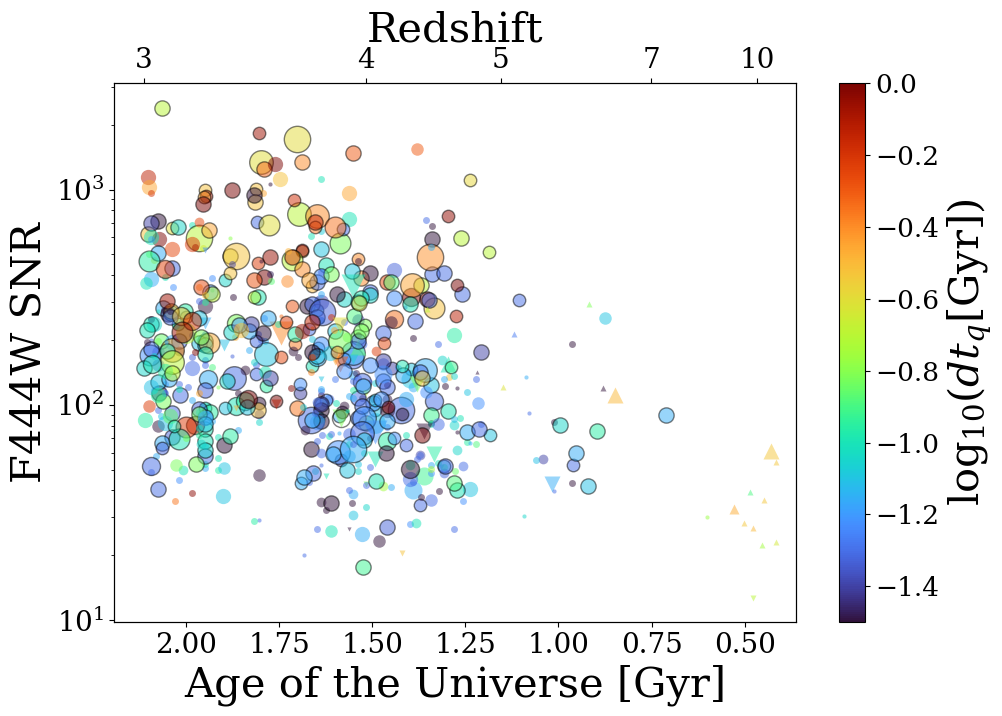}
    \caption{SNR in F444W of the candidates. Symbols are as in Fig. \ref{zm}.}
    \label{detcomp}
\end{figure}

\begin{figure*}
   \centering
   \includegraphics[width=0.49\hsize]{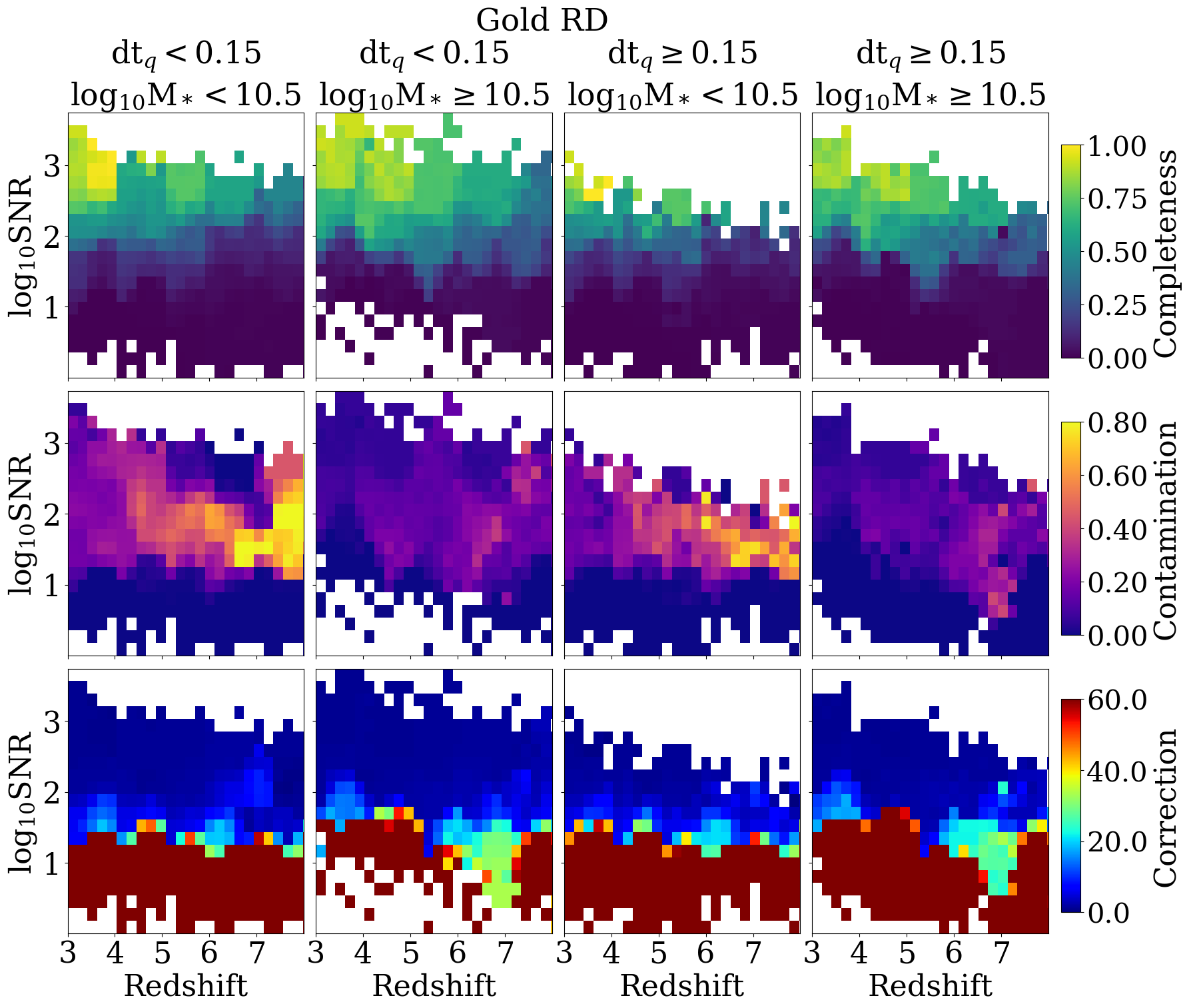}
   \hfill
    \includegraphics[width=0.49\hsize]{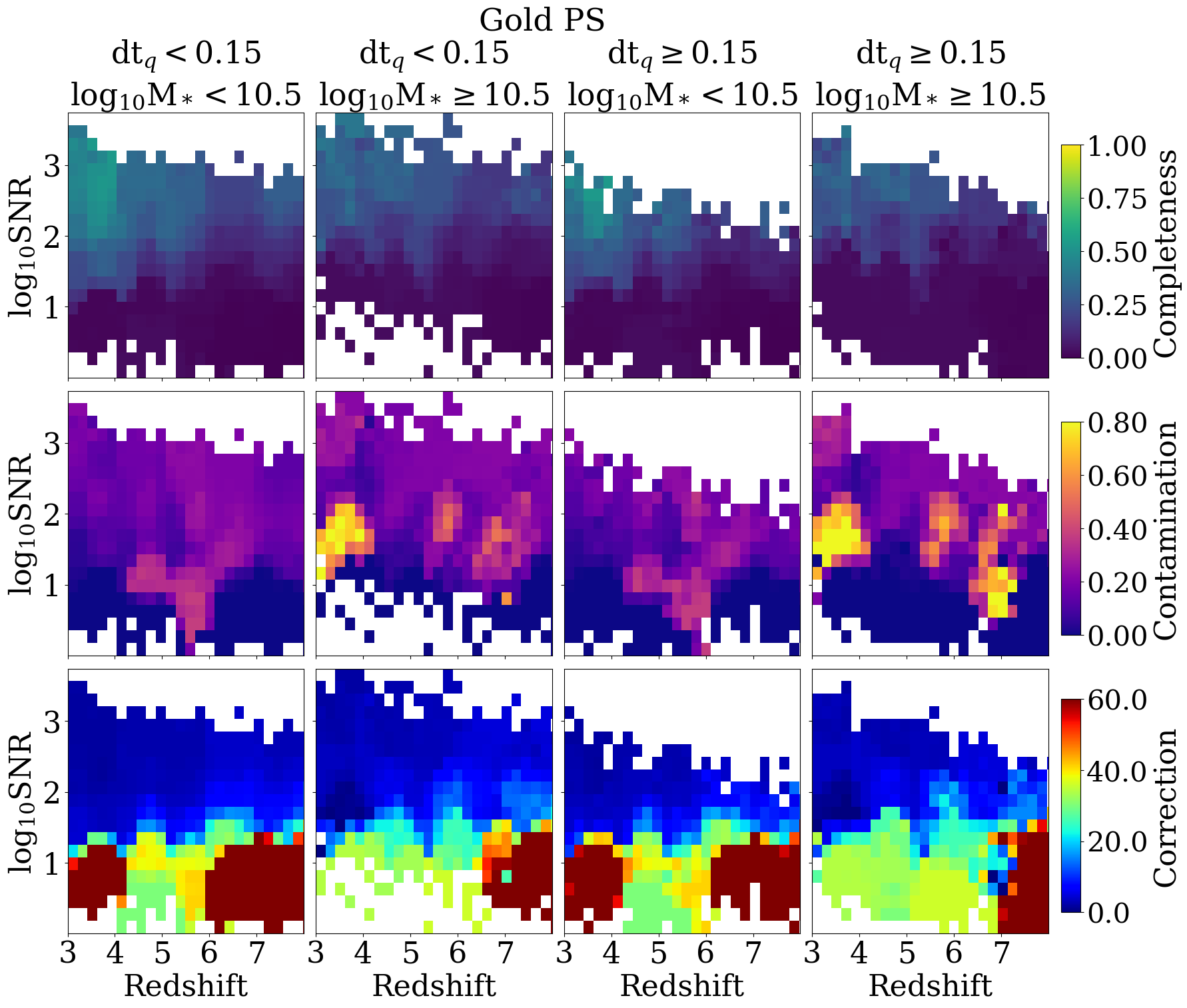}

    \caption{Completeness, contamination, and correction term computed on simulated data (left: RDs, right: PSs), as described in Sect. \ref{simulations}, for the ``gold'' selection criteria. Each sub-panel shows the results in the SNR vs $z$ plane for a given subset of $dt_q$ and M$_*$ parameters, as indicated in the panel titles.}
    \label{sims_gold}
\end{figure*}

\begin{figure*}
   \centering
   
   \includegraphics[width=0.49\hsize]{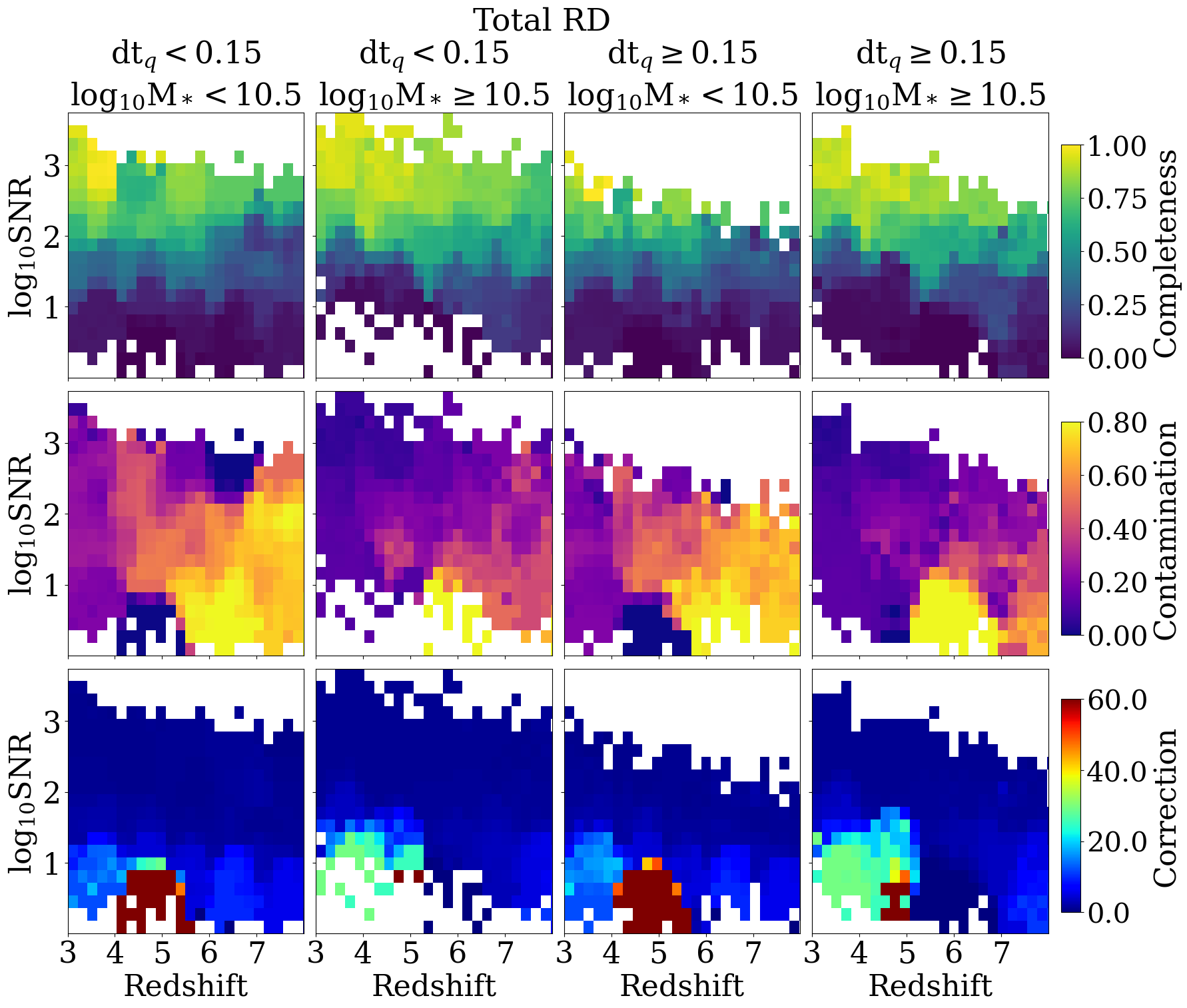}
    \hfill
    \includegraphics[width=0.49\hsize]{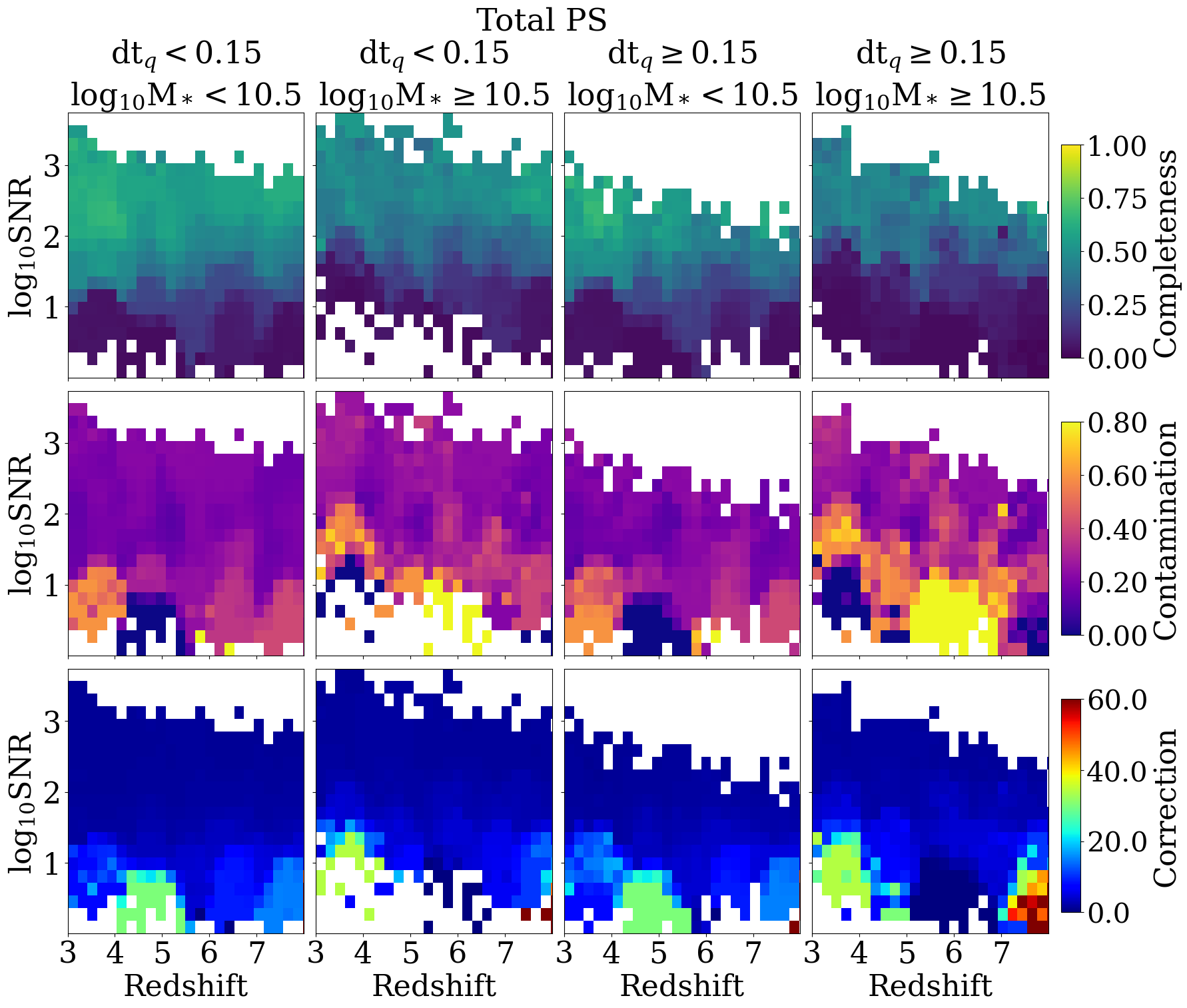}
    \caption{Same as Fig. \ref{sims_gold}, for the total selection criteria.}
    \label{sims_total}
\end{figure*}

\section{Simulations} \label{simulations}


\begin{figure}
   \centering
   \includegraphics[width=\hsize]{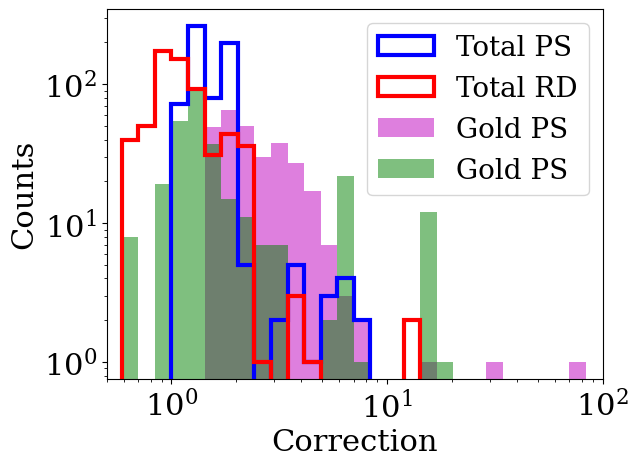}
    \caption{Distribution of the values of the correction terms assigned to the real objects in our selections.}
    \label{corrterms}
\end{figure}

\begin{figure}
   \centering
   \includegraphics[width=\hsize]{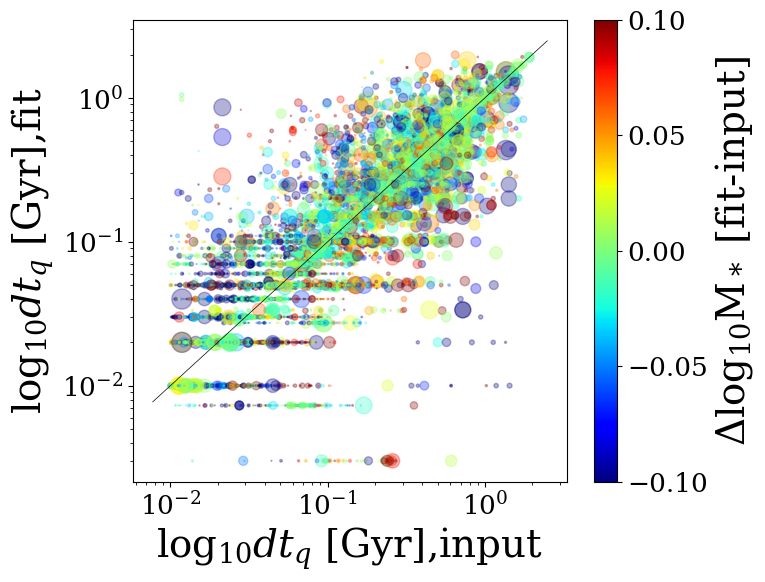}

   \includegraphics[width=\hsize]{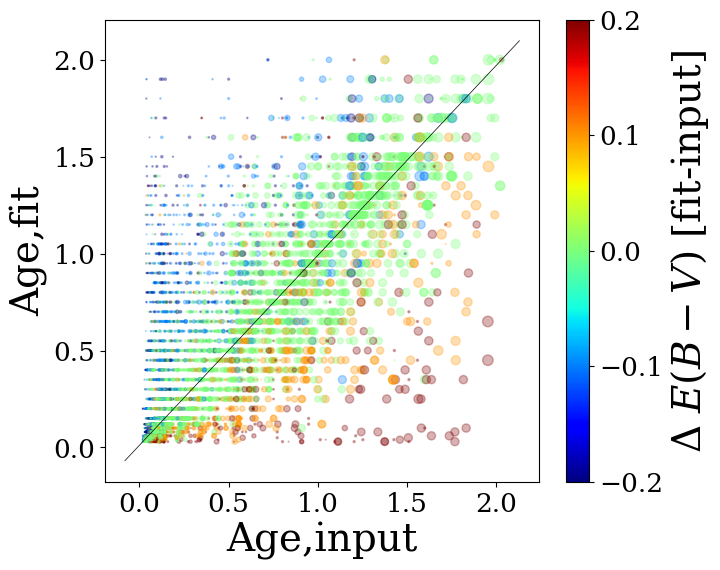}

   \includegraphics[width=\hsize]{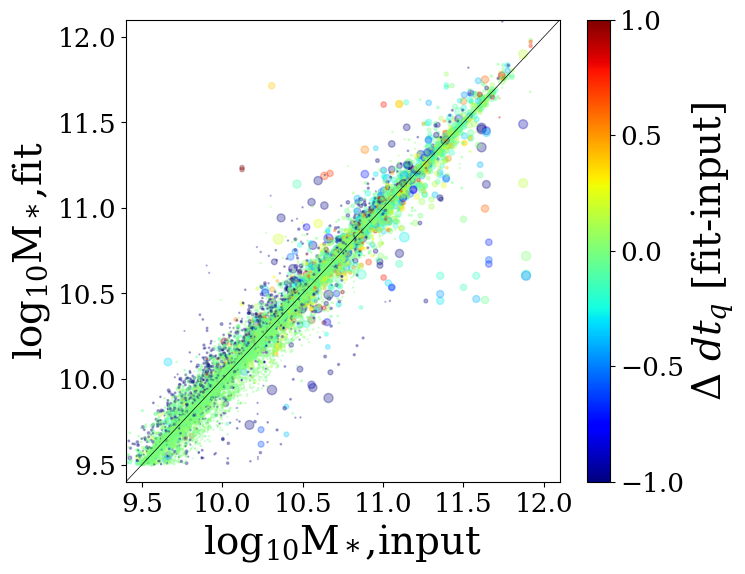}
   \caption{Fitted vs true values of the simulated quiescent sources with F444W SNR $>50$ and lM$_*>9.5$: from top to bottom: $dt_q$, age and stellar mass. Dots are color-coded by lM$_{*,fit}$-lM$_{*,input}$ (top), $E(B-V)_{fit}$ - $E(B-V)_{input}$ (middle), and $dt_{q,fit}$ - $dt_{q,input}$ (bottom); their size is a proxy for lM$_{*,input}$ (top), duration of the SF burst (middle), and $dt_{q,input}$ (bottom). The black line in each panel is the 1:1 relation.}
   \label{checks}
   
\end{figure}

To assess the completeness and the contamination degrees of our procedure we used simulated data, for which the ground truth values are known. Because the detection completeness, using the estimate in M24 for the reference limiting magnitude, is $\approx\!100\%$, we can focus only on the selection technique.
A straightforward idea would be to use the output of cosmological simulations as ground truth. However, the number of quiescent sources in the simulations providing publicly available data is too small, so we proceeded in a different way.

We started by creating a library of theoretical models with \textsc{zphot}, which we want to divide into roughly equally populated bins of physical parameters. Using a random grid of models would not have allowed us to assign meaningful corrective terms for completeness and contamination in each bin, because models corresponding to rare (or non-existent) objects would be given the same weight as models corresponding to common, ubiquitous objects, thus biasing the results; rather, we needed a plausible imitation of the real Universe. We therefore took the best-fits of the SED-fitting process on M24 real data as a starting point. To cope with the fact that there are too few quiescent galaxies to obtain meaningful statistics for all bins (especially at high redshifts), we created four replicas of each galaxy at $3<z<4$, adding integers to its nominal redshift value $z$ to obtain new values at $z+i$ with $i=1,2,3,4$. For each replica we then modified with random perturbations the best-fit physical properties -- namely, stellar mass, redshift and extinction (we did not perturb the metallicity value, as only five discrete values are included in the library) for three times, and we re-created the corresponding $k$-corrected synthetic observed magnitudes. In this procedure we neglected the physical evolution of the galactic populations with redshift. Finally, we assigned to each source the original error budgets in all bands, and perturbed the fluxes consistently for three times, thus creating three ``noisy'' replicas of the very same model (we neglected the fact that, in reality, the error budget of an object is estimated from the pixels of the RMS map, so as the source is redshifted and it becomes fainter, its area becomes smaller, and its error should consistently decrease). Thus, each source in the original $3<z<4$ catalog was replicated 45 times (five redshift bins $\times$ three physical perturbations $\times$ three noisy replicas), totalling $\approx\!130,000$ simulated objects, of which $\approx\!34,000$ were quiescent with $10^{9.5}<\mbox{M}_*<10^{12}$.

We then proceeded to fit again this mock catalog, and to select quiescent candidates with the procedure described in Sect. \ref{sec:meth}. 
We used the \texttt{sklearn.cluster.KMeans} method to create 80 bins of simulated sources with roughly the same number of objects in the $z$, SNR, and $M_*$ space, and computed the values of contamination and completeness for each bin, for the two selection modes (total and ``gold'', corresponding to different $r$ values). It is important to stress that while we considered $z$ and SNR to be known a priori, we used the fitted values of M$_*$ rather than the input values, since in the real world we do not know the ground truth values.
The completeness factor $f_{comp}$(SNR, $z$, M$_*$, selection) was defined as the fraction of true RD (PS) simulated sources in a given bin that also were fitted and selected as RDs (PSs). Conversely, the contamination factor $f_{cont}$(SNR, $z$, M$_*$, selection) was defined as the number of true non-RDs (non-PSs) objects in a given bin, divided by the total number of objects selected as RDs (PSs) in the bin. After testing different threshold values for the reliability parameter $r$ (and considering the results from the test on the spectroscopic sample described in Sect. \ref{specomp}), we found a good balance between completeness and contamination with $r=0.1$ for the total selection and $r=0.4$ for the ``gold'' selection. 
Finally we created a grid of corrective factors $f_{corr}=(1-f_{cont})/f_{comp}$, which were used for the computation of the number densities (Sect. \ref{nd}) assigning each selected candidate to the relevant [SNR,$z$,M$_*$,$dt_q$] bin, and counting it with a value $f_{corr}$ rather than 1. 
Figures \ref{sims_gold} and \ref{sims_total} shows the results of this exercise, which we already summarised in Sect. \ref{simu}: each panel shows the values of completeness, contamination and corrective term for a specified category of sources, and for the ``gold'' and total selections. While in some cases the values of the correction factor $f_{corr}$ are very large (up to $\sim$300), only a few real galaxies are assigned such values, with most of the sources having values between 0.5 and 10 (this is shown in Fig. \ref{corrterms}). We checked that even imposing some reasonable limits on the values of the correction factor, e.g. $0.5 \leq f_{corr} \leq 5$, the overall results for the real number densities discussed in Sect. \ref{nd} do not change significantly. 


As a cautionary remark, we point out that the corrective terms should be considered as approximate for at least two reasons: first, the same libraries and codes were used to both produce the simulated dataset and to fit it (and they do not include templates for AGNs, LRDs or other peculiar sources); second, we did not consider any possible uncertainty in the redshift estimate. 

\subsection{Contamination of the low mass PS sample} \label{pssim}

In Sect. \ref{downsiz} we argued that high mass candidates are mostly RD and low mass candidates are mostly PS. It is important to quantify the degree of possible contamination between the two samples. The top panel of Fig. \ref{checks} shows the fitted vs. input $dt_q$ values of the simulated quiescent sources with F444W SNR $>50$ (but cutting at SNR $>10$ makes little difference), color-coded by the difference between the fitted and the input mass (in logarithmic values). It can be seen that most of the simulated sources lie along the diagonal (green dots), i.e. they are correctly fitted. 
Considering the ``gold'' selection criteria, we find that only 2\% (all values are approximated) of the simulated sources selected as low mass (lM$_*<10.5$) PSs are actually RDs, while 84\% are true PSs, and the rest are SF. Conversely, 1\% of true low mass PS are selected as RDs, and 29\% are selected as PSs (so the completeness is low in this mass regime). Finally, 5\% of true low mass RDs are selected as low mass PSs, and 39\% as RDs. So we can conclude that the selection technique provides remarkably pure samples; almost none of the selected low mass PSs are actually RDs, and a larger fraction of true low mass RDs than of true low mass PSs is correctly selected, strenghtening the conclusions of Sect. \ref{downsiz}.

\subsection{Accuracy of age estimates} \label{agesim}

In Sect. \ref{assembly} we discussed the mass assembly history of our ``gold'' candidates inferred from the parameters of their SED-fitting best solution, finding extreme values of formation redshift and quenching for some massive sources. However, we warned that the estimate of the age can often be less accurate than that of other parameters such as stellar mass and $dt_q$, because of well known degeneracies with metallicity and extinction. The middle panel of Fig. \ref{checks} shows the fitted vs. input age values of the simulated quiescent sources with F444W SNR $>50$ and lM$_*>10.5$, color coded by the difference between the fitted and the input $E(B-V)$ values, and with the sizes of the dots as a proxy fot the duration of the SF burst (which is the difference of age and $dt_q$). It can be seen that while for the majority of sources the age is reasonably well recovered (the dots close to the diagonal, mostly green indicating a good estimate of the extinction), a non-negligible fraction is fitted as older than it actually is, and with under-estimated extinction values (the dots on the upper left part of the diagram, mostly blue-ish indicating an under-estimation of the extinction, to compensate in the physical color of the galaxies for the over-estimation of the age). Quantitatively, 31\% (approximate values) of the simulated quiescent sources with lM$_*>10.5$ and F444W SNR $>50$ have their ages overestimated by up to 100 Myr, 9\% by up to 500 Myr, and 1\% by up to 1 Gyr (conversely, for the underestimation these values are 16\%, 3\%, and 0.5\% respectively). Considering all the simulated sources fitted with lM$_*>10.5$, $z_{quench} > 8$ and $E(B-V)\leq0.1$ (as the potential ``Universe breakers'' discussed in Sect. \ref{assembly}), 40\% actually have lower input $z_{quench}$, with higher extinction. This emphasizes that the results must be taken with caution. 

As a complementary information, we point out that the stellar masses are estimated with good precision, as shown in the bottom panel of Fig. \ref{checks}. While this is a well-established feature of the SED-fitting technique, it must be remembered again that this is a simplified test in which the SEDs of the simulated sources are created using the same models used to fit them. Furthermore, the caveat discussed in Sect. \ref{downsiz} about the possible overestimation of the stellar mass estimates remains valid, because the main cause of the issue is the absence of far-infrared data. 

\begin{table*}
\caption{List of spectroscopically confirmed quiescent galaxies at $z>3$ from the literature having a match in the M24 catalog, and corresponding result in the selection procedure of this work. The references are \citet[][S18]{Schreiber2018b}, \citet[][B24]{Barrufet2024},\citet[][C24]{Carnall2024}, \citet[][D24]{DEugenio2024}, \citet[][J24]{Jin2024}, \citet[][L17]{Luo2017} \citet[][N15]{Nandra2015}, \citet[][N24]{Nanayakkara2024}, \citet[][S24]{Setton2024}, \citet[][dG25]{DeGraaff2025} \citet[][W25]{Weibel2025}, \citet[][B25a]{Baker2025}.} \label{spectrq}
\centering
\begin{tabular}{l l l l l l l l}    
\hline\hline
Field and ID in M24 & RA & DEC & Reference & Redshift & Selection & Comment \\ \hline
ABELL2744 13197 & 3.5628 & -30.3909 & S24 & 3.97 & Gold & \\
CEERS 45474 & 214.8956 & 52.8566 & N24 & 3.25 & Gold & X-ray emitter in N15 \\
CEERS 47665 & 214.8712 & 52.8451 & J24 & 3.442 & Gold & X-ray emitter in N15 \\ 
CEERS 56717 & 214.9051 & 52.8916 & S18 & 3.234 & - & Ambiguous characterisation in S18\\ 
CEERS 58873 & 214.7606 & 52.8453 & S18 & 3.219 & Gold & \\
CEERS 66244 & 214.8661 & 52.8843 & N24 & 3.434 & Gold & \\
CEERS 79522 & 214.9155 & 52.9491 & dG25 & 4.896 & Gold & \\
CEERS 79531 & 214.8132 & 52.8590 & N24 & 3.227 & - & Best fitted as mildly SF in N24\\ 
JADES-GN 7115 & 189.2657 & 62.1684 & B25a & 4.13 & Gold & \\
JADES-GN 23100 & 189.2754 & 62.2141 & B25a & 4.39 & Gold & \\
JADES-GN 41598 & 189.0258 & 62.2605 & B25a & 3.12 & - & Bad photometric data in M24\\
JADES-GS 30294 & 53.0787 & -27.8396 & B24 & 3.47 & Silver & \\
JADES-GS 36729 & 53.1082 & -27.8251 & C24 & 4.658 & Gold & Broad-line AGN in C24\\
JADES-GS 40643 & 53.1653 & -27.8141 & D24 & 3.063  & Gold & X-ray AGN in L17 \\
JADES-GS 63546 & 53.1969 & -27.7605 & D24 & 3.611 & Gold & \\
JADES-GS 12902 & 53.0623 & -27.8751 & B25a & 4.22 & Gold & \\
JADES-GS 35300 & 53.0819 & -27.8288 & B25a & 4.37 & Gold & \\
PRIMER-COSMOS 86447 & 150.1217 & 2.3746 & S18 & 3.481 & - & Best fitted as re-juvenated in S18\\
PRIMER-COSMOS 88150 & 150.0615 & 2.3787 & S18 & 3.7153 & Gold & ``Jekyll'' \citep{Schreiber2018a} \\
PRIMER-COSMOS 95163 & 150.0873 & 2.3960 & S18 & 3.782 & Gold & \\
PRIMER-UDS 26918 & 34.2895 & -5.2698 & N24 & 3.813 & - & \\ 
PRIMER-UDS 31415 & 34.2904 & -5.2621 & N24 & 3.703 & Gold & \\
PRIMER-UDS 42790 & 34.3404 & -5.2413 & N24 & 3.976 & Gold & \\
PRIMER-UDS 46683 & 34.2559 & -5.2338 & N24 & 3.207 & Gold & Discussed in \citet{Glazebrook2024} \\
PRIMER-UDS 47709 & 34.2589 & -5.2323 & N24 & 3.208 & Gold & \\
PRIMER-UDS 50567 & 34.2938 & -5.2269 & N24 & 3.55 & Silver &\\
PRIMER-UDS 89448 & 34.4851 & -5.1578 & N24 & 3.529 & Gold & \\
PRIMER-UDS 95085 & 34.3651 & -5.1488 & C24 & 4.6227 & Gold & \\
PRIMER-UDS 102452 & 34.3996 & -5.1363 & C24 & 4.6194 & Gold & \\
PRIMER-UDS 117643 & 34.4296 & -5.1123 & W25 & 7.287 & Gold & Current redshift record-holder \\
\hline
\end{tabular}

\end{table*}

\section{Comparison with other selections} \label{othersel}

We checked the consistency of our sample with previously published selections. Table \ref{spectrq} lists the spectroscopically confirmed quiescent sources discussed in Sect. \ref{specomp}. 

We also checked photometric selections from the literature. First, we compared our sample with the one in our previous work on CANDELS data. As already mentioned, in M19 we used two SED-fitting libraries including only TH SFH models; using the one without emission lines we selected 102 candidates across the five CANDELS fields, 40 of which were also selected using the one including emission lines. In this work, all our models include emission lines, so a fair comparison can only be made with the M19 ``lines'' selection. Matching the coordinates of the detected sources with a radius of 0.3'', the M24 catalog has 104,282 sources in common with the CANDELS catalogs (because the observed areas do not perfectly overlap, and the detection was performed at very different wavelengths). Also, in some cases the redshift estimate changed, bringing some sources above (below) $z=3$ while they were below (above) in M19. With these premises, we find that 27 M19 ``lines'' candidates are matched in M24. Of these, 20 ($74\%$) are in our ``gold'' selection (and none in the ``silver'' selection). Of the remaining 7 missing ones, two have $z<3$ in M24, one has only F444W data in JWST observations and is therefore excluded, three have a SF best-fit, and one fails the selection because a high-probability SF solution exists.
On the other hand, 233 ``gold'' candidates have a coordinate match in the CANDELS catalogs, and only 20 of them were selected in M19 ``lines'' (27 in the ``no-lines'' selection). This demonstrates how the JWST data allows for finding many more robust candidates than was possible with HST and Spitzer data. 

\citet{Santini2021} used ALMA data to confirm the quiescent nature of 25 M19 candidates, of which 18 are matched in M24. Of these, ten are in our ``gold'' selection and two more are in the ``silver'' selection. Of the remaining ones, one has $z<3$ in M24; the other five, which are all in the GOODS-South field, failed the selection criteria (three are best fitted as SF and two have alternative SF solutions).

Considering more recent, JWST-based selections, \citet[][C23 hereafter]{Carnall2023} identified 15 quiescent candidates in the first patch of the CEERS data (which covered an area of $\approx\!30$ sq.arcmin), 11 of which were dubbed as ``robust''. Thirteen of their total sample are in our ``gold'' selection (no one is in the ``silver'' selection). 
The two missing matches are ID66258 (ID52124 in C23, not``robust'') which is best fitted as a SF galaxy in our selection, and ID73449 (ID17318 in C23, ``robust''), which is identified as a probable brown dwarf star in \citet[][with ID31016]{Holwerda2024}.

\citet[][V23]{Valentino2023} compiled a list of quiescent candidates from eleven JWST fields with publicly available observations collected during the first three months of operations; they used various types of rest-frame color selection. Their $UVJ$ ``strict'' sample included 21 candidates in CEERS and PRIMER-UDS at $z>3$ (their estimate), all of which have a match in M24; 11 of them (52\%) are in our ``gold'' selection, and three more in the ``silver'' selection (for a total of 67\%). Of the other seven, one is best fitted with lM$_*<9.5$ and is therefore excluded, three are best fitted as SF, and three fail the probabilistic selection. 

\citet{PerezGonzalez2023} selected 34 quiescent candidates in CEERS (with SED-fitting and $UVJ$), 19 of which were estimated at $z>3$ and lM$_*>9.5$. Of these, ten are in our ``gold'' selection and two more in the ``silver'' sample. Of the 7 missing, two are in the list of brown dwarfs by \citet{Holwerda2024}, one has bad photometry in some bands in M24, one has $z_{best}<3$ in our estimates, and the remaining three are best fitted as SF (but they have slightly different redshifts with respect to those in their work, two of them being now spectroscopic).

More recently, various studies searched for high-redshift quiescent galaxies using the SSFR $<0.2/t_U(z)$ yr$^{-1}$ criterion (see Sect. \ref{sec:intro}). \citet[][R24]{Russell2024} published a list of 61 candidates in CEERS and JADES-GS fields (plus 28 in the NEP field). Of these, 46 meet our mass selection criteria, but only eleven are in our ``gold'' selection, and three more in the ``silver'' sample. Also, of their 12 candidates at $z>5$ in CEERS only three have lM$_*\geq10.5$ (in their fit); noticeably, all three have $z_{phot}<3$ in M24, and were therefore excluded from our selection.
\citet[][dlV25]{delaVega2025} used the SSFR criterion after a machine-learning pre-selection and singled out 44 candidates in the two JADES fields: 14 are in our ``gold'' selection and 4 more are in our ``silver'' selection. 
Finally, \citet[][B25b]{Baker2025b} published a list of 744 quiescent candidates at $2<z<5$ in CEERS, JADES and PRIMER fields, cutting on the SSFR after pre-selecting with the $UVJ$ diagram. Of these, 597 have a match with M24 objects (the remaining ones are detected outside the M24 FoV), and using our $z_{best}$ values we ended up with 170 sources at $z>3$, of which 98 are in common with our quiescent selections (basically, most of the massive RDs; their selection does not include what we defined as PSs). Of the remaining 72 coordinate matches, 42 are excluded from our selection because of a SF best-fit, and 30 because of alternative high-probability star-forming solutions. 
It might be tempting to assume that the difference in the SSFR criterion plays a relevant role in explaining these differences, the one adopted by these authors being more relaxed at $z>3$ than the hard cut at SSFR=$10^{-10}$ yr$^{-1}$ that we adopted. However, we checked a posteriori that our additional criteria, presented in Sect. \ref{selec}, actually made the details of the SSFR threshold choice substantially irrelevant for our selection, as most of our quiescent candidates are best-fitted as objects having experienced an abrupt, total quenching, making their SFR consistent with zero at the moment of observation, while many candidates included in other selections are excluded from ours because of the existence of reliable SF solutions with SSFR$>0.2/t_U(z)$.

These checks show that most of the massive and long-time quiescent sources tend to be more robustly identified and selected (although it is worth pointing out that, for example, the dlV25 and R24 selections only have 3 candidates in common in the JADES-GS field), while low mass and recently quenched galaxies are more elusive. Although our selection criteria are quite demanding and we are reasonably confident in the results (see also Appendix \ref{simulations}), they are by no means intended to be definitive.

\section{Examples of interesting candidates} \label{objs}

\begin{figure*}
\centering

\includegraphics[width=0.9\hsize]{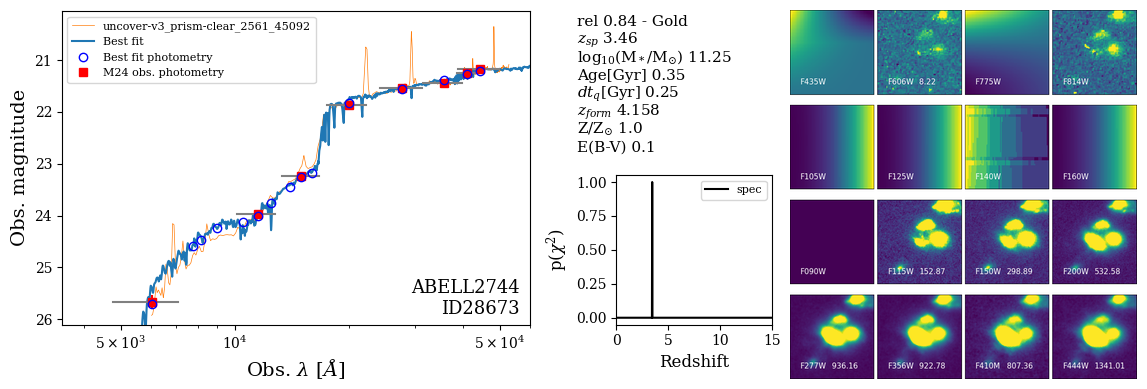}

\vspace{3mm}

\includegraphics[width=0.9\hsize]{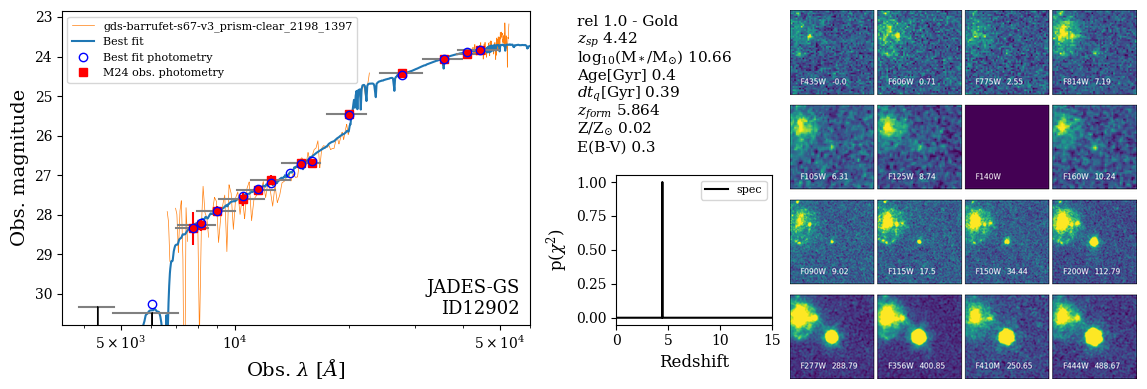}

\vspace{3mm}

\includegraphics[width=0.9\hsize]{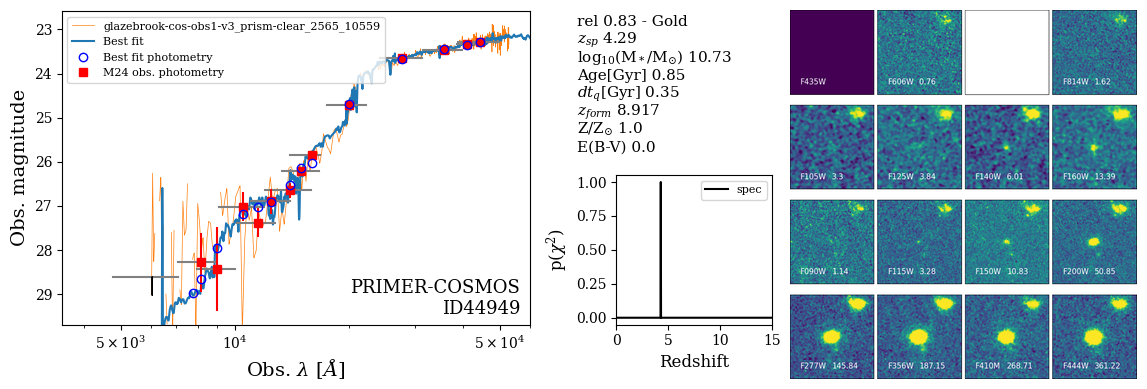}

\vspace{3mm}

\includegraphics[width=0.9\hsize]{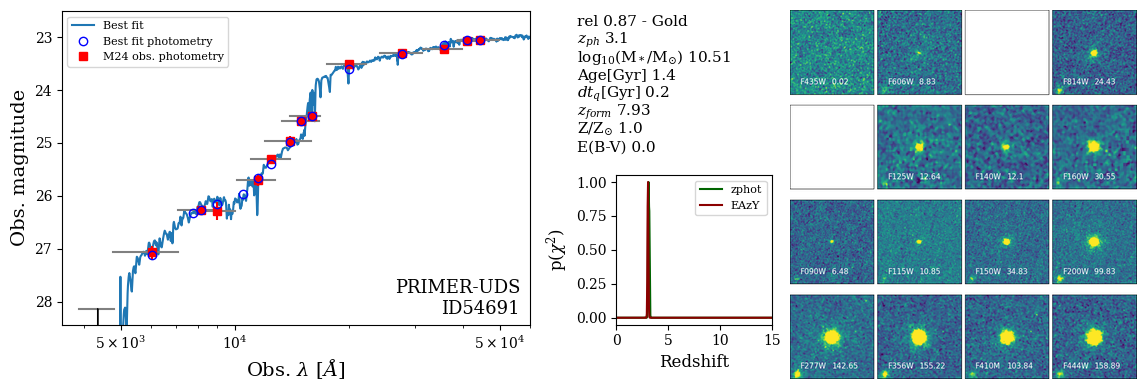}

\caption{Examples of RD ($dt_q\geq150$ Myr) ``gold'' candidates with high reliability ($r\geq0.8$), discussed in Sect. \ref{wow_objs}: observed photometry, observed spectrum when available, and best fit model (left); PDF(z) and physical parameters from the fit (mid); and multiband snapshots (right).}
\label{objs1}
\end{figure*}

\begin{figure*}
\centering

\includegraphics[width=0.9\hsize]{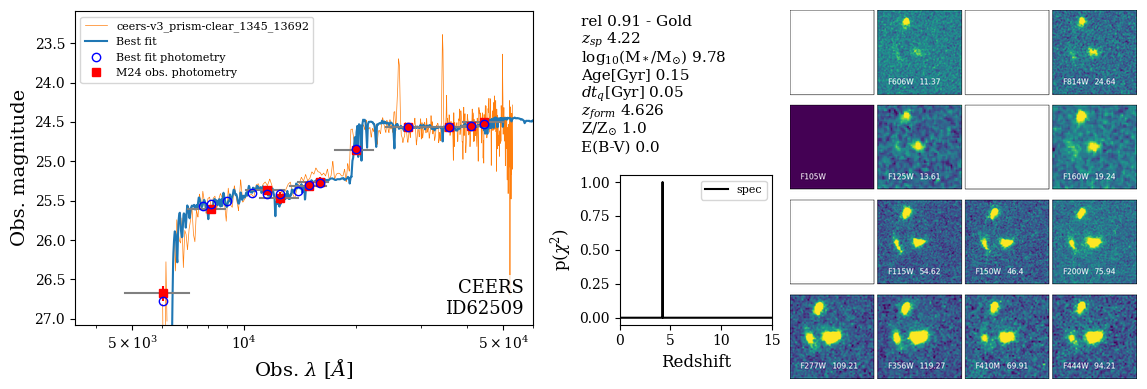}

\vspace{3mm}

\includegraphics[width=0.9\hsize]{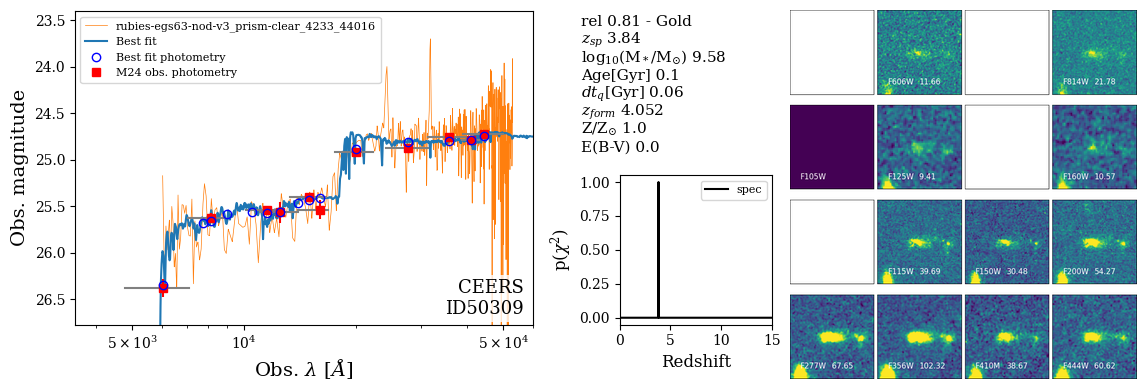}

\vspace{3mm}

\includegraphics[width=0.9\hsize]{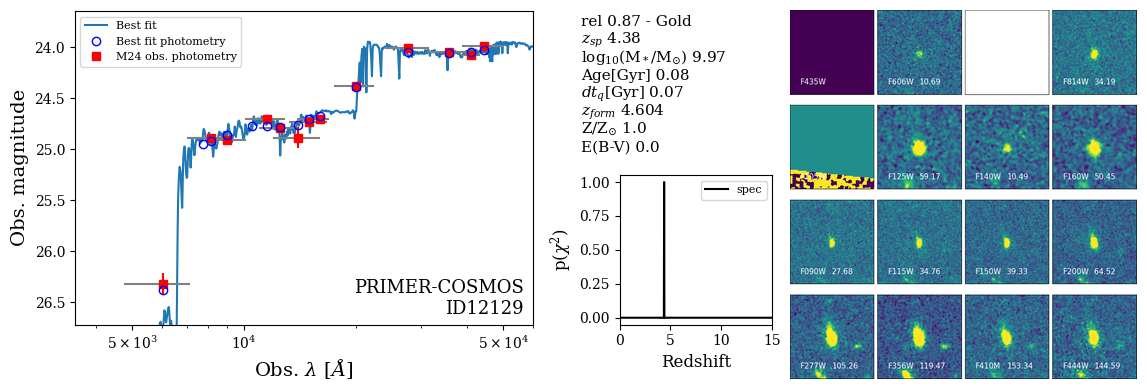}

\vspace{3mm}

\includegraphics[width=0.9\hsize]{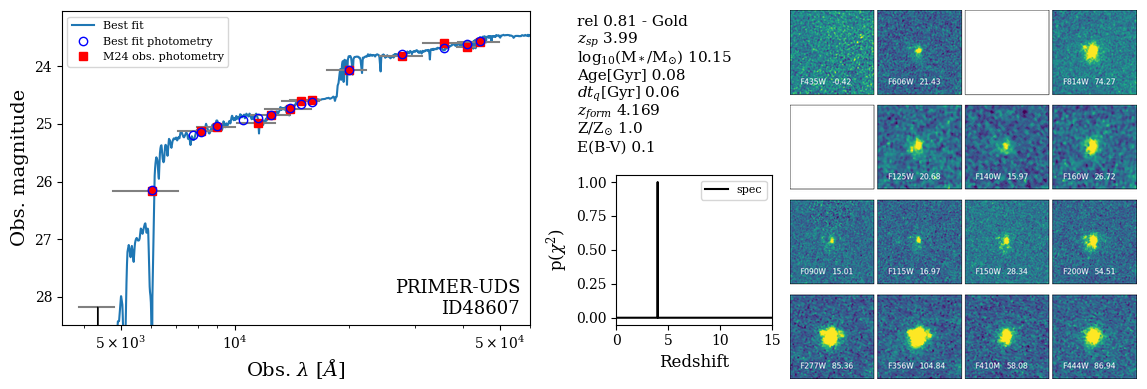}

\caption{Examples of PS ($dt_q<150$ Myr) ``gold'' candidates with high reliability ($r\geq0.8$), discussed in Sect. \ref{wow_objs}: observed photometry, observed spectrum when available, and best fit model (left); PDF(z) and physical parameters from the fit (mid); and multiband snapshots (right).}
\label{objs1b}
\end{figure*}

\begin{figure*}
\centering
\includegraphics[width=0.9\hsize]{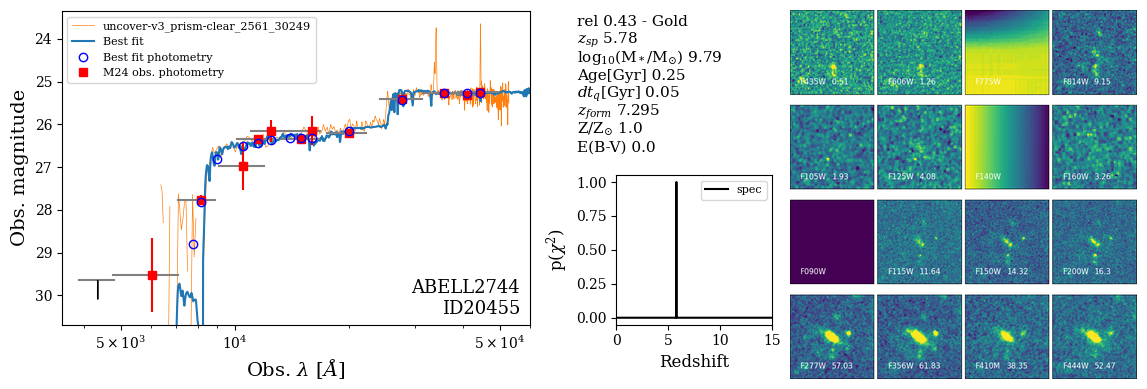}

\vspace{3mm}

\includegraphics[width=0.9\hsize]{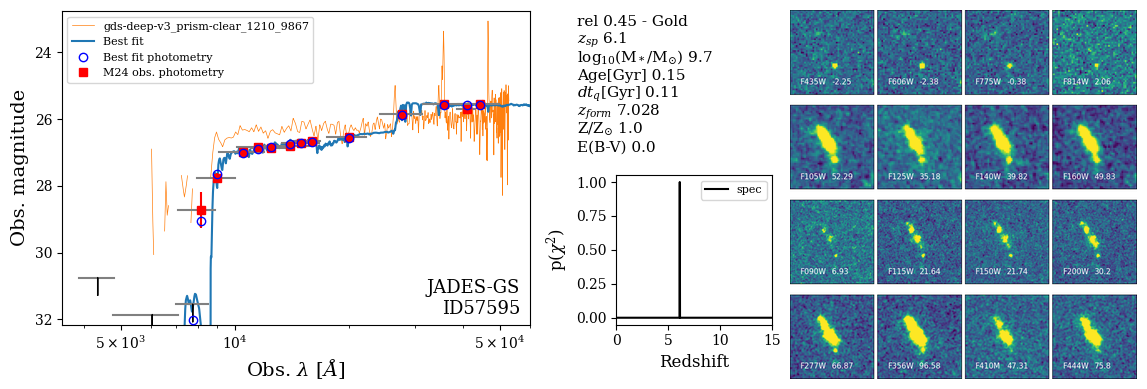}

\vspace{3mm}

\includegraphics[width=0.9\hsize]{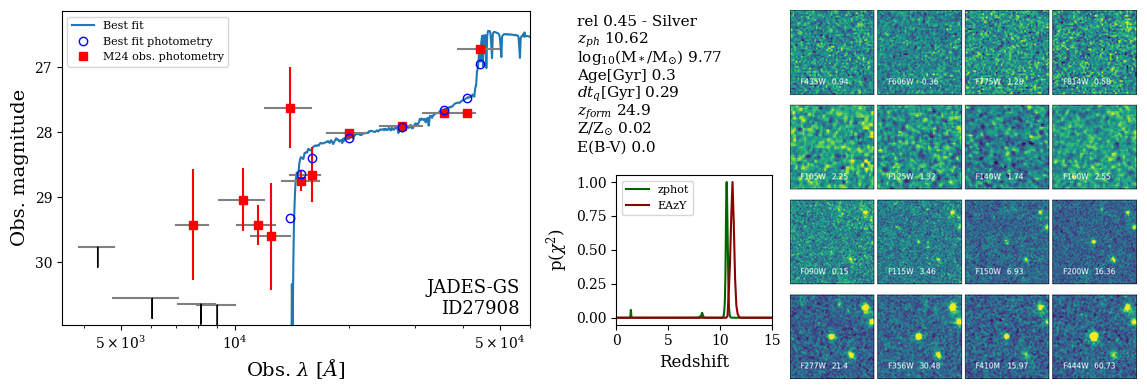}

\vspace{3mm}

\includegraphics[width=0.9\hsize]{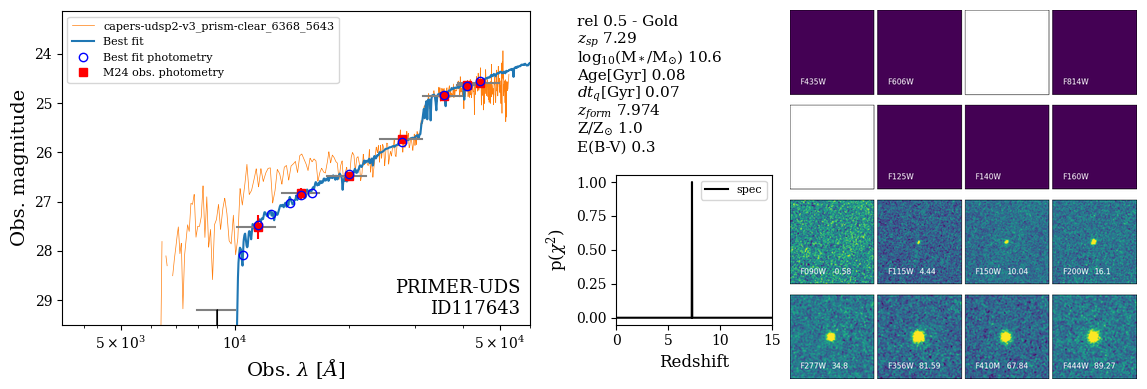}

\caption{Examples of candidates at $z>5$, discussed in Sect. \ref{wow_objs}: observed photometry, observed spectrum when available, and best fit model (left); PDF(z) and physical parameters from the fit (mid); and multiband snapshots (right).}
\label{objs2}
\end{figure*}

\begin{figure*}
\centering
\includegraphics[width=0.9\hsize]{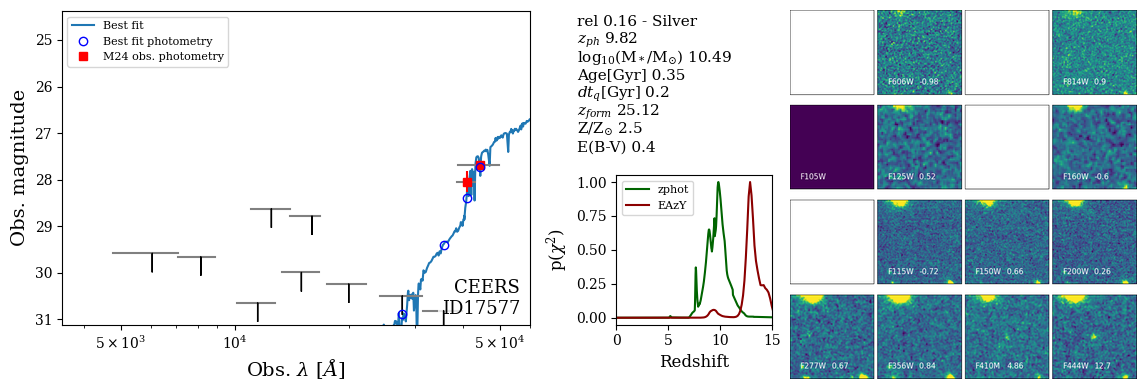}

\vspace{3mm}

\includegraphics[width=0.9\hsize]{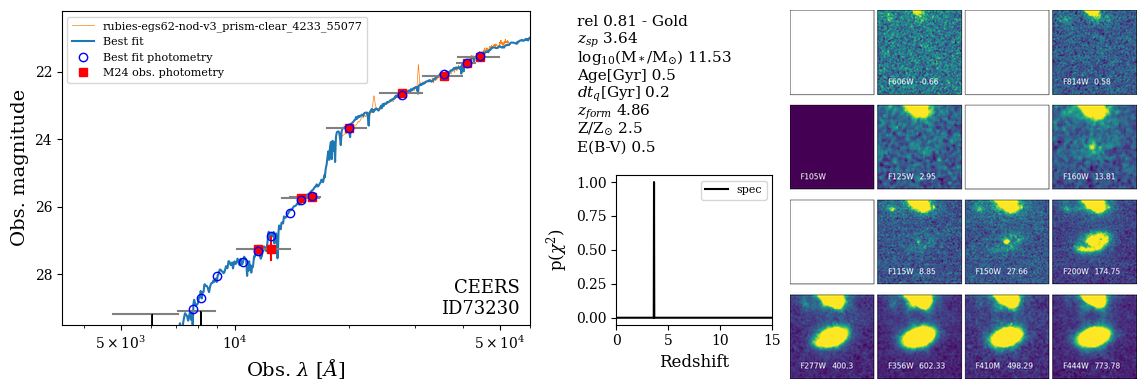}

\vspace{3mm}

\includegraphics[width=0.9\hsize]{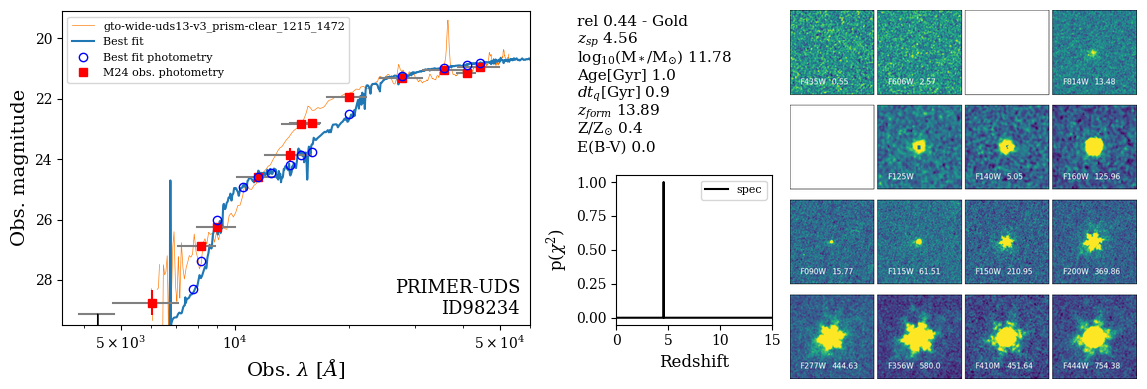}

\vspace{3mm}

\includegraphics[width=0.9\hsize]{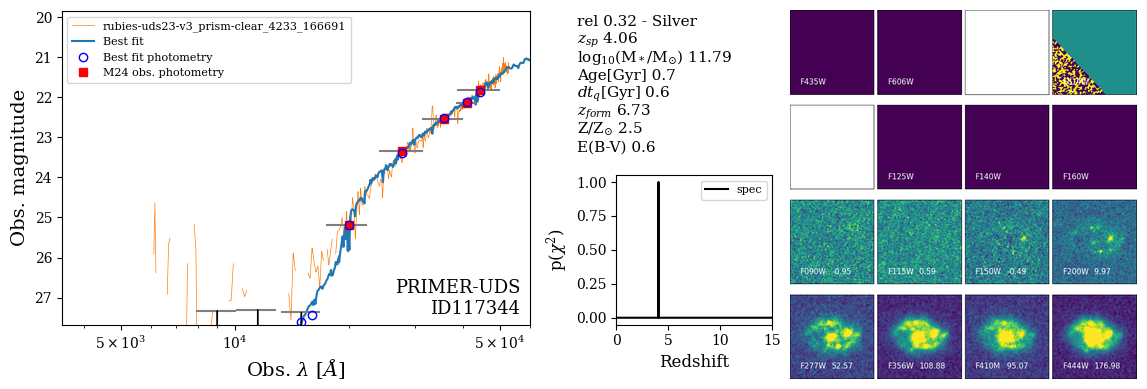}

\caption{Very massive quiescent candidates (lM$_*>11.5$): observed photometry, observed spectrum when available, and best fit model (left); PDF(z) and physical parameters from the fit (mid); and multiband snapshots (right).}
\label{objs3}
\end{figure*}

In Figs. \ref{objs1} to \ref{objs3} we show the multi-band imaging snapshots, the photometric data  (plus spectroscopic data, when present), and the best fit model and PDF($z$), of some interesting candidates among those discussed in Sect. \ref{wow_objs}. They are chosen as examples among the ``gold'' sources with reliability $r\geq0.8$ and not already discussed in the literature (to the best of our knowledge), the candidates at $z>5$ with $r\geq0.4$, and those having lM$_*>11.5$ in the best-fit model. 
In the panels, the blue lines are the best-fit quiescent model, the empty blue circles are the modeled photometry, and the red squares are the observed photometry; when present, the orange lines are the observed spectra as downloaded from the DJA archive, scaled in magnitude to best match the model spectra.

\end{appendix}

\end{document}